\documentstyle[preprint,aps,eqsecnum,epsfig]{revtex}
\tighten
\newdimen\rotdimen
\def\vspec#1{\special{ps:#1}}
\def\rotstart#1{\vspec{gsave currentpoint currentpoint translate
   #1 neg exch neg exch translate}}
\def\rotfinish{\vspec{currentpoint grestore moveto}}
\def\rotr#1{\rotdimen=\ht#1\advance\rotdimen by\dp#1%
   \hbox to\rotdimen{\hskip\ht#1\vbox to\wd#1{\rotstart{90 rotate}%
   \box#1\vss}\hss}\rotfinish}
\def\RFig#1#2#3{\begin{center} \epsfxsize=#1in
  \setbox0=\hbox to\hsize{\hfil\epsfbox{#2}\hfil} \rotr0
  {\sl #3} \end{center}}
\begin{document}
\draft
\title{\bf ASYMPTOTIC DYNAMICS IN  SCALAR  FIELD THEORY:\\
ANOMALOUS RELAXATION}

\author{{\bf D. Boyanovsky$^{(a)}$, C. Destri$^{(b)}$,
H.J. de Vega$^{(c)}$, R. Holman$^{(d)}$, J. Salgado $^{(c)}$}}
\address
{ (a)  Department of Physics and Astronomy, University of
Pittsburgh, Pittsburgh, PA. 15260, U.S.A. \\
 (b) Dipartimento di Fisica, Universit\`a di Milano
          and INFN, sezione di Milano,
           Via Celoria 16, 20133 Milano ITALIA\\
(c)  LPTHE, \footnote{Laboratoire Associ\'{e} au CNRS UA280.}
Universit\'e Pierre et Marie Curie (Paris VI) 
et Denis Diderot  (Paris VII), Tour 16, 1er. \'etage, 4, Place Jussieu
75252 Paris, Cedex 05, France \\
(d) Department of Physics, Carnegie Mellon University, Pittsburgh,
PA. 15213, U. S. A. }
\date{November 1997}
\maketitle
\begin{abstract}
We analyze the dynamics of dissipation and relaxation in the unbroken
and broken symmetry phases of  scalar theory in the {\bf nonlinear}
regime for large initial energy densities, and  
after {\em linear} unstabilities (parametric or spinodal) are shut-off
by the quantum backreaction. A new time scale emerges that separates
the linear from the non-linear regimes. This scale  is non-perturbative in
the coupling and initial amplitude. The non-perturbative 
evolution is studied within the context of the $O(N)$ vector model in
the large $N$ limit. 
A combination of numerical analysis and the implementation of a dynamical
renormalization group resummation via multitime scale analysis reveals
the presence of unstable bands in the nonlinear regime. These are
associated with {\bf power law} growth of quantum fluctuations, that
result in power law  
relaxation and dissipation with {\bf non-universal and
non-perturbative dynamical anomalous exponents}.   
We find that there is substantial particle production during this
non-linear evolution which is of the same order as that in the linear
regime and results in a non-perturbative distribution. The expectation
value of the scalar field vanishes asymptotically transferring all of the
initial energy into produced particles via the non-linear resonances
in the unbroken symmetry phase. The effective mass squared for the
quantum modes tends asymptotically to a constant plus oscillating $
{\cal O}(1/t) $ terms. This slow approach to asymptotia causes
the power behaviour of the modes which become free harmonic modes for
late enough time.
We derive a simple expression for the equation of  state for the
fluid of produced particles that interpolates between radiation-type and
dust-type equations according to the initial value of the order
parameter for unbroken symmetry. For broken symmetry the produced
particles are asymptotically  massless Goldstone bosons with an
ultrarelativistic equation of state. We find the onset of a novel 
form of dynamical Bose condensation in the collisionless regime in the
absence of thermalization.  

\end{abstract}

\section{Introduction and Motivation}

The next generation of high luminosity heavy ion colliders at
Brookhaven and CERN will offer the possibility of probing the dynamics
of states of high energy density and possibly strongly out of
equilibrium. The energy densities attained for central collisions at
central rapidity will 
hopefully allow to study the quark-gluon plasma and also the chiral phase
transition in a situation that parallels that achieved in the very early
stages of the Universe\cite{book1}-\cite{meyer}. Dynamical phenomena,
non-equilibrium and  collective effects are expected to take place on
time scales of a few tens of fm/c and length scales of few fermis.

This unparalleled short time and length scale regime for dynamical phenomena 
soon to be probed experimentally, has sparked a considerable effort to
study the {\em dynamics} of strongly out of equilibrium situations within
the realm of quantum field theory. 

The usual semi-phenomenological framework to study the dynamics
is based on the transport approach in terms of single (quasi) particle
distribution functions with collisional
relaxation\cite{wang}-\cite{geiger}. The best known  
dynamical processes of relaxation are those of (few body) collisions and
dephasing processes akin to Landau
damping\cite{geigmuller}-\cite{landau,blaizot,weldon}. Our
understanding of these 
relaxational processes is usually based on perturbative expansions,
linearized approximations or small departures from equilibrium. The 
validity of these coarse grained descriptions of relaxational dynamics 
within the realm of high energy and high density regimes in quantum
field theory is not clear and a closer scrutiny of relaxational phenomena
is warranted.

Whereas equilibrium phenomena are fairly well understood and there
are a variety of tools to study perturbative and non-perturbative
aspects, strongly out of equilibrium phenomena are not well understood
and require different techniques.  

Our goal is to deal with the out of equilibrium evolution for {\bf
large} energy densities in field theory.   That is, a large number of
particles per 
volume $ m^{-3} $, where $ m $ is the typical mass scale in the
theory.   The most familiar techniques of field theory, based on the S-matrix
formulation of transition amplitudes and perturbation theory apply in
the opposite limit of low 
energy density and since they only provide information on {\em in}
$\rightarrow$ {\em out} matrix elements, are unsuitable for
calculations of time dependent expectation values.

Recently non-perturbative approaches to study particle
production\cite{cooper},  
dynamics of phase transitions\cite{cooper,boyan1} and novel forms of
dissipation\cite{boyan2} have emerged that provide a promising
framework to study the dynamics for large energy densities like in
heavy ion collisions.  

Similar tools are also necessary to describe consistently the dynamical
processes in the Early Universe\cite{boyan2}. In particular it has
been recognized that 
novel phenomena associated with parametric amplification of quantum 
fluctuations can play an important role in the process of reheating and
thermalization\cite{cosmo,boyan2,usfrw}.  It must be noticed
 that the dynamics in cosmological spacetimes is
dramatically different to the dynamics in  Minkowski spacetime. 
Both in fixed FRW\cite{frw2} and de Sitter\cite{De Sitter} backgrounds and in a
dynamical geometry\cite{din} the dynamical evolution is qualitatively and
quantitatively different to the Minkowski case considered in the present
paper.

Our program to study the dynamical aspects of relaxation out
of equilibrium both in the linear and non-linear regime has revealed
new features of relaxation in the collisionless regime in scalar
field theories\cite{boyan2,inhomo}. Recent investigation of scalar
field theories in the non-linear regime, including self-consistently
the effects of quantum backreaction in an energy conserving and 
 renormalizable framework have pointed out to a wealth of interesting
non-perturbative phenomena both in the broken and unbroken symmetry
phases\cite{boyan2}. These new phenomena are a consequence of the
non-equilibrium 
evolution of an initial state of large energy density which results
in copious particle production  leading to non-thermal and
non-perturbative distribution of particles. Our studies have focused
on the situation in which the {\em 
amplitude} of the expectation value of the scalar field is
non-perturbatively large, 
$A \approx \sqrt{\lambda} <\Phi> /m \approx {\cal O}(1)$ ($m$ is the
mass of the scalar field and $\lambda$ the self-coupling) and most of
the energy of the initial 
state is stored in the `zero mode', i.e. the (translational
invariant) expectation value of 
the scalar field $\Phi$. Under these circumstances the initial energy
density $\varepsilon 
\approx m^4/\lambda$. During the dynamical evolution the energy
initially stored in one 
(or few) modes of the field is transferred to other modes resulting in
copious particle production initially either by parametric
amplification of quantum fluctuations in the 
unbroken symmetry phase, or spinodal instabilities in the broken
symmetry phase. This  
mechanism of energy dissipation and particle production results in a
number of produced 
particles per unit volume ${\cal N} \propto m^3/\lambda$ which for
weak coupling is non-perturbatively large\cite{boyan2}.   We call
`linear regime' to this first stage dominated either by parametric or
spinodal unstabilities.

We recognized\cite{boyan2} a new {\em dynamical} time
scale $ t_1 $ where the linear regime ends. By the time  $ t_1 $
the effects of the quantum fluctuation on the
dynamical evolution become of the same order as the classical contribution
given by the evolution of the expectation value of the field.
The `non-linear' regime starts by the time $ t_1 $.
In the case of broken symmetry, this time scale corresponds to the
spinodal scale at which the backreaction of quantum fluctuations
shut-off the spinodal instabilities.  At this
scale non-perturbative physics sets in and the non-linearities 
of the full quantum theory determine the evolution. This time scale $ t_1 $,
which we call the {\em non-linear time}, is a non-universal feature of
the dynamics and depends strongly on the initial state and
non-perturbatively on the coupling, as $ t_1 \propto
\log[\lambda^{-1}] $ for 
weak coupling\cite{boyan2}. 


The purpose of this paper is to carefully analyze the {\bf nonlinear} dynamics
of relaxation after the  time $ t_1 $ in a weakly coupled scalar field
theory within a non-perturbative self-consistent scheme. We  focus on
the asymptotic time regime both in the unbroken and broken symmetry
states. Here we provide refined numerical analysis of the
non-equilibrium evolution that 
reveal the onset of widely separated relaxational time scales. We  use
a {\em dynamical renormalization group} implemented via a multitime
scale analysis to provide an analytic description of the asymptotic
dynamics and establish that relaxation occurs via  power laws with
anomalous dynamical exponents.  

The main results of this article can be summarized as follows: 

\begin{itemize}

\item The hierarchy of separated time scales allow us to implement a 
 dynamical renormalization group resummation via the method of
 multitime scale analysis. The novel result that emerges from this
 combination 
of numerical and dynamical renormalization group analysis is the presence
of {\em non-linear resonances} that lead to  asymptotic  relaxation
 described by {\em non-universal power laws}.  These power laws are
 determined by {\em dynamical anomalous exponents} which depend
 non-perturbatively  on the coupling.

\item The effective mass felt by the quantum field modes
in high energy density situations varies with time   and depends
on the fields themselves reflecting the nonlinear character of the
dynamics. Both for broken and unbroken symmetry the effective mass 
tends asymptotically to a constant. This
constant is non-zero and depends in the initial state for unbroken
symmetry. For broken symmetry,  the effective mass tends to 
zero  corresponding to Goldstone bosons. In both cases the effective
mass approaches its $ t = \infty $ value as $ 1/t $ times oscillating
functions.

\item The fact that the effective mass tends asymptotically to a constant
implies that the modes becomes effectively {\bf free}. Non-resonant
modes oscillate harmonically for times $ t > t_1 $. 
Resonant modes change from non-universal power behaviour to
oscillatory behaviour at a time that depends on the wavenumber of the mode.
Only the modes in the borders of the band resonate indefinitely. 

\item In the unbroken symmetry case, we find that the expectation
value relaxes to zero asymptotically with 
a non-universal power law. The initial energy density which
is non-perturbatively large goes completely into  production of
massive particles.  The asymptotic particle distribution
is localized within a  band  determined by the initial conditions with
non-perturbatively large amplitude $ \sim 1/\sqrt{\lambda}$ which
could be described as a `semiclassical condensate' in the unbroken
phase. 

\item The particle distribution in the condensate is nontrivial. We
establish sum rules that yield explicit values for integrals over such
asymptotic distribution. We derive in this way the  asymptotic
equation of state. For unbroken symmetry, it interpolates between dust
and radiation according to the initial field amplitude.

\item  
In the broken symmetry phase when the initial expectation value of the
scalar field (order parameter) is close to the false vacuum (at the origin) 
we similarly find the onset of non-linear resonances at times larger
than the  non-linear (spinodal) time $ t_1 $. The expectation value of
the order parameter approaches for late times a nonzero
limit that  depends on the initial conditions.


\item
The effective time dependent mass vanishes as $ 1/t $  resulting in that  the 
asymptotic states are Goldstone bosons. We also find a hierarchy of
time scales of which $t_1 \propto \ln[\lambda^{-1}]$ is the first an
another longer time scale $t_2 \propto 1/\sqrt{\lambda}$.  As a
consequence of the non-linear 
resonances  the particle distribution becomes localized  for $t > t_1$ at very
low momentum resulting  again in a `semiclassical condensate' with
non-perturbatively large amplitude $ \sim 1/\sqrt{\lambda}
$. Asymptotically the equation of state  is that of  radiation
although the particle distribution (Goldstone bosons) is non-thermal.  

\item 
For even larger time scales,   $ t \sim \sqrt{V} $ (where $V$ stands
for the volume of the system),  we find for broken symmetry a novel
form of Bose condensation in the collisionless regime that results
from a linear growth in time of homogeneous quantum fluctuations. 

\end{itemize} 

The article is organized as follows: in section II we briefly summarize
the nature of the approximations, the non-equilibrium framework and
some of the previous results for the benefit of the reader and for
coherence. In section III we study the unbroken symmetry case and
distinguish the linear regime of parametric amplification $ t<t_1 $ from
the non-linear regime ($ t>t_1 $ ) in which the backreaction of
quantum fluctuations dominates the evolution. In section IV we study
the  latter regime in the unbroken symmetry case and we find that 
particle production continues beyond
$ t>t_1 $ and non-linear resonances develop leading to power law
relaxation. We provide a full numerical analysis and implement a 
renormalization group resummation of secular terms via a multi-timescale
approach. Asymptotic sum rules and the equation of state are discussed
in detail. In section V we study the dynamics in the broken symmetry
phase establishing a difference between the early and intermediate 
scales dominated by spinodal instabilities and the asymptotically large
time scale dominated by non-linear resonances leading to power law
relaxation. We provide a numerical analysis as well as arguments based
on multitime scale resummation. We find a novel form of Bose
condensation with a quadratic time dependence for the formation of
an homogeneous condensate.

Conclusions and further questions are summarized at the end of the article.  


\section{Preliminaries}

As our previous studies of scalar field theory have
revealed\cite{boyan2}, there are two very important  
parameters that influence the quantum dynamics: the strength of the
coupling constant $\lambda$ and the initial energy density in units
of the scalar field mass $m$. If in the initial state most of the
energy is stored in few modes, the energy density is determined by
the amplitude of the expectation value of those modes $A \approx
\sqrt{\lambda}<\Phi>/m$. The value of this field amplitude determines 
the regime of applicability of perturbation theory methods. Usual
S-matrix theory treatment (in terms of a perturbative expansion) are
valid in the small amplitude regime $A<<1$ even for high
energies. However, when the initial state has a large energy {\em
density} perturbative methods are invalid.   

We shall be concerned here with the {\bf non-perturbative} regime in 
which $ A =  \sqrt{ \lambda} \; <\Phi>/ m \simeq {\cal{O}}(1)$. It is
important to point out that for 
large field amplitude, even for very weakly coupled theories
non-linear effects will be important and must be treated
non-perturbatively. This is 
the case under consideration. Having recognized the non-perturbative
nature of the problem for large amplitudes we must invoke a
non-perturbative, consistent calculational scheme which respects the
symmetries (continuous global symmetries and energy-momentum
conservation), is renormalizable and lends itself to a numerical treatment. 

We are thus led to  consider  the $O(N)$ vector model with quartic
self-interaction\cite{boyan2} and the scalar field in the vector  
representation of $O(N)$. 

The action and Lagrangian density are given by,
\begin{eqnarray}
S  &=&  \int d^4x\; {\cal L},\label{action} \cr \cr
{\cal L}  &=&   \frac{1}{2}\left[\partial_{\mu}{\vec{\Phi}}(x)\right]^2
-V(\vec{\Phi}(x))\; ,
\cr \cr
V(\vec{\Phi})  &=&  \frac{\lambda}{8N}\left(\vec{\Phi}^2+\frac{2N
m^2}{\lambda}\right)^2 - {{N\;m^4}\over { 2 \lambda}} \quad . \label{potential}
\end{eqnarray}

The canonical momentum conjugate to $ {\vec{\Phi}}(x)$ is,
\begin{equation}
\vec{\Pi}(x) = \dot{\vec{\Phi}}(x), 
\end{equation}
and the Hamiltonian is given by,
\begin{equation}
H(t) = \int d^3x\left\{
\frac12 \vec{\Pi}^2(x)+\frac12\;[\nabla\vec{\Phi}(x)]^2+
\;V(\vec{\Phi})\right\}.
\end{equation}

The calculation of expectation values requires the study of a density
matrix whether or not the initial state is pure or mixed.
Its time evolution in the Schr\"odinger picture  is determined by the
quantum Liouville equation
\begin{equation} \label{liouqua}
i{{\partial  {\hat \rho}}\over { \partial t}} = [ H ,  {\hat \rho}]\; .
\end{equation}
The expectation value of any physical magnitude $ {\cal A} $ is given
as usual by
\begin{equation}
< {\cal A} >(t) = {\rm Tr}[   {\hat \rho}(t)\,  {\cal A}] \; .
\end{equation}
The time evolution of all physical magnitudes is unitary as we
see from eq. (\ref{liouqua}).

In the present case we will restrict ourselves to a translationally
invariant situation, i.e. the density matrix commutes with the total
momentum operator. In 
this case  the order parameter $< {\vec \Phi}(\vec{x}, t) >$ 
will be independent of the spatial coordinates $ \vec{x} $ and only
depends on time. 

We write the field $\vec{\Phi}$ as $\vec{\Phi} = (\sigma, \vec{\pi} )$
where $\vec{\pi}$ represents the 
$N-1$ `pions', and choose the coupling
 $\lambda$ to remain fixed in the large $N$ limit. In what follows, we
will consider two different cases of the 
potential (\ref{potential}) $ V(\sigma, \vec{\pi} ) $, with ($ m^2 < 0 $)
or without ($ m^2 > 0 $) symmetry breaking.

We can decompose the field $\sigma$ into its expectation value and fluctuations
$\chi( \vec{x},t )$ about it:
\begin{equation}
\sigma  (\vec{x},t ) = \phi(t)\; \sqrt{N}+ \chi ( \vec{x},t) \; ,
\end{equation}
with $ \phi(t) $ being a c-number of order one in the $ N \to \infty $
limit and $\chi$ an operator.

To leading order the large $N$-limit is implemented by considering a
Hartree-like factorization (neglecting $ 1/N $terms) and assuming
$O(N-1)$ invariance by  writing
\begin{equation}
\vec{\pi}(\vec x, t) = \psi(\vec x, t)
\overbrace{\left(1,1,\cdots,1\right)}^{N-1}
\end{equation}
 where  $ \psi(\vec x, t) $ is a  quantum operator\cite{boyan2}.
Alternatively the large $ N $ expansion is systematically implemented
by introducing an auxiliary field\cite{cooper}. To leading order the
two methods are equivalent.  

The generating functional of real time non-equilibrium Green's functions can be
written in terms of a path integral along a complex contour in time,
corresponding to forward and backward time evolution and if the
initial density matrix describes a state of local thermodynamic
equilibrium at finite temperature  a branch down the imaginary time axis. 
This requires doubling the
number of fields 
which now carry a label $\pm$ corresponding to forward ($+$), and backward
($-$) time evolution\cite{hu,noneq}. 

We shall not rederive here the field evolution equations for 
translationally invariant quantum states the reader is referred to the
literature.  (See refs. \cite{hu,noneq,eri96}). In the leading order
in the large $ N $ approximation the theory becomes Gaussian at the
expense of a self-consistent condition\cite{cooper,boyan2,eri96}, this in turn
entails that the  Heisenberg field operator $\psi(\vec x,t)$ can be
written as 
\begin{equation} \label{campoH}
\psi(\vec x , t) = \int {{d^3 k} \over {(2\pi)^3}}
\frac{1}{\sqrt{2 }}\left[  \;
a_{ \vec k} \; f_k(t) \; e^{i \vec k \cdot \vec x} + a^{\dagger}_{ \vec
k} \; f^*_k(t) \;  e^{-i \vec k \cdot \vec x}  \; \right],
\end{equation}
where $a_k \, , a^{\dagger}_k$ are the canonical creation and annihilation
operators, the mode functions $f_k(t)$ are solutions of the Heisenberg
 equations of motion\cite{cooper,boyan2,eri96} to be specified below
for each case.  

Our choice of initial conditions on the density matrix is that of
the vacuum for the instantaneous modes of the Hamiltonian at the
initial time\cite{boyan2,eri96}. Therefore we choose the initial
conditions on the mode functions to represent positive energy particle
states of the instantaneous Hamiltonian at $t=0$, which is taken to be
the initial time. That is, 
\begin{equation} \label{condI}
f_k(0)= \frac{1}{\sqrt{W_k}} \; ; \; \dot{f}_k(0) = -i \sqrt{W_k} \; \; ; \; \;
W_k= \sqrt{k^2+M^2_0}\; \; ,  
\end{equation}
 where the mass $M_0$ determines the frequencies $\omega_k(0)$ and will be
defined explicitly later as a function of $ \phi(0) $ [see
eqs.(\ref{initialfreqs2}) and (\ref{stafrequ})].

With these boundary conditions, the mode functions
$f_k(0)$ correspond to positive frequency modes (particles) of the
instantaneous quadratic Hamiltonian for oscillators of mass $M_0$.

We point out that the behaviour of the system depends mildly on the
initial conditions on the mode functions as we have found by varying
eqs.(\ref{condI}) within a wide range. In 
particular, the various types of linear and nonlinear resonances 
are independent of these initial conditions\cite{boyan2,eri96}.

It proves convenient to introduce the following dimensionless quantities:
\begin{eqnarray}
&& \tau = |m|\, t \; \; , \; \;  q = \frac{k}{|m| }
\; \; , \; \; \Omega_q= \frac{W_k}{|m|} \; \; ,\label{tauomega} \\
&&\eta^2(\tau) = \frac{\lambda }{2 |m|^2 }\; \phi^2(t)  \; \; , 
\label{etadef} \\ 
&& g \Sigma(\tau) = 
\frac{\lambda}{2|m|^2 } \left[ \langle \psi^2(t) \rangle_R- \langle
 \psi^2(0) \rangle_R  
\right]  \; \; , \; \; \left( \; \Sigma(0) = 0  \;\right) \label{sigma} \\
&& g = \frac{\lambda}{8\pi^2}  \; \; , \; \;
\varphi_q(\tau) \equiv \sqrt{|m|} \; f_k(t)  \; \; .\label{gren}
\end{eqnarray}
Here  $ \langle \psi^2(t) \rangle_R $ stands for the renormalized
composite operator [see eq.(\ref{sigmafin}) for an explicit expression].

In the $ N = \infty $ limit the field $  \chi ( \vec{x},t) $ decouples
and does not contribute to the equations of motion of either the
expectation value or the transverse fluctuation modes.

\section{Unbroken Symmetry}
\subsection{Evolution equations in the large $N$ limit}

In this case $M^2_R = |M_R|^2$, and in terms of the dimensionless variables
introduced above  the renormalized equations of motion are found to be
(see references\cite{boyan2,eri96}):

\begin{eqnarray}
& & \ddot{\eta}+ \eta+
\eta^3+ g \;\eta(\tau)\, \Sigma(\tau)  = 0 \label{modo0} \\
& & \left[\;\frac{d^2}{d\tau^2}+q^2+1+
\;\eta(\tau)^2 + g\;  \Sigma(\tau)\;\right]
 \varphi_q(\tau) =0 \; , \label{modok}\label{conds1} \\
&& \varphi_q(0) = {1 \over {\sqrt{ \Omega_q}}} \quad , \quad 
{\dot \varphi}_q(0) = - i \; \sqrt{ \Omega_q} \nonumber \\
&&\eta(0) = \eta_0  \quad , \quad {\dot\eta}(0) = 0 \label{conds2}
\end{eqnarray}
Hence, 
\begin{equation} \label{masef}
{\cal M}^2(\tau) \equiv 1+\;\eta(\tau)^2 + g\;  \Sigma(\tau) 
\end{equation} 
plays the r\^ole of a (time dependent)   effective mass squared.  

As mentioned above, the choice of $\Omega_q$ determines the initial state. We
will choose these such that at $t=0$ the quantum fluctuations are in the ground
state of the oscillators at the initial time. Recalling that by definition
$g\Sigma(0)=0$, we choose the dimensionless frequencies to be

\begin{equation}
{\Omega_q}= \sqrt{q^2+1 +  \eta^2_0}. 
\label{initialfreqs2}
\end{equation}

The Wronskian of two solutions of (\ref{modok}) is given by
\begin{equation}
\label{wrfi} 
{\cal{W}}\left[ \varphi_q, {\bar  \varphi_q}\right] = 2 i \; ,
\end{equation}
while $g \Sigma(\tau)$ is given by the self-consistent
condition\cite{boyan2,eri96} 
\begin{eqnarray}
g \Sigma(\tau) & = & g \int_0^{\infty} q^2
dq \left\{ 
\mid \varphi_q(\tau) \mid^2 
-\frac{1}{\Omega_q}   \right. \nonumber \\
 &   & \left. + \frac{\theta(q-1)}{2q^3}\left[ 
 -\eta^2_0 + \eta^2(\tau) + g \; \Sigma(\tau) \right] \right\} \;
. \label{sigmafin} 
\end{eqnarray}
We thus see that the effective mass at time $ \tau $ contains all
$q$-modes and the zero mode at the same time $ \tau $. The evolution
equations are then nonlinear but local in time in the infinite $N$ limit.

\subsection{Particle Number}

Although the notion of particle number is ambiguous in a time dependent
non-equilibrium situation, a suitable definition can be given with respect to
some particular pointer state. We consider two particular definitions that
are physically motivated and relevant as we will see later. The first
corresponds to 
defining particles with respect to the initial Fock vacuum state, while the
second corresponds to defining particles with respect to the
instantaneous adiabatic vacuum state.

In the former case we write the spatial Fourier transform of the
fluctuating field 
$\psi(\vec x,\tau)$ in   (\ref{campoH}) and its
canonical momentum $ \Pi(\vec x, \tau)$ as
\begin{eqnarray}
\psi_q(\tau) & = & \frac{1}{\sqrt{2}}\left[a_q \; \varphi_q(\tau) +
a^{\dagger}_{-q} \; \varphi^*_q(\tau) \right]  \label{psioft} \nonumber \\
\Pi_q(\tau)  & = &   \frac{1}{\sqrt{2}}\left[a_q  \;\dot{\varphi}_q(\tau) +
a^{\dagger}_{-q} \;  \dot{\varphi}^*_q(\tau) \right]  \nonumber
\end{eqnarray}
with the {\em time independent} creation and annihilation operators, such that
$a_q$ annihilates the initial Fock vacuum state. Using the initial conditions
on the mode functions, the Heisenberg field operators are written as
\begin{eqnarray}
\psi_q(\tau) & = & {\cal{U}}^{-1}(\tau) \; \psi_q(0) \; {\cal{U}}(\tau) = 
 \frac{1}{\sqrt{2 \Omega_q}}\left[ \tilde{a}_q(\tau)+
 \tilde{a}^{\dagger}_{-q}(\tau) 
\right] 
\nonumber \\
\Pi_q(\tau)  & = & {\cal{U}}^{-1}(\tau) \; \Pi_q(0) \; {\cal{U}}(\tau) = 
-i \sqrt{\frac{\Omega_q}{2}}
\left[ \tilde{a}_q(\tau)- \tilde{a}^{\dagger}_{-q}(\tau)
\right]  \nonumber \\
\tilde{a}_q(\tau)    & = & {\cal{U}}^{-1}(\tau) \; a_q \; {\cal{U}}(\tau)
 \label{akoft}  \nonumber
\end{eqnarray}
with ${\cal{U}}(\tau)$ the time evolution operator with the boundary condition
${\cal{U}}(0)=1$.  The Heisenberg operators $\tilde{a}_q(\tau)\ ,
\tilde{a}^{\dagger}_q(\tau)$ are related to $a_q, a^{\dagger}_q$ by a
Bogoliubov (canonical) transformation (see
reference\cite{boyan2,eri96} for details). 

The particle number  with respect to the initial Fock vacuum state is
defined in term of the dimensionless variables introduced above as
\begin{eqnarray}
N_q(\tau) &=& \langle \tilde{a}^{\dagger}_q(\tau) \tilde{a}_q(\tau) \rangle 
\cr \cr
&=& \frac{1}{4}\left[
\Omega_q |\varphi_q(\tau)|^2+\frac{|\dot{\varphi}_q(\tau)|^2}{\Omega_q}
\right] -\frac{1}{2}\; . \label{partnumber}
\end{eqnarray}
We consider here zero initial temperature so the 
 occupation number  vanishes at $ \tau = 0 $.

In order to define the particle number with respect to the adiabatic vacuum
state we note that the mode equations (\ref{modok},\ref{modokR}) are those of
harmonic oscillators with time dependent squared frequencies
\begin{equation}
\omega^2_q(\tau)= q^2 + 1+\eta^2(\tau)+g\Sigma(\tau). 
\label{timedepfreqs2}
\end{equation}
 When the frequencies are real (as is the case for unbroken symmetry),
the adiabatic 
modes can be introduced in the following manner:
\begin{eqnarray}
\psi_q(\tau) & = & 
 \frac{1}{\sqrt{2\omega_q(\tau)}}\left[ \alpha_q(\tau) \;
e^{-i\int_0^t\omega_q(\tau') d\tau'}
+\alpha^{\dagger}_{-q}(\tau) \; e^{i\int_0^t\omega_q(\tau') d\tau'} 
\right]  
\label{psiad} \\
\Pi_q(\tau)  & = &-i \sqrt{\frac{\omega_q(\tau)}{2}}
\left[ \alpha_q(\tau) \;  e^{-i\int_0^t \omega_q(\tau') d\tau'}
- \alpha^{\dagger}_{-q}(\tau) \;  e^{i\int_0^t \omega_q(\tau') d\tau'}
\right]   \label{piad}
\end{eqnarray}
where now $\alpha_q(\tau)$ is a canonical operator that annihilates
the adiabatic 
vacuum state, and is related to $a_q, \; a^{\dagger}_q$ by a Bogoliubov
transformation. This expansion diagonalizes the instantaneous Hamiltonian in
terms of the canonical operators $\alpha(\tau) \; , \alpha^{\dagger}(\tau)$.  
The adiabatic particle number is given by
\begin{eqnarray}
N^{ad}_q(\tau) &=& \langle \alpha^{\dagger}_q(\tau) \alpha_q(\tau) \rangle  
 \label{adpart} \\
&=& 
\frac{1}{4}\left[ \omega_q(\tau) \;
|\varphi_q(\tau)|^2+\frac{|\dot{\varphi}_q(\tau)|^2}{\omega_q(\tau)} 
\right] -\frac{1}{2}\; . \nonumber
\end{eqnarray}

 These adiabatic modes and the
corresponding adiabatic particle number have been used previously within the
non-equilibrium context\cite{cooper} and will be very
useful in the analysis of the energy below. Both definitions coincide
at $ \tau = 0 $
because $\omega_q(0)= \Omega_q$.
(For non-zero initial temperature see refs.\cite{boyan2,eri96,cooper}).

It is the  adiabatic definition (\ref{adpart}) of particle number that will
be used in what follows.

The total number of produced particles $ {\cal N}^{ad}(\tau) $ per volume
$|M_R|^3$ is given by:
\begin{equation}\label{nadto}
{\cal N}^{ad}(\tau) \equiv \int {{d^3 q}\over {(2\pi)^3}} \;
N^{ad}_q(\tau) \; .  
\end{equation}
The asymptotic behaviour of the mode functions ensures that this integral
converges\cite{boyan2,eri96}.
 
\subsection{The early time evolution: parametric resonance}

Let us briefly review the dynamics in the weak coupling regime and for times
small enough so that the quantum fluctuations, i.e. $ g\Sigma(\tau) $, are not
large compared to the `tree level' quantities\cite{boyan2,eri96}. 

Since $\Sigma(0) = 0 $, the back-reaction term $ g \Sigma(\tau) $ is
expected to 
be small for small $ g $ during an interval say $ 0 \leq \tau < \tau_1
$. This time $\tau_1$, to be determined below, will be called the
nonlinear time and it determines the time scale when the backreaction 
effects and therefore
the quantum fluctuations and non-linearities become important. 

During the interval of time in which the back-reaction term $ g \Sigma(\tau) $
can be neglected  eq.(\ref{modo0}) reduces to the classical equation of 
motion (in dimensionless variables)
\begin{equation}
 \ddot{\eta}+ \eta+ \eta^3  = 0 \; . \label{classical}
\end{equation}
The solution of this equation with the initial conditions (\ref{conds2})
can be written  in terms of elliptic functions with the result:
\begin{eqnarray}\label{etac}
\eta(\tau) &=& \eta_0\; \mbox{cn}\left(\tau\sqrt{1+\eta_0^2},k\right)
\cr \cr
k &=& {{\eta_0}\over{\sqrt{2( 1 +  \eta_0^2)}}}\; , 
\end{eqnarray}
where cn stands for the Jacobi cosine.  Notice that $ \eta(\tau) $ has period $
4 \omega \equiv {{ 4 \, K(k)}\slash {\sqrt{1+\eta_0^2}}} $, where $ K(k) $ is
the complete elliptic integral of first kind. In addition we note that since
\begin{equation}
 \eta(\tau + 2 \omega ) =  - \eta(\tau) \; , \label{odd}
\end{equation}

$ 1 + \eta^2(\tau) $ is periodic with period $2\omega$.  

Inserting this form for 
$\eta(\tau)$ in eq.(\ref{modok}) and neglecting $ g \Sigma(\tau) $ yields
\begin{equation}\label{modsn}
 \left[\;\frac{d^2}{d\tau^2}+q^2+1+  \eta_0^2\;
\mbox{cn}^2\left(\tau\sqrt{1+\eta_0^2},k\right) \;\right]
 \varphi_q(\tau) =0 \; .
\end{equation}

This is the Lam\'e equation for a particular value of the coefficients that
make it solvable in terms of Jacobi functions \cite{boyan2,eri96}.  

Since the coefficients of eq.(\ref{modsn}) are periodic with period $ 2 \omega
$, the mode functions can be chosen to be quasi-periodic (Floquet type) with
quasi-period $ 2 \omega $.
\begin{equation}\label{floq}
 U_q(\tau + 2  \omega) =   e^{i F(q)} \; U_q(\tau),
\end{equation}
where the Floquet indices $ F(q) $ are independent of $\tau$.  In the allowed
zones, $ F(q) $ is  real  and the functions $  U_q(\tau) $ 
are bounded with a constant
maximum amplitude. In the forbidden zones $ F(q) $ has a non-zero imaginary
part and the amplitude of the solutions either grows or decreases
exponentially.
The mode functions  $\varphi_q(\tau)$ obey the  boundary conditions
eq.(\ref{conds1}) and they are not Floquet solutions. However, they
can be expressed as linear combinations of Floquet solutions
\cite{boyan2,eri96} as follows,
\begin{equation}
\varphi_q(\tau)={1 \over {2 \sqrt{\Omega_q}}}\;\left[
\left(1 - {{2i\Omega_q}\over {{\cal{W}}_q}} \right)\;  U_q(-\tau)+
\left(1 + {{2i\Omega_q}\over {{\cal{W}}_q}} \right)\; U_q(\tau)
\right]\; , \label{combo}
\end{equation}
where $ U_q(0) = 1 $ and
\begin{equation}
{\cal{W}}_q = - 2 q \sqrt{{ {{\eta_0^2}\over 2}+1 + q^2 }\over {
{{\eta_0^2}\over 2}- q^2} }\; . \label{calomega}
\end{equation}

We find {\em two} allowed bands and {\em two} forbidden
bands \cite{boyan2,eri96} for eq.(\ref{modsn}).  In the physical region $q^2>0$ the allowed band 
corresponds to
\begin{equation}\label{bandaq}
 {{\eta_0^2}\over 2} \leq q^2 \leq +\infty \;\; ,
\end{equation}
and the forbidden band  to
\begin{equation}\label{bandap}
0  \leq q^2 \leq  {{\eta_0^2}\over 2} \, .
\end{equation}
The modes in the   forbidden band, $ 0 < q < \eta_0/\sqrt2 $,
 grow exponentially with time (parametric resonance) while those 
in the allowed band, $ \eta_0/\sqrt2 < q < \infty $, oscillate in time
 with constant amplitude. Analytic expressions for all modes were
 given in \cite{boyan2,eri96}. The modes from the forbidden band $ 0 < q <
 \eta_0/\sqrt2 $ dominate $ \Sigma(\tau) $. For $ 0 < \tau < \tau_1 $,
  $ \Sigma(\tau) $ oscillates with an exponentially growing
 amplitude. This amplitude (envelope)  $ \Sigma_{env}(\tau) $
can be represented to a very good approximation by the formula\cite{boyan2,eri96}
\begin{equation}\label{polenta}
 \Sigma_{env}(\tau) = { 1 \over { N \, \sqrt{\tau}}}\; e^{B\,\tau}\; ,
\end{equation}
where $ B $ and $ N $ are functions of $ \eta_0 $ given by
\begin{eqnarray}\label{ByN}
B(\eta_0) &=&  \displaystyle{
8\, \sqrt{1+\eta_0^2}\; {\hat q } \; (1 - 4 {\hat q }) +  O(
{\hat q }^3) }\; , \cr \cr
N(\eta_0)  &=& {4 \over {\sqrt{ \pi}}} \; \sqrt{  {\hat q }}\;
{{ ( 4 + 3 \, \eta_0^2) \, \sqrt{  4 + 5 \, \eta_0^2}}\over{
 \eta_0^3 \, (1+\eta_0^2)^{3/4}}} \left[ 1 + O ({\hat q })\right]\; \; .
\end{eqnarray}
and the elliptic nome $ {\hat q } $ can be written as a function of $
\eta_0 $ as  
\begin{equation}\label{qaprox}
{ \hat q }(\eta_0) =  \frac12 \;  {{ (1+\eta_0^2)^{1/4} 
-  (1+\eta_0^2/2)^{1/4}}
\over { (1+\eta_0^2)^{1/4} +  (1+\eta_0^2/2)^{1/4}}}  \; .
\end{equation}
 with an error smaller than $\sim 10^{-7} $.

Using this estimate for the quantum fluctuations $ \Sigma(\tau) $, we
can now estimate  the value of 
the nonlinear time scale $\tau_1$ at which the back-reaction becomes
comparable 
to the classical terms in the differential equations. Such a time is defined by
$ g \Sigma(\tau_1) \sim (1+\eta^2_0/2) $. From the results presented above, we
find 
\begin{equation}
\tau_1 \approx {1 \over B(\eta_0)} 
\, \log\left[{N(\eta_0)\,(1+\eta^2_0/2) \over { g \,\sqrt{
B(\eta_0)}}}\right]\; .\label{maxtime} 
\end{equation}
 
The time interval from $\tau=0$ to $\tau\sim \tau_1$ is when most of
the particle production takes place. After $\tau \sim \tau_1 $ the quantum
fluctuation become large enough to begin shutting-off the growth of the modes
and particle production slows down dramatically. This dynamical time
scale separates two distinct types of dynamics, for $\tau < \tau_1$ the 
evolution of the quantum modes $ \varphi_q(\tau) $ is essentially linear, 
the backreaction effects are small and particle production proceed via
parametric amplification. Recall that the zero mode $ \eta(\tau) $, obeys the
nonlinear evolution equation (\ref{classical}).
For $ \tau > \tau_1 $ the quantum backreaction
effects are as important as the tree level term $ \eta(\tau)^2 $ and
the dynamics is fully  non-linear.   

We plot in fig. 1 $\tau_1 $ as a function of the initial amplitude $
\eta_0 $ for different values of the coupling $ g $.

The growth of the unstable modes in the forbidden band shows that
particles are created copiously ($\sim 1/g $ for $ \tau \sim
\tau_1 $). Initially, ($\tau = 0$) all the energy is in the classical zero
mode (expectation value). Part of this energy is rapidly transformed
into particles 
through parametric resonance during the interval  $ 0 < \tau < \tau_1$.
At the same time, the amplitude of the expectation value decreases as
is clearly displayed in  fig. 2. 
 We plot in fig. 3 the adiabatic number of  particles [as defined 
by eq.(\ref{nadto})] as a function of time.

The momentum distribution of the produced particles
follows the Floquet index and is peaked at $ q \approx \frac12 \eta_0
\,( 1 - {\hat q})$\cite{boyan2,eri96} this is shown in fig. 4. 

\section{Asymptotic nonlinear evolution}
\subsection{Numerical Analysis}
 
In the previous section we have summarized the dynamical evolution in
the {\em linear} 
regime in which the backreaction effects can be neglected and
estimated the {\em first} 
new, non-perturbative dynamical time scale $\tau_1$ as that beyond
which the dynamics is fully non-linear.

In this section we present the time evolution {\bf after} the
nonlinear time $\tau_1$. That is, when the backreaction $ g \Sigma(\tau) $
is important and  the full solutions to the  non-linear equations
(\ref{modo0}-\ref{conds2}) is needed. We have implemented a refined
numerical treatment   for a wide
range of initial amplitudes and couplings. The numerical method uses a
fourth order 
Runge-Kutta algorithm and 16-point Gauss integrations for the
integrals  over $ q $ and is appended with a FFT analysis to determine
the frequency spectrum 
of the oscillatory component. The precision of our results is better than one
part in $ 10^5 $.

To begin with, we observe that  $ g \Sigma(\tau) $ and $ \eta^2(\tau) $
oscillate with the same frequency and {\bf opposite} phase. Thus, a
remarkably cancellation takes place between these two terms in the
effective mass squared. This phase opposition is analogous to  
Landau damping\cite{landau}. One sees such cancellation comparing fig. 2 for $
\eta(\tau) $, fig. 5 for $ g \Sigma(\tau) $ and figs. 6 for  $ {\cal
M}^2(\tau) $.

Moreover, we see that  $ {\cal M}^2(\tau) $ tends to a constant value
for $ \tau \to \infty $. We find numerically that this value turns out to be
\begin{equation}
{\cal M}^2_{\infty} = 1 + {{\eta_0^2}\over 2}
\label{minfi}   
\end{equation}
for the values of $ g $ and $ \eta_0 $ considered in figs. 1-12.
(up to corrections of order $ g $ that are beyond our numerical precision).
It must be noticed that $ {\cal M}^2_{\infty} $
 coincides with the lower border of the allowed band (\ref{bandaq}). 

Furthermore $ {\cal M}^2(\tau) $ approaches its asymptotic limit (\ref{minfi})
oscillating with decreasing amplitude. 
More precisely, using a detailed numerical analysis of the asymptotic behavior
and fast Fourier transforms we find from our
numerical results for $\tau > \tau_1$ [see figs.6a and 6b],
\begin{equation}\label{masin2}
 {\cal M}^2(\tau) = {\cal M}^2_{\infty} + {{p_1(\tau)}\over {\tau}} +
 {\cal O}\left({1 \over {\tau^2}}\right) \label{masasimpto}
\end{equation}
with 
\begin{equation}
p_1(\tau) = K_1 \cos[ 2 \,{\cal M}_{\infty} \; \tau 
+ 2 a_2 \log(\tau/\tau_1) + \gamma_1 ] +
K_2 \cos[ 2 \, {\cal M}_0\; \tau +  2 b_2 \log(\tau/\tau_1)+\gamma_2 ]
\; ,\label{funp1} 
\end{equation}
where $ K_1, \, K_2 , \,
  \gamma_1 $ and $ \gamma_2 $ are constants and an excellent numerical
fit for the coefficients $ a_2 $ and $ b_2 $ is given by

\begin{eqnarray}\label{coefa}
a_2 &\approx& 0.16 \ln{1 \over g} + 0.6 \nonumber \\ 
b_2 &\approx & 0.6-0.16 \ln{1 \over g}
\end{eqnarray} 
within a wide range of (weak) couplings and initial values of $\eta(0)$.
We also find that the coefficients $K_1;\; K_2$ vary linearly with
$\ln({1 \over g})$ a result that will be obtained self-consistently
below. 

\subsubsection{The evolution of the expectation value and mode functions}

Since the effective mass tends asymptotically to the constant value 
$ {\cal M}_{\infty} $, the expectation value $ \eta(\tau) $ oscillates
with frequency $ {\cal M}_{\infty} $ and the $q$-modes
$\varphi_q(\tau)$ with frequency  
\begin{equation} 
\omega(q) \equiv \sqrt{ q^2 + {\cal M}^2_{\infty}} \; \; . \label{asintofreq}
\end{equation}
[Notice that $ \omega(q) = \omega_q(\tau=+\infty) $].
These oscillation frequencies are confirmed by the numerical analysis
of the evolution of the expectation value and the $q$-modes. 
Figs. (7 a-c) display the momentum distribution of the created
particles at different times. One of the noteworthy features is that
whereas up to time $\tau \approx \tau_1$ the distribution only has
one peak at the value of maximum Floquet exponent, for larger times 
the backreaction effects introduce new structure and oscillations, 
keeping the   {\em borders} of the band fixed
 throughout the evolution in the non-linear regime. 

The numerical results displayed in figs. (7 a-c) show that the
position of the  main peak $ q_0(\tau) $, 
decreases with time. We performed a numerical fit for the time
dependence of the peak position and found its behavior to be well
described by the estimate: 
\begin{equation}\label{estpico}
q_0^2(\tau) \approx  {{K_1}\over {\tau}} \;.
\end{equation}
with the constant $K_1$ introduced in equation (\ref{funp1}) above. 
We will provide an analytic, self-consistent description of this
behavior below.

$q$-modes above and below $ q_0 $ behave quite differently. Modes
with $ q > q_0(\tau) $  oscillate in time with constant amplitude. Modes
with    $ q < q_0(\tau) $ also oscillate but with increasing
amplitude. 

Eq.(\ref{estpico}) shows that 
the peak position in $ q $ decreases monotonically as $ \sim
1/\sqrt{\tau} $. As time evolves, more and more $q$-modes cross the
peak and become purely oscillatory. Only the amplitude of the $ q
\equiv 0 $ mode 
(which is not to be confused with the expectation value  $ \eta(\tau)
$), keeps  growing. As we shall discuss in detail  below, there is a 
band of {\bf non-linear} unstability for $ 0 < q < q_0(\tau) $. 
We also find numerically, that a second non-linear resonance band
appears just below $ q = 
\eta_0/\sqrt2 $ for $ q_1(\tau) < q <
\eta_0/\sqrt2 $, and we find numerically that
\begin{equation}
q^2_1(\tau) \approx \frac{\eta^2_0}{2}- \frac{\sqrt{2}\;K_2}{\tau}\; . 
\label{funq2}
\end{equation}
However, the growing modes in this upper band give a much less important 
contribution to the physical magnitudes than the first band. 

The modes in between, $  q_0(\tau) < q <  q_1(\tau) $ ,
oscillate for times $ \tau > \tau_1 $ with stationary amplitude
$M_q(\tau) $. 

This crossover behaviour of the modes can be expressed by introducing a 
$q$-dependent time scale beyond which the modes become oscillatory. 
Such scale is given by
\begin{equation}\label{taub1}
\tau_I(q) =  {{K_1}\over {q^2}}
\end{equation}
for the lower nonlinear band and
\begin{equation}\label{taub2}
\tau_{II}(q) =  {{K_2}\over{\frac{\eta^2_0}{2}- q^2}}
\end{equation}
for the upper nonlinear band.

Both $  \varphi_{q=0}(\tau) $ and $ \eta(\tau) $ obey
the equation (\ref{modo0})and are linearly independent solutions,
their difference arising from the initial conditions. 
 $  \varphi_{q=0}(\tau) $ has growing amplitude  while the amplitude
of $ \eta(\tau) $ decreases with time. Since these are linearly
independent solutions of the same equation   their Wronskian is a
non-vanishing constant. Therefore  if one solution 
grows, the other independent solution must decrease in order to
respect the wronskian condition (\ref{wrfi}). Since the total energy
is conserved $ \eta(\tau) $ must  necessarily be a decreasing solution.

The fact that the zero-mode amplitude $ \eta(\tau) $ vanishes for 
$\tau = \infty $  implies that {\bf all } the available energy
transforms into particles for $ \tau = \infty $. This conclusion which will
be further clarified in what follows is a consequence of the
non-linear dynamics. 
It is the more remarkable because the particles produced are {\em massive} and
therefore there is a threshold to {\em perturbative } particle production. It
will be seen in detail below that the particle production in this regime is
a truly non-perturbative phenomenon associated with non-linear resonances. 

\bigskip

For $\tau >\tau_1, \tau_I(q),\tau_{II}(q) $ the effective mass squared 
tends to a constant [see
 eq.(\ref{minfi})], therefore, the asymptotic behavior of $\varphi_q(\tau) $ is
given by
\begin{equation}\label{fiAB}
\varphi_q(\tau) \buildrel{\tau \to \infty}\over= 
A_q \; e^{i \omega(q) \, \tau} + B_q \;
e^{-i \omega(q) \, \tau} + {\cal O}\left(\frac{1}{\tau}\right)
\end{equation}
with $\omega(q)$ given by eq.(\ref{asintofreq}).
It is then convenient to define the following functions 
\begin{eqnarray}\label{defiAB}
A_q(\tau) \equiv \frac12\; e^{-i \omega(q) \, \tau} \; \left[ 
\varphi_q(\tau) - {i \over { \omega(q)}}\; {\dot \varphi}_q(\tau)
\right] \; ,\cr \cr
B_q(\tau) \equiv \frac12\; e^{+i \omega(q) \, \tau} \; \left[ 
\varphi_q(\tau) + {i \over { \omega(q)}}\; {\dot \varphi}_q(\tau)
\right] \; .
\end{eqnarray}
which for $\tau > \tau_1$ are slowly varying functions of $\tau$, with the
asymptotic limits
\begin{equation}
\lim_{\tau \to \infty}A_q(\tau)=A_q \; \; ; \; \; 
\lim_{\tau \to \infty}B_q(\tau)=B_q \; . \label{asintolimits}
\end{equation}

We can thus express the mode functions $ \varphi_q(\tau) $ in terms of
$ A_q(\tau) $ and $ B_q(\tau) $ as follows,
\begin{equation}
\varphi_q(\tau) = A_q(\tau) \; e^{i \omega(q) \, \tau} + B_q(\tau) \;
e^{-i \omega(q) \, \tau} \; . \label{modeagain}
\end{equation}

We obtain from eq.(\ref{defiAB}) for the square modulus of the modes, 
\begin{equation}\label{modfi1}
|\varphi_q(\tau)|^2 = |A_q(\tau)|^2 + |B_q(\tau)|^2 + 2 |A_q(\tau)\;
B_q(\tau)| \; \cos\left[ 2 \, \omega(q) \, \tau + \phi_q(\tau) \right]
\end{equation}
where we have set
\begin{equation}
A_q(\tau)\;B_q(\tau)^* = |A_q(\tau)\; B_q(\tau)|\; e^{i \phi_q(\tau)}\;.
\end{equation}
The wronskian relation (\ref{wrfi}) implies that the functions 
$A_q(\tau) $ and  $ B_q(\tau) $ are related  asymptotically through 
\begin{equation}\label{bmenosa}
|B_q(\tau)|^2 - |A_q(\tau)|^2 = { 1 \over {\omega(q)}} \; .
\end{equation}

plus terms that vanish asymptotically. 
The virtue of introducing the amplitudes $A_q(\tau) \; ; \; B_q(\tau)$ is that
their variation in $\tau$ is slow, because the rapid variation of the mode
functions is accounted for by the phase.  

Figs. 8-11 show the (scaled) modulus , 
\begin{equation}\label{defiAm}
M_q(\tau) \equiv \sqrt{g}\; \sqrt{|A_q(\tau)|^2 + |B_q(\tau)|^2}
\end{equation}
and $ \phi_q(\tau) $ for some relevant cases. As shown in these figures 
$ M_q(\tau) $ and $ \phi_q(\tau) $ do not exhibit rapid
oscillations with period $ 2 \pi/\omega(q) $ and 
$ 2\pi/{\cal M}_{\infty} $ which are present  in $ \varphi_q(\tau) $ and $
\eta(\tau) $, respectively.  That is as anticipated above, $ M_q(\tau)
$ and $ \phi_q(\tau)$ vary {\bf slowly} with $ \tau $. 

For small coupling $ g $, $ |\varphi_q(\tau)|^2, \; |B_q(\tau)|^2 $
and $ |A_q(\tau)|^2 $ are of order $1/g$ for $ q $ in the forbidden
band and times later than $ \tau_1 $ \cite{boyan2,eri96}. Therefore, 
$M_q(\tau) $ becomes  of order one after the non-linear time scale for
modes inside the band, and is perturbatively small for modes outside the band. 
Moreover, eq.(\ref{bmenosa})
implies that $ |B_q(\tau)|^2 = |A_q(\tau)|^2 [ 1 + O(g) ] $ and for
modes inside the band we can 
approximate eq.(\ref{modfi1}) as follows,
\begin{equation}\label{aprofi}
g |\varphi_q(\tau)|^2 =  M_q(\tau)^2 \left\{ 1 + 
\; \cos\left[ 2 \, \omega(q) \, \tau + \phi_q(\tau) \right]\right\}[ 1
+ O(g) ] \; ,
\end{equation}
for $ 0 < q < \eta_0/\sqrt2 $. This expression is very illuminating because
it displays a separation between the short time scales in the argument of
the cosine, and the long time scales in the modulus $ M_q(\tau) $
and phase $ \phi_q(\tau) $.

We now introduce  slowly varying coefficients for the order parameter
$ \eta(\tau) $. Let us define
\begin{eqnarray}
D(\tau) & \equiv &  \sqrt{ \eta(\tau)^2 + {\dot \eta}(\tau)^2/{\cal
M}_{\infty}^2 } \label{etaampli} \\
\phi(\tau) & \equiv & - {\cal M}_{\infty}\, \tau - \arctan\left[{{ {\dot
\eta}(\tau)}\over { \eta(\tau)}}\right]  . \label{etaphase}
\end{eqnarray}
Using the result that asymptotically the effective time dependent mass
reaches the asymptotic limit ${\cal M}_{\infty}$ and the fact that $\eta(\tau)$
is a real function we write  
\begin{equation}\label{etaDfi}
\eta(\tau) = D(\tau) \; \cos\left[ {\cal M}_{\infty}\, \tau + \phi(\tau)
\right] \left[ 1 +  {\cal O}\left(\frac{1}{\tau}\right) \right] \; .
\end{equation}
We plot in figs. 12 $ D(\tau) $ and $ \phi(\tau) $ as functions of 
$ \tau$.

\bigskip

We begin our numerical analysis  by considering the $ q=0 $ mode function.
After an exhaustive analysis we have found that  $ A_{q=0}(\tau) $ and
$ B_{q=0}(\tau) $ exhibit power behaviour for $ \tau > \tau_1 $ [see
figs. 8-10]. To 
our numerical precision these power laws can be fit by the following form
\begin{equation}\label{modoC}
 A_{q=0}(\tau) \approx \tau^{ia_2}\left[
C_1 \; \tau^{a_1} + C_2 \; \tau^{-a_1} \right]
\quad , \quad B_{q=0}(\tau) \approx \tau^{-ia_2}\left[
C'_1 \; \tau^{a_1} + C'_2 \; \tau^{-a_1} \right]
\; ,
\end{equation}
where  the numerical results yield for the {\em anomalous dynamical
exponents} 
\begin{equation}\label{coefc}
a_1 \approx 0.27 \; \;
\end{equation}
while $ a_2 $ is the {\em same} as in equations (\ref{funp1})-(\ref{coefa}).

The behaviour (\ref{modoC})  appears also in the 
evolution of the expectation value $ \eta(\tau) $ but with the growing
power of $ \tau $ absent ( $ C_1 =  C'_1 \equiv 0 $ ) resulting in
that the order parameter decreases with time exhibiting a logarithmic
phase: 
\begin{equation}
\eta(\tau) = D_0 \; \left({{\tau}\over{\tau_1}}\right)^{-a_1} \; \cos
\left[{\cal M}_{\infty} \tau + a_2 \log(\tau/\tau_1) + f_0 \right]
\left[ 1 +  {\cal O}\left(\frac{1}{\tau}\right) \right] \label{ordpararela}
\end{equation}
where $ f_0 $ is a small constant. Therefore, comparing with eq.(\ref{etaDfi})
we find the remarkable result
that the {\em amplitude} of the expectation value relaxes with a
dynamical power law exponent: 

\begin{equation}\label{dtau}
D(\tau) = D_0 \; \left({{\tau}\over{\tau_1}}\right)^{-a_1} \left[ 1 +
{\cal O}\left(\frac{1}{\tau}\right) \right] \; ,
\end{equation}
and a logarithmically varying phase

\begin{equation}\label{fitau}
\phi(\tau) = a_2 \log \left({{\tau}\over {\tau_1}}\right) + f_0 + {\cal
O}\left(\frac{1}{\tau}\right) \; .
\end{equation}
The time $ \tau_1 $ appears here since it is the natural time scale
for the non-linear phenomena. 

\bigskip

The  $q \neq 0$-modes also grow with time with a power-like behaviour
for $ 0 < q < q_0(\tau)  $ but with a larger power than the $ q = 0 $
mode. [See figs. 8-10]. Such growth is definitely milder than the exponential
increase of the modes inside the forbidden band in parametric resonance.
Our interpretation of this phenomenon is that $ 0 < q < q_0(\tau)  $
is a {\bf non-linear} resonant band. It is not a resonance in a linear
differential equation  (as it is parametric resonance), but a {\bf
new} nonlinear effect which is a consequence of the backreaction of
the quantum fluctuations through $g\Sigma(\tau)$.  

A second non-linear resonance band appears just below $ q =
\eta_0/\sqrt2 $ for $ q_1(\tau) < q <
\eta_0/\sqrt2 $. However, we find numerically that the contribution
from  this upper band to the physical quantities such as particle
production, is much smaller  than the 
first band. The modes in between, $  q_0(\tau) < q <  q_1(\tau) $ ,
oscillate for times $ \tau > \tau_1 $ with stationary amplitude $
M_q(\tau) $. 

The nonlinear resonant bands become narrower as a function of time,
i.e.  $  q_0(\tau) $ and 
$ \eta_0/\sqrt2 - q_1(\tau) $ decrease with time ($ q_1(\tau) $ increases).

The growth of the amplitudes $ M_q(\tau) $ in the nonlinear resonant
bands for a fixed  $ q $ stops when  $ q $ crosses the borders 
 $ q_0(\tau) $ or $ q_1(\tau) $. After that time, such $q$-modes
oscillate with constant   
amplitude this behavior is displayed in  fig. 8d. There is a crossover
for  $ q \sim 
q_0(\tau) $ and for $ q \sim q_1(\tau) $ from monotonic growth to
oscillatory behavior.  

The phase $ \phi_q(\tau) $ exhibits an analogous behaviour (see fig. 9d).  
 
The particle distribution exhibits marked peaks at $ q \approx  
q_0(\tau) $ and at $ q \approx q_1(\tau) $, which are clearly displayed
in  fig. 7. Notice that the peak near $ q \approx q_1(\tau) $ has
a much smaller amplitude. 

The mode exactly at $ q = \eta_0/\sqrt2 $ has an   analogous behaviour
to the $ q = 0 $ mode. [Compare figs. 13 with figs. 8]. We find to 
our numerical accuracy that the amplitudes behave as 
\begin{equation}\label{moder2}
 A_{q=\eta_0/\sqrt2}(\tau) \approx  \tau^{ib_2}\left[E_1 \; \tau^{b_1} + E_2 \;
 \tau^{-b_1}\right]\quad  , \quad B_{q=\eta_0/\sqrt2}(\tau) \approx 
\tau^{-ib_2}\left[E'_1
 \tau^{b_1} + E'_2 \;  \tau^{-b_1}\right]\; .
\end{equation}
The numerical calculations yield for the dynamical exponents 
\begin{equation}\label{coefd}
b_1 \approx 0.19 
\end{equation}
and $b_2$ is the {\em same} exponent as in eqs.(\ref{funp1})-(\ref{coefa}). 

\bigskip

The growth of the $q$-modes in both nonlinear resonant bands
leads to particle
production. We see from fig. 3 that the number of particles continues
to grow after the nonlinear time $ \tau = \tau_1 $. Although this growth 
is {\bf much slower} than before  $ \tau = \tau_1 $, the total  number
of particles produced  {\bf after} the time  $\tau_1$ is substantial
and  turns to be of the same order of magnitude than those produced
before $\tau_1$.  

The adiabatic number of produced particles $ N_q^{ad}(\tau) $, can be
expressed for late times in terms of the mode amplitudes $ M_q(\tau) $
as follows:
\begin{equation}\label{nadinfi}
N_q^{ad}(\tau) = \frac1{2g} \; \omega(q) \; M_q(\infty)^2 +   {\cal
 O}\left(\frac{1}{\tau}\right) +  {\cal O}\left(g\right)
\end{equation}
where we used eqs.(\ref{adpart}), (\ref{fiAB}) and (\ref{defiAm}).
Notice that asymptotically the adiabatic particle number depends solely on the long time scale as the terms containing the  fast oscillating function $
\cos\left[ 2 \, \omega(q) \, \tau + \phi_q(\tau) \right] $ 
cancels out  to  order  $ \tau^0 $
for large $ \tau $. This is one of the important advantages of this
definition of the particle number. 

For $\tau >> \tau_1$ the total number of produced particles approaches its asymptotic value ${\cal N}^{ad}(\infty) $ as 
\begin{equation}\label{ocnum}
  {\cal N}^{ad}(\tau) = {\cal N}^{ad}(\infty) - {G \over {\tau}} +  {\cal
 O}\left(\frac{1}{\tau^2}\right) \; ,
\end{equation}
where 
\begin{equation}\label{ninfi}
g\,{\cal N}^{ad}(\infty)= {1 \over {4 \pi^2}} \int_0^{\eta_0/\sqrt2} q^2 \, dq
\; \omega(q) \;  M_q(\infty)^2 +  {\cal O}\left(g\right)
\end{equation}
and $ G $ is positive. 

The numerical analysis show that $g\, {\cal N}^{ad}(\infty) $ and $g\,
G $ depend very 
little on $ g $ for small $ g < 10^{-3} $. Both 
 $ {\cal N}^{ad}(\infty) $
and $ G $ grow with $ \eta_0 $. Precise numerical fits yield the behavior 
\begin{equation}
  g\, {\cal N}^{ad}(\infty) \sim 0.007 \; \eta_0^{2.8} 
\end{equation}
for a wide range of couplings and $\eta_0$.

At this point we summarize the results from the numerical analysis for
the unbroken symmetry case: 
\begin{itemize}
\item The effective time dependent mass reaches a finite asymptotic
value $ {\cal M}_{\infty} $ in the form given by equations 
(\ref{masasimpto},\ref{funp1})-(\ref{coefa}). This in turn
means that the modes become {\bf free} asymptotically with plane
wave behavior and the non-linear self-consistent coupling between modes
vanishes. The large $ N $ limit yields free modes in the infinite time limit.

\item For weak coupling and for $\tau > \tau_1$ there is a separation
of time scales, with a short time scale corresponding to the oscillations
with frequencies corresponding to a mass $\approx {\cal M}_{\infty}$ and
a longer time scale that depends on $\tau_1$. 

\item The amplitude of the expectation value relaxes with a power  law 
with {\em non-universal} dynamical exponents and logarithmic phases that
vary solely on the long time scale. The expectation value vanishes
asymptotically, despite the fact that its energy is dissipated into 
massive particles for which there are perturbative thresholds for production. 
The relaxation mechanism is {\em non-linear} and clearly 
non-perturbative even at long times.  

\item For $\tau > \tau_1$ there are non-linear resonant bands which
form at the edges of the original band for parametric amplification.
The width of these non-linear resonant bands vanishes asymptotically 
resulting in that all  modes oscillate harmonically for asymptotically 
large time. For $\tau=\infty $ both  unstable non-linear bands
 ( $ 0 < q < q_0(\tau) $ and 
$ q_1(\tau) < q < \eta_0/\sqrt2 $ shrink to zero). The crossover from
power-like to oscillatory behaviour takes place at the $q$-dependent time
scales given by eqs.(\ref{taub1})-(\ref{taub2}).
 \item 
The particle distribution $ N^{ad}_q(\tau) $ has a finite and nontrivial
limit for $ \tau \to \infty $. In particular, a consequence of the
non-linear resonant bands is that   $ N^{ad}_q(\infty) $ will be
peaked at $ q = 0 $. The asymptotic form of the distribution is a function of
the initial conditions and the 
coupling $ g $. In particular, $ N^{ad}_q(\infty) $ is of order $ 1/g $ for
$ q < \eta_0/\sqrt2 $ and it is of order one for $ q > \eta_0/\sqrt2 $.
That is, the support of the particle distribution valid for short
times $ \tau < \tau_1 $ survives for all times including $ \tau =
\infty $. Furthermore, for weak coupling the large number of particles
inside this band allows us to interpret this asymptotic state as a
non-perturbative semiclassical condensate in the unbroken symmetry phase 
that has formed dynamically through the relaxation of the initial energy.

\end{itemize}

\subsection{Asymptotic Analysis I: Perturbation Theory}
In the previous section we presented an exhaustive numerical study of
the evolution of the mode functions and the expectation value. 
In this section we provide an analytic perturbative approach to explain and understand the numerical results. 

In order to study analytically the asymptotic behaviour for late times, 
it is convenient to write the equations for the expectation value and
mode functions as follows:
\begin{eqnarray}
\left[\;\frac{d^2}{d\tau^2}+q^2+
\;{\cal M}^2_{\infty} + w(\tau) \;\right] \varphi_q(\tau) &=&0 \; , 
\label{modmas2} \cr \cr
\left[\;\frac{d^2}{d\tau^2}+
\;{\cal M}^2_{\infty} + w(\tau) \;\right] \eta(\tau) &=&0
\end{eqnarray}
where
\begin{equation}
 w(\tau) \equiv {\cal M}^2(\tau) - {\cal M}^2_{\infty}
={{p_1(\tau)}\over {\tau}} + {\cal O} \left({1 \over {\tau^2}}\right)
\end{equation}
and $p_1(\tau)$ given by eq. (\ref{funp1}) will be treated as a small
perturbation for $\tau >> \tau_1$. 

These equations can be written as integral equations using the proper
Green's function. 
That is,
\begin{eqnarray}\label{eqinte}
\varphi_q(\tau) &=&  A_q \; e^{i \omega(q) \, \tau} + B_q \;
e^{-i \omega(q) \, \tau} \cr \cr
&-&\int_{\tau}^{\infty}d\tau'\; {{\sin{\omega(q)(\tau'-\tau)}}\over {
\omega(q)}} \;  w(\tau') \; \varphi_q(\tau')  \; .
\end{eqnarray}
Here we used the advanced Green's function  that obeys
\begin{equation}
\left[\;\frac{d^2}{d\tau^2}+q^2+
\;{\cal M}^2_{\infty}  \;\right] \left\{ \theta(\tau' -\tau)
{{\sin[{\omega(q)(\tau'-\tau)}]}\over {
\omega(q)}}\right\} = \delta(\tau -\tau') \; .
\end{equation}

Since $  w(\tau) =  {\cal O} \left({1 \over {\tau}}\right) $ we can generate 
the asymptotic expansion for $ \varphi_q(\tau) $ just by iterating 
eq.(\ref{eqinte}). We find,
\begin{eqnarray}\label{eqinte1}
\varphi_q(\tau) &=&   A_q \; e^{i \omega(q) \, \tau} + B_q \;
e^{-i \omega(q) \, \tau} \cr \cr
&-&\int_{\tau}^{\infty}d\tau'\; {{\sin{\omega(q)(\tau'-\tau)}}\over {
\omega(q)}} \; {{p_1(\tau')}\over {\tau'}} 
\; \left[  A_q \; e^{i \omega(q) \, \tau'} + B_q \;
e^{-i \omega(q) \, \tau'} \right] \; d\tau' \cr \cr
&+& {\cal O} \left({1 \over {\tau^2}}\right)\; .
\end{eqnarray}
The integrals here can be performed in closed form up to terms of
$ {\cal O} \left({1 \over {\tau^2}}\right) $ by
using eq.(\ref{funp1}) for $ p_1(\tau) $. 
The result is given by 
\begin{eqnarray}\label{AyBpertu}
\varphi_q(\tau) &=&  \left( A_q\left[ 1 + 
{{K_1 \sin\Psi_1(\tau)}
\over{4\,i\,{\cal M}_{\infty}\; \omega(q)\; \tau}}
+{{K_2 \sin\Psi_2(\tau)}
\over{4\,i\,{\cal M}_0\; \omega(q)\; \tau}}\right]\right.
\cr \cr
&+& {{B_q}\over {8\, \omega(q)\; \tau}}\left\{
K_1 \left[ {{e^{i\left( \Psi_1(\tau)-2 \, \tau\, \omega(q)\right)}}
\over{\omega(q)-{\cal M}_{\infty}}} +
 {{e^{-i\left( \Psi_1(\tau)+2 \, \tau\, \omega(q)\right)}}
\over{\omega(q)+{\cal M}_{\infty}}} \right] \right. \cr \cr
 &+&\left. \left. K_2 
\left[ {{e^{i\left( \Psi_2(\tau)-2 \,\tau\, \omega(q)\right)}} 
\over{\omega(q)-{\cal M}_0}} +
{{e^{-i\left( \Psi_2(\tau)+2 \, \tau\, \omega(q)\right)}}
\over{\omega(q)+{\cal M}_0}} \right] \right\} \right) \; 
e^{i \omega(q) \, \tau}\cr \cr   
&+& \left( B_q\left[ 1 - {{K_1 \sin\Psi_1(\tau)}
\over{4\,i\,{\cal M}_{\infty}\; \omega(q)\; \tau}}
-{{K_2 \sin\Psi_2(\tau)}
\over{4\,i\,{\cal M}_0\; \omega(q)\; \tau}}\right]\right.
\cr \cr
&+& {{A_q}\over {8\, \omega(q)\; \tau}}\left\{
K_1 \left[ {{e^{-i\left( \Psi_1(\tau)-2 \, \tau\, \omega(q)\right)}}
\over{\omega(q)-{\cal M}_{\infty}}} +
 {{e^{i\left( \Psi_1(\tau)+2 \, \tau\, \omega(q)\right)}}
\over{\omega(q)+{\cal M}_{\infty}}} \right] \right. \cr \cr
 &+&\left.\left. K_2 \left[ 
{{e^{-i\left( \Psi_2(\tau)-2 \, \tau\, \omega(q)\right)}}
\over{\omega(q)-{\cal M}_0}} +
 {{e^{i\left( \Psi_2(\tau)+2 \, \tau\, \omega(q)\right)}}
\over{\omega(q)+{\cal M}_0}} \right] \right\}\right)\;e^{-i \omega(q) \, \tau}
\cr \nonumber \\
&+&{\cal O} \left({1 \over {\tau^2}}\right) \; \; . 
\end{eqnarray}
where
$$
\Psi_1(\tau) \equiv 2\,{\cal M}_{\infty} \; \tau +  2 a_2 
\log{{\tau}\over {\tau_1}} + \gamma_1 \; ,
$$
and
$$
\Psi_2(\tau) \equiv 2\,{\cal M}_0 \; \tau + 2 b_2 \log{{\tau}\over {\tau_1}}
+ \gamma_2 \; .
$$

These expressions display resonant denominators for
 $ \omega(q) ={\cal M}_{\infty} $
and $ \omega(q) = {\cal M}_0 $. These resonances correspond to
$ q = 0 $ and $ q = \eta_0/\sqrt2 $, respectively. This perturbative
 approach is  expected to be valid when the first order correction is
 smaller than the zeroth order. A necessary condition for its validity
 is given by  
\begin{equation}
{{K_1}\over{\omega(q)\; \tau \; [ \omega(q)-{\cal M}_{\infty}] }}<1 \; .
\end{equation}

This implies for $ q $ significatively smaller than $ {\cal M}_{\infty} $,
\begin{equation}\label{cond}
 q^2 > {{ K_1}\over {\tau}} \; .
\end{equation}
where, we approximated 
\begin{equation} 
\omega(q) \simeq {\cal M}_{\infty} + {{q^2}\over { 2 \;  {\cal
M}_{\infty}}} \; .
\end{equation}

Thus in the regime where eq.(\ref{cond}) holds  the behavior of the
mode functions 
is {\it oscillatory} and given by (\ref{AyBpertu}). This is in
agreement with the 
numerical results discussed in sec. IV.A. The equality sign in eq.(\ref{cond}) 
yields the  eq.(\ref{estpico}) for the 
peak position which was found numerically and therefore now interpreted as
the result of a resonance condition. Such peaks can be seen also in
the mode functions amplitudes displayed in figs. (8,11). 

The resonance at $ q =  \eta_0/\sqrt2 $ can be treated analogously. 
The necessary condition for the validity of the perturbative approach
is then 
$$
{{K_2}\over{\omega(q)\; \tau \; [ \omega(q)-{\cal M}_0 ] }}<1 \; .
$$
Near   $ q  = \eta_0/\sqrt2 $ we can write,
$$
\omega(q) \simeq {\cal M}_0 + {{\eta_0(q - \eta_0/\sqrt2)  }\over 
{ \sqrt2 \;  {\cal M}_0}} \; . 
$$
Therefore, the oscillatory  behaviour (\ref{AyBpertu}) applies for
\begin{equation}\label{cond2}
{{\eta_0}\over {\sqrt2}} - q >  {{ K_2}\over {\eta_0\;\tau}} \; .
\end{equation}
The equality sign in eq.(\ref{cond2})
yields the peak positions of the mode functions amplitudes near such
resonance:
\begin{equation}\label{q1tau}
q_1(\tau) = {{\eta_0}\over {\sqrt2}} - {{ K_2}\over {\eta_0\;\tau}}
\; .
\end{equation}

These results are in remarkable agreement with our numerical calculations
[see figs. 8,11]. 

The position of the main peak in the particle distributions precisely
 corresponds to the situation where $ q^2 $ is balanced
by the amplitude in the `potential' $ - {{p_1(\tau)}\over {\tau}} $. 
Namely, for $ q^2 > K_1/\tau $, we have oscillating modes and for 
$ q^2 <  K_1/\tau $, resonant (growing) modes. The same argument
 applies to the secondary peak.

\subsection{\bf Asymptotic Analysis II: Multitime Scales }
The perturbative analysis took us a long ways towards understanding
the presence 
of the non-linear resonances associated with the backreaction effects
and revealed the position of these resonances in complete agreement
with the numerical study. 

However to describe the evolution of the modes {\em inside} these bands the
perturbative approach is insufficient and a non-perturbative method of
resumming the potential secular terms associated with the resonances
must be implemented. 

The main observation from the numerical analysis is that for weak
coupling there 
are two widely separated time scales, the short time scale associated with
oscillations with frequency determined by the asymptotic value of the
effective mass, and a long scale associated with the non-linear time
$\tau_1$. This suggests to implement a multitime scale
analysis\cite{nayfeh} which resums the secular 
terms and results in a {\em uniform} expansion. This method implements a
{\em dynamical renormalization group} resummation which was already implemented
successfully in non-equilibrium evolution in quantum field
theory\cite{salgado} and 
previously applied to quantum mechanical problems\cite{betten,egus}.

In this section we implement the method of multitime scales to the
equations for the expectation value and the $q-modes$ which we write
in the following form   

\begin{eqnarray}
&&\left[\frac{d^2}{d\tau^2} + q^2 + 
{\cal M}^2(\tau)\right]\varphi_q(\tau) = 0
\label{newqmodes} \\
&&\left[\frac{d^2}{d\tau^2} + {\cal M}^2(\tau)\right]\eta(\tau) = 0
\label{newzeromode}
\end{eqnarray}
and use the asymptotic behavior of the effective mass squared
obtained from the detailed numerical analysis
${\cal M}^2(\tau)$ for $\tau > \tau_1$  
\begin{eqnarray}
{\cal M}^2(\tau) & = & {\cal M}^2_{\infty} + 
\frac{p_1(\tau)}{\tau} + {\cal O}({1\over{\tau^2}}) \label{asintomass} \\
p_1(\tau) & = & K_1 \cos\left[2{\cal M}_{\infty} \tau 
+ 2 a_2 \ln(\tau/\tau_1) + \gamma_1\right]  \nonumber \\
& + &  K_2 \cos\left[2{\cal M}_{0} \tau + 2 b_2 \ln(\tau/\tau_1) 
+ \gamma_2\right] \label{p1oft}\\
{\cal M}^2_{\infty} & = & 1+ \frac{\eta^2_0}{2}\quad ; \quad 
{\cal M}^2_{0}=  1+ {\eta^2_0}\label{massinfty}\nonumber \\ \nonumber
\end{eqnarray} 
with $\tau_1 \approx \ln\left[{1 \over g} \right]$ [see eq.(\ref{maxtime})]
being the nonlinear time scale. 

As emphasized above,  the non-perturbative dynamics  has generated a
new time scale $\tau_1$ and for weak coupling there are at least two
widely separated time scales, the short time scale corresponding  
to oscillatory behavior  of ${\cal O}({\cal M}^{-1}_{\infty})$ and
the long time scale for non-linear relaxation  of the order of $\tau_1$.  

In order to implement the multitime scale analysis it is
convenient to introduce the small quantitity $\epsilon$ and the following 
two time variables ($T_0 $ and $T_1$) by
\begin{eqnarray}
\epsilon & = & \frac{1}{\tau_1} \quad ; \quad \label{epsipara} 
T_0  =   \tau \quad ; \quad T_1 = \epsilon T_0 =
\frac{\tau}{\tau_1}\nonumber \\
\frac{d}{d \tau} & = & D_0 + \epsilon D_1  \quad ; \quad D_n = 
\frac{d}{dT_n} \quad ; \quad n=0,1 \label{diffop}\nonumber
\end{eqnarray}
and to write $p_1(\tau)/\tau$ in a manner that displays at once the
dependence on the short and long time scales
\begin{eqnarray}
\frac{p_1(\tau)}{\tau} & = & \epsilon \; \Gamma(T_0,T_1)
\label{newp}\nonumber \\
\Gamma(T_0,T_1) & = & \frac{K_1}{T_1} \cos\left[2{\cal M}_{\infty} T_0 
+ 2 a_2 \ln(T_1) + \gamma_1\right]  \nonumber \\
& + &  \frac{K_2}{T_1} \cos\left[2{\cal M}_{0} T_0 + 2 b_2 \ln(T_1) + 
\gamma_2\right] \label{p1oftnew1}
\end{eqnarray}

To ${\cal O}(\epsilon)$ the multitime scale analysis of the asymptotic
time dependence of $\eta \; ; \; \varphi_q$ begins by proposing the
following {\em uniform} perturbative expansion for the solution
\begin{eqnarray}
\eta(T_0,T_1)  & = & \eta^{(0)}(T_0,T_1) + \epsilon \;
\eta^{(1)}(T_0, T_1) + \cdots \nonumber \\
\varphi_q(T_0,T_1) & = &  \varphi_q^{(0)}(T_0,T_1) + \epsilon \;
\varphi_q^{(1)}(T_0, T_1) + \cdots  \label{unifexp}
\end{eqnarray}

\subsection{\bf $q=0$ modes: $ \eta $ and $ \varphi_{q=0}$}

We generically call $f(T_0,T_1)$ both  $\eta$ and
$\varphi_{q=0}$. The only difference between these is that whereas
$\eta$ is always real, $\varphi_{q=0}$ is complex, 
this difference will be accounted 
for in the final form below. Comparing powers of $\epsilon$, we find the following equations for the $q=0$ modes to first order in $\epsilon$

\begin{eqnarray}
&& \left[D^2_0 + {\cal M}^2_{\infty}\right]f^{(0)}(T_0,T_1) = 0 
\label{zerord}\nonumber \\
&& \left[D^2_0 + {\cal M}^2_{\infty}\right]f^{(1)}(T_0,T_1) = 
-\left[2D_0D_1 + \Gamma(T_0,T_1)\right]f^{(0)}(T_0,T_1)  
 \label{firstord}
\end{eqnarray} 
The solution to (\ref{zerord}) is
obviously
\begin{equation}
f^{(0)}(T_0,T_1) = A(T_1)\; e^{i {\cal M}_{\infty}T_0} +B(T_1)\;
e^{-i {\cal M}_{\infty}T_0} \label{solzero} 
\end{equation}
where for $\eta$ the reality condition implies $B(T_1)=A^*(T_1)$. If
the solution of (\ref{firstord}) is sought in terms of the Green's 
function of the differential operator on the left hand side, one finds
that the term proportional to $\cos\left[2{\cal M}_{\infty}T_0\right]$
in $ \Gamma(T_0,T_1) $ would give rise to  secular terms. Therefore 
the condition for a uniform expansion requires that the coefficients of
these secular terms vanish. This  leads to the following differential 
equations for the dependence of the coefficients on the {\em long} time 
scale $T_1$,
\begin{eqnarray}
&& D_1 A -\frac{i\, K_1}{4{\cal M}_{\infty}T_1}
\; e^{i2 a_2\ln(T_1)+i\gamma_1}\;B = 0 
\label{firsteqn}\nonumber \\
&& D_1 B +\frac{i\, K_1}{4{\cal M}_{\infty}T_1}\; e^{-i2
a_2\ln(T_1)+i\gamma_1}\;A = 0  
\label{secondeqn}
\end{eqnarray}

We find the solutions,
\begin{eqnarray}
A(T_1)  & = &  a_{\pm}\; e^{\frac{i}{2}\left[2 a_2 \ln(T_1)+\gamma_1\right]}
\; T^{\pm a_1}_1  \quad , \quad
B(T_1)   =   b_{\pm}\; e^{-\frac{i}{2}\left[2 a_2 \ln(T_1)+\gamma_1\right]}
\;T^{\pm a_1}_1 \label{coeffB} \nonumber\\
a_1 & = & \sqrt{\left( \frac{K_1}{4{\cal M}_{\infty}}\right)^2-
 a_2^2} \label{a1expo} \quad , \quad
b_{\pm}  =  e^{\mp i \delta} a_{\pm} \label{propor} \\
\tan\delta & = & \frac{a_1}{a_2} \label{angle} \nonumber
\end{eqnarray}

This solution confirms the {\em power law} relaxation found numerically and
provides the consistency condition

$$
K^2_1= 16 {\cal M}^2_{\infty}(a^2_1+a^2_2)
$$
  
This condition is verified numerically to our level of precision. In addition
for weak coupling the numerical evidence gives $a^2_1 << a^2_2$ (see
equations (\ref{coefa}) and (\ref{coefc}) leading to  

\begin{equation}
K_1 \approx 4 {\cal M}_{\infty} \; a_2 \label{esti}
\end{equation}

The final form of the solution is  given by a linear
combination of the two independent solutions above, yielding

\begin{eqnarray}
f(T_0,T_1) & = &  T^{a_1}_1 \; C_+ \; \cos\left[{\cal M}_{\infty} T_0+ 
a_2\ln(T_1)+\frac{\gamma_1}{2}+\frac{\delta}{2}\right] 
\nonumber \\
& + & T^{-a_1}_1 \; C_- \; \cos\left[{\cal M}_{\infty} T_0+ 
a_2\ln(T_1)+\frac{\gamma_1}{2}-\frac{\delta}{2}\right]+
{\cal O}(\epsilon) \nonumber
\end{eqnarray}
where for $\eta$ the coefficient $C_{-}$ is real and $C_{+}=0$,
whereas they are 
complex and both $\neq 0$ for $\varphi_{q=0}$. The ${\cal
O}(\epsilon)$ correction quoted 
above is {\em bound} as a function of time as it arises from the
perturbative solution without secular terms.

Therefore we quote the final form of the solutions
\begin{eqnarray}
\eta(\tau)&  = &  D_0\left(\frac{\tau}{\tau_1}\right)^{-a_1} 
\cos\left[{\cal M}_{\infty} T_0+ 
a_2\ln(T_1)+\frac{\gamma_1}{2}-\frac{\delta}{2}\right] 
\nonumber \\
\varphi_{q=0}(\tau) & = &  C_+ \left(\frac{\tau}{\tau_1}\right)^{a_1}  
\cos\left[{\cal M}_{\infty} T_0+ a_2\ln(T_1)+\frac{\gamma_1}{2}
+\frac{\delta}{2}\right] \nonumber \\
&+  &  C_- \left(\frac{\tau}{\tau_1}\right)^{-a_1}  
\cos\left[{\cal M}_{\infty} T_0+ a_2\ln(T_1)+\frac{\gamma_1}{2}
-\frac{\delta}{2}\right] \nonumber
\end{eqnarray}
 In addition, we have checked the constancy of 
the Wronskian
$$
\varphi_{q=0}(\tau)
\dot{\varphi}^*_{q=0}(\tau)-\dot{\varphi}_{q=0}(\tau)\varphi^*_{q=0}(\tau)=
2i = {\cal M}_{\infty}  
\sin\delta \left[ C_+ C^*_- - C^*_+ C_- \right]
$$
leading to the conclusion that neither of the coefficients $C_{\pm}$ can
vanish for $\varphi_{q=0}$. It is a matter of straightforward algebra
to find that these are indeed  solutions of the equations
(\ref{newqmodes}) (for $q=0$) and (\ref{newzeromode}) up to terms that
fall off faster in time. Furthermore, a {\em uniform} perturbative 
expansion in $\epsilon$ can now be carried out to the next order. 

\subsection{\bf $q\neq 0$ modes}
Proposing a uniform $\epsilon$- expansion for the mode functions and
keeping only up to ${\cal O}(\epsilon)$ we find the equation
\begin{equation}
\left[D^2_0 + 2\epsilon D_0D_1 + q^2 + {\cal M}^2_{\infty} + \epsilon
\Gamma(T_0,T_1) \right]\left[\varphi^{(0)}_q(T_0,T_1)+\epsilon \;
\varphi^{(1)}_q(T_0,T_1)\right]=0 \label{qmulti}
\end{equation}

Since $\Gamma$ contains the oscillating factors
$ \cos\left[2{\cal M}_{\infty} T_0\right] $ and $ 
\cos\left[2{\cal M}_{0} T_0\right] $ a naive perturbative expansion in
$\epsilon$ will produce secular terms for $\omega_q \approx {\cal
M}_{\infty}\; , {\cal M}_{0}$, i.e. for $ q \approx 0 \;   
\; , \; {\cal M}^2_{\infty}-{\cal M}^2_{0}= \eta^2_0/2 $, these were
the values for which the resonant denominators in the perturbative
expansion vanish. Therefore for these values of $q$ we must implement
a multi-time scale analysis to resum the secular terms.  

\subsubsection{Small but nonzero $q^2$}
A consistent expansion in $\epsilon$ can be implemented by writing
$q^2=\epsilon \; q^2_1$ with $q_1 \approx {\cal O}(1)$. 
This is a nonrelativistic approximation since then $q^2$ is  much smaller
than ${\cal M}_{\infty}^2$.
The zeroth order solution is clearly
$$
\varphi^{(0)}_q(T_0,T_1) = {\cal A}_q(T_1)\; e^{i {\cal M}_{\infty} T_0} + 
{\cal B}_q(T_1)\; e^{-i{\cal M}_{\infty} T_0} 
$$

Secular terms in a naive perturbative expansion will arise from the
term proportional to $K_1$ in $p_1(\tau)$.

It is convenient to define the coefficients
\begin{eqnarray}
a_q(T_1) & = & e^{-i\left[a_2 \ln(T_1)+\frac{1}{2}\gamma_1\right]} 
\; {\cal A}_q(T_1) \quad , \quad
b_q(T_1)  =  e^{i\left[a_2 \ln(T_1)+\frac{1}{2}\gamma_1\right]} \;
{\cal B}_q(T_1) \label{coefB}
\end{eqnarray}  
Requesting that the coefficients of the secular terms in the perturbative
solution vanish we obtain the following differential equations
\begin{eqnarray}
D_1a_q +  \frac{i a_2 a_q}{T_1}-
\frac{i q_1^2}{2{\cal M}_{\infty}}a_q-\frac{i K_1}{4{\cal M}_{\infty}T_1}b_q & 
= & 0 \nonumber\\
D_1b_q - \frac{i a_2 b_q}{T_1}+
\frac{i q_1^2}{2{\cal M}_{\infty}}b_q+\frac{i K_1}{4{\cal M}_{\infty}T_1}a_q & 
= & 0 \nonumber
\end{eqnarray}
These equations simplify considerably by  introducing the variable $ z $ as
\begin{equation}\label{defz}
 z \equiv { {  q^2 \; T_1} \over { 2\, {\cal M}_{\infty}}} \; .
\end{equation}
Then, eqs.(\ref{coefB}) can be rewritten as two decoupled second order
differential equations:
\begin{eqnarray}
\left[ z {{d^2}\over{dz^2}} + {d \over {dz}} +z -i-2a_2 - {{
a_1^2}\over z} \right]a_q(z) &=& 0 \\ \cr
 \left[ z {{d^2}\over{dz^2}} + {d \over {dz}} +z +i-2a_2 - {{
a_1^2}\over z} \right]b_q(z) &=& 0 \quad . \nonumber
\end{eqnarray}
These are confluent hypergeometric equations with the solutions:
\begin{equation}\label{whit}
{\cal A}_q(z) = z^{i\,a_2 - \frac12}\; 
M_{i\,a_2 - \frac12, \pm a_1}(2iz) \quad ,
\quad {\cal B}_q(z) = z^{-i\,a_2 - \frac12}\; 
M_{-i\,a_2 - \frac12, \pm a_1}(2iz)
\end{equation}
where $ M_{\lambda, \mu}(z) $ stands for a Whittaker function
\cite{gr}. 
We find the asymptotic behavior to be given by\cite{gr} 
$$
{\cal A}_q(z)\buildrel{z \to 0}\over= (2i)^{\frac12 \pm a_1}\; z^{i\,a_2 \pm
a_1} \; ,  
$$
$$
{\cal B}_q(z)\buildrel{z \to 0}\over=  (2i)^{\frac12 \pm a_1}\;
z^{-i\,a_2 \pm a_1} \; , 
$$
and
\begin{eqnarray}
{\cal A}_q(z)&\buildrel{z \to \infty}\over=& (2i)^{-ia_2+\frac12 }\; 
{{\Gamma(1 \pm 2 \, a_1)}\over {\Gamma(1 -ia_2\pm  a_1)}}\; 
e^{i z} \; , \cr \cr
{\cal B}_q(z)&\buildrel{z \to \infty}\over=&(2i)^{ia_2+\frac12 }\; 
{{\Gamma(1 \pm 2 \, a_1)}\over
{\Gamma(1 +ia_2\pm  a_1)}}\; e^{-i z} \; .\nonumber
\end{eqnarray}
In figs. 15 we plot $ | {\cal A}_q(z) | $ 
and the phase of $ e^{-i z} \; {\cal A}_q(z) $ as a
function of $ z $. One sees that the behaviour of the numerically
computed modes in figs. 8c, 8d, 9c and 9d is accurately reproduced.

Using the integral representation for the solutions\cite{gr}, we can
find where the functions $ {\cal A}_q(z) $ and  
$ {\cal B}_q(z) $ oscillate
with $ z $ and where they do not.
$$
 {\cal A}_q(z) = k \; z^{i\,a_2 \pm a_1} \; 
\int_0^1 {{dt}\over t} \; e^{i\left\{2zt +
 a_2\log[{{1-t}\over t}] \right\}} \; \left[t(1-t) \right]^{\pm a_1} 
$$
where $k$ is a constant. This integral has stationary points at:
$$
t = \frac12 \; \left[1 \pm \sqrt{1 - {{2a_2}\over z}} \right] \; .
$$

For $ z > 2a_2 $ the stationary points are real indicating an
oscillatory behaviour whereas they are complex for  $ z < 2a_2 $
implying a non-oscillatory behaviour. We see from eqs.(\ref{defz}) and
(\ref{esti}) that  
the condition $ z =  2a_2 $ precisely corresponds to the peak position 
(\ref{estpico}).

\subsubsection{$q^2 \approx {\cal M}^2_0-{\cal M}^2_{\infty}$}

In this region of momentum we write $q^2 = {\cal M}^2_0-{\cal M}^2_{\infty} 
+ \epsilon q^2_2$ in order to implement a multi-time scale analysis. 
In this case, secular terms in the perturbative expansion
in $\epsilon$ will arise from the term proportional to $K_2$ in $p_1(\tau)$, 
i.e. the term $\cos[2{\cal M}_0 T_0]$. 

Using the same notation as in the previous subsection, the zeroth
order solution is now

\begin{equation}
\varphi^{(0)}_q(T_0,T_1) = {\cal A}_q(T_1)\; e^{i {\cal M}_{0} T_0} + 
{\cal B}_q(T_1)\; e^{-i{\cal M}_{0} T_0} \label{massqso}
\end{equation}

Defining the coefficients
\begin{eqnarray}
a_q(T_1) & = & e^{-\frac{i}{2}\left[2 b_2 \ln(T_1)+
\gamma_2\right]} {\cal A}_q(T_1) \nonumber \\
b_q(T_1) & = & e^{\frac{i}{2}\left[2 b_2 \ln(T_1)+
\gamma_2\right]}  {\cal B}_q(T_1) \nonumber
\end{eqnarray}  
Requesting that the coefficients of the secular terms in the perturbative
solution vanish we obtain the following differential equations
\begin{eqnarray}
D_1a_q + i b_2 \frac{a_q}{T_1}-i \frac{q_2^2}{2{\cal M}_{0}}a_q-
i\frac{K_2}{4{\cal M}_{0}T_1}b_q & = & 0 \nonumber \\
D_1b_q - i b_2 \frac{b_q}{T_1}+i \frac{q_2^2}{2{\cal M}_{0}}b_q
+i\frac{K_2}{4{\cal M}_{0}T_1}a_q & = & 0 \nonumber
\end{eqnarray}

As in the previous case these coupled first order differential equations
reduce to two decoupled confluent hypergeometric equations with
solution:
$$
{\cal A}_q(y) = y^{i\,b_2 - \frac12}\; M_{i\,b_2 - 
\frac12, \pm b_1}(2iy) \quad ,
\quad {\cal B}_q(y) = y^{-i\,b_2 - \frac12}\; 
M_{-i\,b_2 - \frac12, \pm b_1}(2iy)
$$
where
$$
y  \equiv  { {  (\eta_0^2/2 -q^2) } \over { 2 {\cal M}_0}} \;\tau\quad .
$$

The transition from non-oscillatory to oscillatory behaviour takes
place here at $ y = 2 b_2 $ implying that the modes with
$$
q^2 < \frac{\eta^2_0}{2} - 
\frac{4\, b_2\,{\cal M}_{0}}{\tau}
$$
will grow  in time.

The solutions for the mode with $q=\eta_0/\sqrt{2}$ correspond to
exponents
\begin{eqnarray}
\varphi_{q=\frac{\eta}{\sqrt{2}}} & = & \tau^{\pm b_1+ib_2}\nonumber\\
b_1 & = &    \sqrt{\left( \frac{K_2}{4{\cal M}_{0}}\right)^2-
b_2^2} \nonumber
\end{eqnarray}

This analysis leads to the conclusion that as a consistency condition, the constants $ K_1 $ and  $ K_2 $ can be expressed as
follows,
\begin{equation}\label{K1K2}
K_1 \approx 4\,\sqrt{1 + \eta_0^2/2} \; ( 0.16 \; \log g^{-1} + 0.6)
\quad , \quad 
K_2 \approx 4\,\sqrt{1 + \eta_0^2}\; ( -0.16 \; \log g^{-1} + 0.6) \; ,
\end{equation}
where we used eqs.(\ref{coefa}), (\ref{coefc}) and (\ref{coefd}). 

Therefore
 the effective mass squared behaves as
\begin{eqnarray}
{\cal M}^2(\tau) &=& {\cal M}^2_{\infty} + {4 \over {\tau}}\;
\left\{ {\cal M}_{\infty}\; a_2 \; \cos[ 2 \,{\cal M}_{\infty} \; \tau 
+ 2 a_2 \log(\tau/\tau_1) + \gamma_1 ] \right.\cr \cr
&+& \left. {\cal M}_0 \; b_2 \;  
 \cos[ 2 \, {\cal M}_0\; \tau +  2 b_2 \log(\tau/\tau_1)+\gamma_2 ]
\right\} + {\cal O}({1 \over {\tau^2}})\nonumber
\end{eqnarray}

We have confirmed these results numerically within our  precision. 

These results are noteworthy, by implementing a multitime scale analysis
which is the dynamical equivalent of a renormalization group 
resummation\cite{salgado} we have obtained power law relaxation for
the expectation value with {\em non-universal dynamical anomalous 
dimensions}. The logarithmic phases are clearly a consequence of the
$1/\tau$ fall off of the potential $ w(\tau) $ 
 in the mode equations (\ref{modmas2}) just as
in the Coulomb problem. The power laws originate from the resummation
of the secular terms arising from the non-linear resonances, 
a non-perturbative result.  

\subsection{Energy and  Pressure}

The energy-momentum tensor for this theory in Minkowski spacetime is given by
\begin{equation}
T^{\mu \nu} = \partial^{\mu}\vec{\phi}\cdot \partial^{\nu}\vec{\phi} 
 - g^{\mu \nu} \left[ \frac12 \, \partial_{\alpha}\vec{\phi}\cdot
\partial^{\alpha}\vec{\phi} -V(\vec{\phi} \cdot \vec{ \phi})\right]
\label{tmunu} 
\end{equation}

Since we consider translationally as well as  rotationally invariant
states, the expectation value of $ T^{\mu \nu} $ takes the fluid form

\begin{eqnarray}
E &=& { 1 \over {N{\cal{V}} } }
< T^{00}(x)> = { 1 \over {N{\cal{V}} } } <  \frac12 \,  {\dot
{\vec \phi}}^2 +  \frac12 \,(\nabla 
{\vec \phi} )^2  + V(\phi) > \cr \cr
 N{\cal{V}} \; P(\tau)  &=&  < T^{11}(x)> = < T^{22}(x)> = < T^{33}(x)>
=  < \frac13  \, 
(\nabla {\vec \phi} )^2 +  {\dot {\vec \phi}}^2 -  T^{00}(x)> \; , \nonumber
\end{eqnarray}
with all off-diagonal components vanishing. 

Hence,
$$
P(\tau) + E =  { 1 \over {N{\cal{V}} } } <  \frac13  \, (\nabla{\vec
\phi} )^2 +  {\dot {\vec \phi}}^2 > 
$$
takes a particularly simple form.

Both $ E $ and $ P(\tau) + E $ can be expressed in terms of the zero
mode and the $q$-modes. All derivations including the renormalization
procedure can be found in refs.\cite{boyan2,eri96,baacke}. We just quote the final results in the unbroken symmetry case referring the reader to the above references for details.
\begin{eqnarray}
E_{ren} & = & \frac{2 |M_R|^4}{\lambda_R} \left\{
\frac12\dot{\eta}^2 +\frac12 ( 1 + \eta^2){\cal M}^2(\tau)
- \frac{{\cal{M}}^4(\tau)+1}{4} + g\, \left[\varepsilon_F(\tau)
+\frac12 J^{+}(\eta_0)\, {\cal M}^2(\tau) \right.
\right.  \nonumber \\ 
&+&   \left. \left. \frac{{\cal{M}}^4(\tau)}{32}
 +  {{ {\cal M }^4(\tau)}\over 8}\;
\ln\left[\frac12{\cal{M}}(\tau)\right]
+ {\cal C}^{+}(\eta_0) 
\right] \right\} \; ,
\label{renorenergy}
\end{eqnarray}
where\cite{boyan2,eri96},
\begin{eqnarray}
 \varepsilon_F(\tau) & = & 
2 \int_{0}^{\infty} q^2 dq \;{\omega}_q(\tau) \; {N}^{ad}_q(\tau)
\cr \cr
 J^{+}(\eta_0) &=& -\frac{1 + \eta_0^2}4 \left[ 1 + \log \left({\frac{1 +
\eta_0^2}4}\right)\right] \cr \cr
{\cal C}^{+}(\eta_0) &=& -\frac34\, (1 + \eta_0^2)\;  J^{+}(\eta_0) \;
 . \nonumber
\end{eqnarray}

Since energy is conserved (as can be verified explicitly by using the
equations of motion), it is equal to the initial value, given by
\begin{equation}\label{enicial}
E_{ren} =  \frac{2|M_R|^4}{\lambda_R} \;   \varepsilon =
\frac{2|M_R|^4}{\lambda_R} \left\{
\frac12  \eta_0^2 \; \left[   1 +  \frac12 \eta_0^2
\right]\right\} \; .
\end{equation}

The renormalized energy plus pressure takes the form\cite{boyan2,eri96},
\begin{eqnarray}\label{preMen}
P(\tau)_{ren} + E_{ren} &=&  \frac{2|M_R|^4}{\lambda_R} [\varepsilon +
p(\tau) ] =
\frac{2 |M_R|^4}{\lambda_R}\left\{
{\dot \eta}^2 +  g\, \int_0^{\infty} q^2 \; dq \; 
\; \left( \; \mid  {\dot\varphi}_q(\tau) \mid^2  
+ \frac13 \, q^2 \,   \mid {\varphi}_q(\tau) \mid^2   \right.
\right. \cr \cr
-\frac43 \, q &-&  \left. \left. {{{\cal M}^2(\tau)} \over {3 q}}
+ \frac{\theta(q-K)}{12\,  q^3}{{d^2}\over {d\tau^2}}\left[{\cal M }^2(\tau)
\right] \right) \right\} \; .\nonumber
\end{eqnarray}

For times after $ \tau_1 $ we can restrict ourselves to the
contribution from the expectation value $\eta$ and the modes in 
the band $ 0 < q < \eta_0/\sqrt2 $. Modes with $q >  \eta_0/\sqrt2 $
only yield perturbatively small corrections $ {\cal O}(g) $.  
Using eq.(\ref{defiAB}) we can write the integrands in
eqs.(\ref{renorenergy})-(\ref{preMen}) as follows:
$$
g |\varphi_q(\tau)|^2 =  M_q(\tau)^2 \left\{ 1 + 
\; \cos\left[ 2 \, \omega(q) \, \tau + \phi_q(\tau) \right]\right\}[ 1
+ O(g) ] \; ,
$$
$$
g |{\dot \varphi}_q(\tau)|^2 =   \omega(q)^2 \; M_q(\tau)^2 \left\{ 1 - 
\; \cos\left[ 2 \, \omega(q) \, \tau + \phi_q(\tau) \right]\right\}[ 1
+ O(g) ] \; ,
$$
Inserting these expressions in eqs.(\ref{renorenergy})-(\ref{preMen})
yields
$$
\varepsilon = \frac12\dot{\eta}^2 +\frac12 ( 1 + \eta^2){\cal M}^2(\tau)
- \frac{{\cal{M}}^4(\tau)+1}{4} + \int_0^{\eta_0/\sqrt2} q^2 \, dq \;
\left[q^2 +{\cal M}^2(\infty) \right] \; M_q(\tau)^2 + {\cal O}(g)\; .
$$
Taking now the $ \tau \to \infty $ limit yields,
\begin{equation}\label{eninfi}
 \varepsilon = - \frac1{16}\eta_0^4 + \int_0^{\eta_0/\sqrt2} q^2 \, dq \;
\left[q^2 +{\cal M}^2(\infty) \right]  \; M_q(\infty)^2+ {\cal O}(g)\; .
\end{equation}
We analogously find for $ \varepsilon + p(\tau) $,
\begin{eqnarray}
\varepsilon + p(\tau) &=& \int_0^{\eta_0/\sqrt2} q^2 \, dq \;
\left[\frac43 q^2 +{\cal M}^2(\infty) \right] \; M_q(\tau)^2\cr \cr
&-&
\int_0^{\eta_0/\sqrt2} q^2 \, dq \; \cos\left[ 2 \, \omega(q) \, \tau +
\phi_q(\tau) \right] 
\left[\frac23 q^2 +{\cal M}^2(\infty) \right] \; M_q(\tau)^2+ {\cal
O}(g)\; . \nonumber
\end{eqnarray}
For large $ \tau $ the integral containing the oscillating cosinus dies
off. We thus obtain combining both expressions:
\begin{equation}\label{prinfi}
p(\infty) = \frac13\; \int_0^{\eta_0/\sqrt2} q^4 \, dq \;
M_q(\infty)^2+ \frac1{16}\eta_0^4 + {\cal O}(g)\; . 
\end{equation}

\subsection{Sum Rules and the Equation of State}

Although we do not know the analytic form of the particle distribution
for late times [see eq.(\ref{nadinfi})], we are able to compute its
first two moments in the following way.

First, we can express the quantum fluctuations $ g \Sigma(\tau) $ in
terms of the modes using eqs.(\ref{sigmafin}) and (\ref{aprofi}),
$$
 g \Sigma(\tau) = \int_0^{\eta_0/\sqrt2} q^2 \, dq \; M_q(\tau)^2
\left\{ 1 +  \cos\left[ 2 \, \omega(q) \, \tau +
\phi_q(\tau) \right] \right\}  + {\cal O}(g)\; .
$$
For large $ \tau $ the integral containing the oscillating cosinus dies
off. Using now that $ \eta(\infty) = 0 $, ${\cal M}^2(\tau) =
1+\;\eta(\tau)^2 + g\;  \Sigma(\tau) $ and eq.(\ref{minfi})  we obtain
the first sum rule 
$$
 \int_0^{\eta_0/\sqrt2} q^2 \, dq \; M_q(\infty)^2 = \frac12 \eta_0^2
 + {\cal O}(g)\; . 
$$
Furthermore, equating the expression for the energy for $ \tau =
\infty $ [eq.(\ref{eninfi})] with its initial value
[eq.(\ref{enicial})], yields the second sum rule:
$$
\int_0^{\eta_0/\sqrt2} q^4 \, dq \;
M_q(\infty)^2 = \frac1{16}\eta_0^4 + {\cal O}(g)\; . 
$$
 
Combining the sum rules with the expressions for the energy and
pressure, eqs.(\ref{eninfi}) and  (\ref{prinfi}) yields
$$
p(\infty) = \frac1{12}\eta_0^4 + {\cal O}(g)\; . 
$$
and
$$
{{p(\infty)}\over {\varepsilon}}= { 1 \over {3 \;\left( 1 +
\frac2{\eta_0^2}\right)}} + {\cal O}(g)\; . 
$$
We see that the bath of produced particles  does not
behave asymptotically  either as radiation or as nonrelativistic matter but
their equation of state interpolates between these two limits as a
function of the initial amplitude of $\eta$.  

For large $ \eta_0 $, we find as expected, radiation behaviour, 
$$
{{p(\infty)}\over {\varepsilon}} \buildrel{\eta_0 \to \infty}\over= \frac13
+ {\cal O}\left({1 \over {\eta_0^2}}\right) \; .
$$
For small  $ \eta_0 $, we find a cold matter behaviour,
$$
{{p(\infty)}\over {\varepsilon}} \buildrel{ \eta_0 \to
0}\over=\frac16\; \eta_0^2
+ {\cal O}\left(\eta_0^4\right) \to 0 \; .
$$

We notice that the energy and the asymptotic pressure
can be expressed in a form that suggests a two component fluid formed
by nonrelativistic and massless particles:
$$
p(\infty) = 0\, .\, {{\eta_0^2}\over 2} + \frac13 \, .\,  {{\eta_0^4}\over 4}
$$
$$
\varepsilon = {{\eta_0^2}\over 2}+ {{\eta_0^4}\over 4}
$$
$  {{\eta_0^2}\over 2} $ and $ {{\eta_0^4}\over 4} $ may be
interpreted as the contributions of massless and massive particles to
the total energy. 

The pressure approaches its asymptotic limit $ p(\infty) $
oscillating with decreasing amplitude. We find from our numerical
analysis [see figs. 14a and 14b]
that this  amplitude falls off as $ \sim 1/\tau $. More
precisely we find for times later than $ \tau_1 $,
$$
p(\tau) =  p(\infty) +  {{q_1(\tau)}\over {\tau}} +
 {\cal O}\left({1 \over {\tau^2}}\right)
$$
Here, $ q_1(\tau) $ oscillates with time with the same frequencies $ 
 2 \,{\cal M}_{\infty} $ and $ 2 \, {\cal M}_0 $ as the effective mass
 squared [see eq.(\ref{funp1})]. 

\section{Broken Symmetry}

In the case of broken symmetry $M^2_R=-|M^2_R|$ and the field equations in the
$N = \infty$ limit become\cite{boyan2,eri96,brokemil}:

\begin{eqnarray}\label{modo0R}
& & \ddot{\eta}- \eta+
\eta^3+ g \;\eta(\tau)\, \Sigma(\tau)  = 0  \\
& & \left[\;\frac{d^2}{d\tau^2}+q^2-1+
\;\eta(\tau)^2 + g\;  \Sigma(\tau)\;\right]
 \varphi_q(\tau) =0 \label{modokR}
\end{eqnarray}
where $ \Sigma(\tau)$ is given in terms of the mode functions $\varphi_q(\tau)$
by the same expression of the previous case, (\ref{sigmafin}). Here,
 $ {\cal M}^2(\tau) \equiv -1+\;\eta(\tau)^2 + g\;  \Sigma(\tau) $ plays
the r\^ole of a  (time dependent)  renormalized effective  mass squared.

The choice
of boundary conditions is more subtle for broken symmetry. The
situation of interest is when 
$0<\eta^2_0 << 1$, corresponding to the situation where the expectation value
rolls down the potential hill beginning very close to the origin. The
modes with $q^2 < 1- 
\eta^2_0$ are unstable and thus do not represent simple harmonic oscillator
quantum states. Therefore one {\em must} chose a different set of boundary
conditions for these modes. Our choice will be that corresponding to the ground
state of an {\it upright} harmonic oscillator. This particular initial
condition corresponds to a quench type of situation in which the initial
state is evolved in time in an inverted parabolic potential (for early times
$t>0$). Thus we shall use the following initial conditions for the mode
functions:
\begin{eqnarray}
&& \varphi_q(0) = {1 \over {\sqrt{ \Omega_q}}} \quad , \quad 
{\dot \varphi}_q(0) = - i \; \sqrt{ \Omega_q} \label{conds12} \\
&&\Omega_q= \sqrt{q^2 +1 +\eta^2_0} \quad {\rm for} \; \; q^2 <
q_u^2 \equiv 1-\eta^2_0 \label{unsfrequ} \cr \cr
&&\Omega_q=  \sqrt{q^2 -1 + \eta^2_0}  \quad {\rm for} \;  \; q^2
>q_u^2 \quad  ; 
 \quad 0 \leq \eta^2_0 <1 \label{stafrequ} \; . 
\end{eqnarray}
along with the initial conditions for the expectation value given by
eq.(\ref{conds2}). Furthermore because the adiabatic frequencies
cannot be defined for the modes in the spinodal band, we use the
definition eq.(\ref{partnumber}) for the particle number.  

\subsection{The early time evolution: spinodal unstabilities}

As in the unbroken case, for $ g << 1 $ we can neglect $ g \Sigma(\tau) $ in
eqs. (\ref{modo0R}-\ref{modokR}) until the spinodal
time $ \tau_1 $ defined to be the time scale at which the
quantum fluctuations become 
 comparable to the `tree level' terms.
In addition, when the initial value of  $ \eta_0 $ is zero or
much smaller than one, we can neglect $ \eta(\tau)^2 $ in
the mode  equations, which simplify to 
\begin{equation}\label{eqespino}
\left[\;\frac{d^2}{d\tau^2}+q^2-1\;\right] \varphi_q(\tau) =0
\end{equation}
The solution of eqs.(\ref{eqespino}) for the initial conditions
(\ref{conds12}) takes the form
$$
\varphi_q(\tau) = {1 \over {2 \sqrt{1 - q^4}}}\left\{ \left[ 
\sqrt{1 - q^2}-i (1+q^2) \right]\; e^{\tau \sqrt{1 - q^2}} + 
 \left[ \sqrt{1 - q^2}+i (1+q^2) \right]\; e^{-\tau \sqrt{1 - q^2}}
\right\} \; .
$$
for $ 0 < q^2 < 1 $. The modes with higher wavenumber $q^2>1$ do not
grow and their contribution to $g\Sigma(\tau)$ is subleading at long
times. 

This solution exhibits the typical exponential growth of spinodal
unstabilities. Very soon, the decreasing exponential can be neglected
and we can set,
\begin{equation}\label{fiapro}
\varphi_q(\tau) \approx {1 \over {2 \sqrt{1 - q^4}}} \left[ 
\sqrt{1 - q^2}-i (1+q^2) \right]\; e^{\tau \sqrt{1 - q^2}}
\end{equation}
to a very good approximation. 

Using such mode functions we can estimate the quantum fluctuations $ g
\Sigma(\tau) $ for short times. Inserting $ \varphi_q(\tau) $ given by
eq.(\ref{fiapro}) in eq.(\ref{sigmafin}) the integral is dominated by
$ q = 0 $ for growing times $ \tau $,
\begin{equation}
\Sigma(\tau) \approx \frac12 \int_0^{1-0} {{q^2 \, dq}\over { 1 - q^4}} \;
\left[ 1 + \frac{q^2}2(1 + q^2) \right] \; e^{2\tau \sqrt{1 - q^2}}
 \approx {{\sqrt{\pi}}\over { 8 \; \tau^{3/2}}} \; e^{2\tau} \left[ 1 +
{\cal O}\left( \frac1{\tau} \right)\right] \; .
\end{equation}
Using this estimate for the quantum fluctuations $ \Sigma(\tau) $, we
can now find  the value of 
the spinodal time scale $ \tau_1 $ at which the back-reaction becomes
comparable 
to the classical terms in the differential equations. Such a time is defined by
$ g \Sigma(\tau_1) \sim 1  $. From the results presented above, we
find 
\begin{equation}\label{tiesp}
\tau_1 \approx \frac12 \log\left[ \sqrt{8 \over {\pi}}\frac1{g}\right]
+ \frac34  \log \log\left[ {8 \over {\sqrt{\pi}\;g}}\right] + \ldots \; .
\label{tspin} 
\end{equation}
 
The time interval from $\tau=0$ to $ \tau\sim \tau_1 $ is when most of
the particle production takes place. After $\tau \sim \tau_1 $ the quantum
fluctuation shut-off the exponential growth of the modes
and particle production slows down dramatically.

We list here  the values of the  spinodal time according to
eq.(\ref{tiesp}) and
the corresponding position of the maximum of $g\Sigma(\tau)$ for
different values of $ g=10^{-n} $ and a fixed $\eta_0=10^{-4}$.
\begin{center}
\begin{tabular}{|c|c|c|}\hline
 $n=-\log_{10}g$ & Spinodal time ($\tau_1$)& Numerical Result \\
  & estimate  & for $\tau_1$ \\ \hline\hline
 5 & 7.85 & 7.87 \\ \hline
 6 & 9.14  & 9.14 \\ \hline
 7 & 10.40 & 10.40 \\ \hline
 8 & 11.65 & 11.40 \\ \hline
 9 & 12.89 & 12.87 \\ \hline
 10 & 14.14& 14.10 \\ \hline
 12 & 16.55 & 16.90 \\ \hline
 14 & 19.00 & 19.29 \\ \hline
\end{tabular}
 \end{center}

We see that the two values are equal up to $2\%$.

\subsection{Asymptotic nonlinear evolution  I: numerical analysis}

We present in this section the numerical analysis for the time
evolution {\bf after} the spinodal time $ \tau_1 $ when the quantum
backreaction from $ g \Sigma(\tau) $
becomes important and a full solution of the  non-linear equations
(\ref{modo0R}-\ref{modokR}) is required. We have carried out a numerical
integration of the integro-differential equations for a wide
range of initial amplitudes and couplings implementing the same
algorithms and with the same precision as in the unbroken symmetry
case before.  In the broken symmetry case we find numerically that the
effective mass squared  
\begin{equation}
{\cal M}^2(\tau) \equiv -1+ \;\eta(\tau)^2 + g\;  \Sigma(\tau)
\end{equation} 
vanishes for $ \tau \to \infty $. 

A detailed analysis that includes FFT, reveals that $ {\cal M}^2(\tau) $ tends to zero oscillating with decreasing amplitude. 
For $\tau > \tau_1$ a very precise fit is obtained in the form 
[see fig. 15], 
\begin{equation}\label{masinR}
 {\cal M}^2(\tau)  =\frac A{\tau}\;s(\tau)\;\cos\left[2\tau+2 \,
c_2 \, s(\tau) \, \log\left(\frac{\tau}{\tau_1}\right) +\gamma\right]
+ {\cal O}(\frac{1}{\tau^2})
\end{equation}
where $ A ,\, c_2 $ and $ \gamma $ are constants and 
$ s(\tau) $ is a slowly decreasing function with $ s(\tau_1) \approx 1 $.
This slowly varying function can be well approximated by:
\begin{equation}
s(\tau) \approx e^{-(\tau/\tau_2)^x} \; .
\end{equation}
For small initial amplitudes ($ \eta_0 < 10^{-3} $), $ A $ and $ \tau_0 $ are
independent of  $ \eta_0 $ and their dependence  on $ g $ can be
numerically fit as follows, 
\begin{equation}\label{newfits}
x \approx 0.25\quad , \quad \tau_0 \sim 1/\sqrt{g}
\quad , \quad A \approx
0.64\,\log{\frac1{g}} - 1.9 \; , \;
c_2  \approx 0.6 - 0.16 \, \log\frac1{g} \; .
\end{equation}
for small couplings ($ g < 10^{-4} $).

These expressions reveal that a {\em second} dynamical time scale
 $ \tau_2 $ emerges in the broken symmetry case. Furthermore, this new
 scale is widely separated from the non-linear scale, i.e. $ \tau_2  
\gg \tau_1 $ for weak coupling and small initial amplitude.

We also notice that the frequency of the oscillation of the effective
mass is given by the initial mass (for very small $\eta_0$), which
is in agreement with the situation in the unbroken phase [see
eq.(\ref{funp1})] since in this broken symmetry case the effective
mass vanishes asymptotically. 

Power law relaxation was also reported in\cite{brokemil} for the
broken symmetry phase.  

\subsection{The evolution of the mode functions}

Since the effective mass vanishes asymptotically,
the $q$-modes $\varphi_q(\tau)$ oscillate as
\begin{equation}\label{fiABR}
\varphi_q(\tau) \buildrel{\tau \to \infty}\over= 
A_q \; e^{i q \, \tau} + B_q \;
e^{-i q \, \tau} + {\cal O}\left(\frac{1}{\tau}\right)
\end{equation}
behavior that is confirmed in our numerical analysis. 

As for the unbroken symmetry case, it is convenient to introduce
slowly varying amplitudes $ A_q(\tau) $ and $ B_q(\tau) $, 
\begin{eqnarray}\label{defiABR}
A_q(\tau) \equiv \frac12\; e^{-i q \, \tau} \; \left[ 
\varphi_q(\tau) - {i \over { q}}\; {\dot \varphi}_q(\tau)
\right] \; ,\cr \cr
B_q(\tau) \equiv \frac12\; e^{+i q \, \tau} \; \left[ 
\varphi_q(\tau) + {i \over { q}}\; {\dot \varphi}_q(\tau)
\right] \; .
\end{eqnarray}
to separate the evolution on short time scales from that on the long
time scales.  
We can thus express the mode functions $ \varphi_q(\tau) $ in terms of
$ A_q(\tau) $ and $ B_q(\tau) $ as follows,
\begin{equation}
\varphi_q(\tau) = A_q(\tau) \; e^{i q \, \tau} + B_q(\tau) \;
e^{-i q \, \tau} \; .
\end{equation}
We find from eqs. (\ref{fiABR})-(\ref{defiABR}),
\begin{equation}
\lim_{\tau \to \infty}A_q(\tau)=A_q \; ,
\end{equation}
\begin{equation}
\lim_{\tau \to \infty}B_q(\tau)=B_q \; .
\end{equation}
We obtain from eq.(\ref{defiABR}) for the square modulus of the modes, 
\begin{equation}\label{modfi1R}
|\varphi_q(\tau)|^2 = |A_q(\tau)|^2 + |B_q(\tau)|^2 + 2 |A_q(\tau)\;
B_q(\tau)| \; \cos\left[ 2 \, q \, \tau + \phi_q(\tau) \right]
\end{equation}
where we set
\begin{equation}
A_q(\tau)\;B_q(\tau)^* = |A_q(\tau)\; B_q(\tau)|\; e^{i \phi_q(\tau)}\;.
\end{equation}
The wronskian relation (\ref{wrfi}) and eq.(\ref{fiABR}) imply that
the functions $ A_q(\tau) $ and  $ B_q(\tau) $ are related through
\begin{equation}\label{bmenosaR}
|B_q(\tau)|^2 - |A_q(\tau)|^2 = { 1 \over q} \; .
\end{equation}
plus terms that vanish fast asymptotically. 
We plot in fig. 18 the (scaled) modulus as a function of time
\begin{equation}\label{defiAmR}
M_q(\tau) \equiv \sqrt{g}\; \sqrt{|A_q(\tau)|^2 + |B_q(\tau)|^2}
\end{equation}
and $ \phi_q(\tau) $ for some relevant cases. This figure shows that
as anticipated, $ M_q(\tau) $ and $ \phi_q(\tau) $ vary {\bf slowly}
with $ \tau $. 

For small coupling $ g $, $ |\varphi_q(\tau)|^2, \; |B_q(\tau)|^2 $
and $ |A_q(\tau)|^2 $ are of order $1/g$ for $ q $ in the spinodal
band $ 0 \leq q \leq 1 $
and times later than $ \tau_1 $ \cite{boyan2,eri96}. Therefore, 
$M_q(\tau) $ is in that case of order one. Moreover, eq.(\ref{bmenosaR})
implies that $ |B_q(\tau)|^2 = |A_q(\tau)|^2 [ 1 + O(g) ] $ and we can
approximate eq.(\ref{modfi1R}) as follows,
\begin{equation}\label{aprofiR}
g |\varphi_q(\tau)|^2 =  M_q(\tau)^2 \left\{ 1 + 
\; \cos\left[ 2 \, q \, \tau + \phi_q(\tau) \right]\right\}[ 1
+ O(g) ] \; ,
\end{equation}
for $ 0 < q < 1 $.

We see from the numerical results (fig. 16) that the momentum
distribution of produced particles evolves with time. In particular,
the position of its peak $ q_B(\tau) $, decreases with time. From
the  numerical
results we can fit  the peak position by the form:
\begin{equation}\label{estpicoR}
q_B(\tau) \approx  {{r_1}\over {\tau-\tau_B}}  \quad , \; \tau >  \tau_1 \; .
\end{equation}
where $ r_1 \approx 3.8 $ for $ g < 10^{-4} , \; \eta_0 < 10^{-5} $
and $ \tau_B \sim \tau_1 $.

Since $  {\cal M}^2(\infty) = 0 $, the asymptotic equation
for the zero mode, $ {\ddot  \eta}(\tau) = 0 $, only admits a constant
as bounded solution. This is precisely what the 
numerical analysis yields.

We find two non-linear resonant bands for $ \tau>\tau_1 $. The first
for $ 0 < q < q_B(\tau) $ and the second for $ q_C(\tau) < q < \sqrt{1 -
\eta_0^2} \approx 1 $. The value of $ q_C(\tau) $ follows by
perturbation analysis [see next section]:
\begin{equation}\label{estpico2}
q_C(\tau)  \approx 1 - { A \over { \tau-\tau_C}}\; s(\tau)
 \quad , \; \tau >  \tau_1 \; .
\end{equation}
where $ \tau_C   \sim \tau_1 $.

We see from eqs.(\ref{estpicoR}) and (\ref{estpico2}) that both unstable
bands shrink to zero for $ \tau \to \infty $.
The modes in these nonlinear resonant bands grow as powers of 
time. The mode exactly at $ q = 0 $ is not resonant. $\eta(\tau)$
tends to a constant value for $ \tau \to \infty $ while $
\varphi_{q=0} (\tau) $ grows linearly with the time. 

The order parameter  $\eta(\tau)$ tends for late times to a nonzero
limit that  depends on the initial conditions.

The modes between the unstable bands [$q_B(\tau) < q <  q_C(\tau) $]
oscillate in time with constant amplitude [see figs. 19].

Most of the particles are produced just before the time $ \tau_1 $
(see fig. 17). Then, the particle number oscillates with decreasing
amplitude around its asymptotic limit. This limiting value is
proportional to $ g^{-1} $ and independent of $ \eta_0 $ for small $
\eta_0 $ and $ g $ and approximately given by:
\begin{equation}
{\cal N}(\infty)\approx {{0.330}\over{4 \, \pi^2 \, g} }
\end{equation}
In the present case the particle number saturates around $ \tau_1 $.

The vanishing of the effective mass in the $ \tau \to \infty $ limit
implies that the $q$-modes become effectively free. As for unbroken
symmetry [see sec. IIIA], the system tends asymptotically to a limit
cycle. All  modes oscillate harmonically
for large enough
times, since for $\tau=\infty $ both  unstable bands ($ 0 < q <
q_B(\tau) $ and $ q_C(\tau) < q < 1 $ shrink to zero). In addition,
the expectation value  tends to a small constant value $ \eta(\infty)=
\eta_{\infty} $.

In the present broken symmetry case we find that the mode $ q = 0 $
 grows linearly in time, whereas modes with 
 $ q = \sqrt{1 - \eta_0^2} \approx 1 $, grow with a power law for $
 \tau >\tau_1 $ approximately given by 

\begin{equation}\label{modoq1}
A_{q=1}(\tau) \approx \tau^{ic_2}\left[
D_1 \; \tau^{c_1} + D_2 \; \tau^{-c_1} \right]
\quad , \quad B_{q=1}(\tau) \approx  \tau^{-ic_2}\left[
D'_1 \; \tau^{c_1} + D'_2 \; \tau^{-c_1} \right] \; , \label{brokenampli}
\end{equation}
(see fig. 18) with
\begin{equation}
c_1 \approx 0.24 \; \; ; \; \; 
c_2  \approx 0.6 - 0.16 \, \log\frac1{g} \; .
\end{equation}
A more  precise numerical fit yields that the dependence of $ c_1 $ 
on $ g $ can be bounded by
\begin{equation}
{{\Delta c_1}\over{\Delta \log{g^{-1}}}} < 4. \, 10^{-3} \; .
\end{equation} 

\subsection{Asymptotic nonlinear evolution  II: perturbative and
multitime scale analysis} 

As in the unbroken symmetry case, we can determine analytically the
position of the resonances by performing a perturbative analysis in a
similar manner, in terms of the advanced Green's function. 

The effective mass $ {\cal M}^2(\tau) $  vanishes for $ \tau \to
\infty $ in the broken symmetry case [see eq.(\ref{masinR})]. Because
the function $s(\tau)$ varies on even longer time scales of order
$1/\sqrt{g}$ we can perform a perturbative expansion on intermediate
asymptotic scales, between $\tau_1$ and the much longer time scale 
associated with $s(\tau)$. In this intermediate asymptotic regime we
can consider $s(\tau)$ to be constant. 

We can thus study the mode equations
\begin{equation}
\left[\;\frac{d^2}{d\tau^2}+q^2+ {\cal M}^2(\tau)\;\right]
 \varphi_q(\tau) =0
\end{equation}
for late times considering $ {\cal M}^2(\tau) $ as a small
perturbation. That is, we can write the integral equation
including the boundary conditions (\ref{fiABR})
\begin{eqnarray}\label{eqinteR}
\varphi_q(\tau) &=&  A_q \; e^{i q \, \tau} + B_q \;
e^{-i q \, \tau} \cr \cr
&-&\int_{\tau}^{\infty}d\tau'\; {{\sin{q(\tau'-\tau)}}\over {q}} \;
{\cal M}^2(\tau') \; \varphi_q(\tau')  \; . 
\end{eqnarray}
Iterating this integral equation and using eq.(\ref {masinR})
yields after calculation if we take the function $ s(\tau) $ as a constant,

\begin{eqnarray}\label{ABperR}
\varphi_q(\tau) &=&  A_q \; e^{i q \, \tau} + B_q \;e^{-i q \, \tau} 
 \cr \cr
&-&\frac{A\;e^{-(\tau/\tau_0)^x}}{8\,\tau }\left\{ A_q \; e^{i q \, \tau}
\left[\frac{e^{i\psi(\tau)}}{q+1}-
\frac{e^{-i\psi(\tau)}}{q-1} \right]\right.\cr \cr
&-&\left.B_q \;e^{-i q \, \tau}
\left[\frac{e^{i\psi(\tau)}}{q-1}-
\frac{e^{-i\psi(\tau)}}{q+1} \right]\right\}
\end{eqnarray}

where $ \psi(\tau) = 2 \, \tau + 2 \, c_2 \, s(\tau) \,
 \log\left(\frac{\tau}{\tau_1}\right) +\phi $.
 
These expressions are singular at  $ q = \pm 1 $ revealing the
the resonant band below $ q = 1 $. Actually, these
singularities disappear when the decrease of $ s(\tau) $ with $ \tau
$ is taken into account on time scales $\tau >> \tau_2$. 

Notice that there is no resonance at $ q = 0 $ for broken symmetry.

For $q^2 \approx 1$ the perturbative expansion breaks down for the
intermediate asymptotic time scales $\tau_1 << \tau <<\tau_2$.

In order to understand the behavior of the mode functions in the
resonant regions, we implement a multitime scale analysis in the same
manner as in the unbroken symmetry case introducing the parameter 
$\epsilon = 1/\tau_1$ and introducing the long time scale
$T_1=\epsilon T_0$ with $T_0=\tau$. The analysis is carried out in
exactly the same 
manner as before. The result is that the amplitudes $A_{q=1}\; ; \; 
B_{q=1}$ are {\em exactly} given by eq.(\ref{brokenampli}). Furthermore
a consistency condition arising from the multitime scale analysis 
is that
\begin{equation}
A = 4 \; \sqrt{c_1^2 + c_2^2}\approx 4 \; c_2 \;. \label{brokecons}
\end{equation}
 
This condition is well verified in the numerical calculations.

\bigskip
Since the effective squared mass vanishes asymptotically and 
there are no resonances at $q=0$ for sufficiently late times the 
$ q = 0 $ mode behaves as
\begin{equation}
\varphi_0(\tau) = L + M \, \tau
\end{equation}
where  $ L $ and $ M $ are complex coefficients that can only be obtained 
from the full time evolution. The wronskian relation (\ref{wrfi}) implies that 
\begin{equation}
Im[ L \, M^* ] = 1 \; ,
\end{equation}
showing that $ M \neq 0 \neq L $. 

It follows from eq.(\ref{partnumber}) that the number of $ q = 0 $ 
quanta grows as $ \tau^2 $ for  asymptotically large  $ \tau $,
\begin{equation}
N_0(\tau) \buildrel{\tau \to \infty}\over= \frac14\, \sqrt{1 + \eta_0^2 }\;
|M|^2 \; \tau^2
\end{equation}
Notice that the total number of particles tends to a constant for 
$ \tau \to \infty $ because $ N_0(\tau) $ does not contribute to the
total due to the vanishing of the phase space factor as $ q^2 $ at low
momentum. This linear growth of the homogeneous quantum mode was also noticed 
in\cite{brokemil}.

This situation is similar to Bose-Einstein condensation
 in which case the excess number of particles 
at a fixed temperature goes into the condensate, whereas the total number 
of particles outside the condensate is fixed by the temperature at zero 
chemical potential. The $q=0$ mode does not contribute to the energy, the
 pressure and the total particle number but it will 
 become macroscopically occupied for $ \tau \approx
\sqrt{V} $ 
with $V$ the volume of the system, in which case the number of particles in
the zero momentum mode becomes of the order of the spatial volume. In this
case this mode must be isolated and studied separately from the 
$q \neq 0$ 
modes because its contribution to the momentum integral will be cancelled 
by the small phase space at small momentum much the same way as in usual
 Bose-Einstein condensation. 

The presence of a Bose condensate through this macroscopic zero
momentum mode  signals spontaneous symmetry breaking even  
when the order parameter remains zero. Therefore we identify the {\em
linear growth} in time of the $q=0$ mode as the onset of a novel form
of Bose condensation of Goldstone bosons and symmetry breaking in the
collisionless regime and in 
the absence of thermalization. This form of Bose condensation of 
Goldstone bosons in a collisionless regime  is similar to that
reported 
recently within a different context\cite{tsu}. 

The time scale for the formation of the condensate ($ \tau \approx
\sqrt{V} $ ), is obviously much larger than the nonlinear scale 
$ t_1 $ and  the thermalization scale when corrections  of order
  $ 1/N $  are included. Therefore, the formation of the Bose
condensate including $ 1/N $ corrections, which will include
collisions, will require a further understanding of the dynamics and
of the time scales involved. 

\subsection{Energy and  Pressure for Broken Symmetry: sum rules}

As shown in refs.\cite{boyan2,eri96} the energy (\ref{renorenergy}) and the
pressure (\ref{preMen})  can be rewritten for broken symmetry as follows

\begin{eqnarray}
E_{ren} & = & \frac{2 |M_R|^4}{\lambda_R}\; \varepsilon\; ,\cr \cr
\varepsilon & = &
\frac{\dot{\eta}^2}{2}+\frac{1}{4}\left(\eta^2 -1\right)^2 +
2g \int_0^{q_u}q^2dq\; \Omega_q \; N_q(\tau) \cr \cr
&+& \frac{g}{2}\Sigma(\tau)\left[-1-\eta^2_0+{\cal{M}}^2(\tau)-
\frac{g}{2}\Sigma(\tau)\right] +{\cal{O}}(g)
\label{enebrok}\cr \cr
P_{ren}(\tau) & = & \frac{2 |M_R|^4}{\lambda_R}\; p(\tau)\; ,\cr \cr
 p(\tau)& = &
g\int_0^{1}q^2 dq \left[ \frac{q^2}{3} |\varphi_q(\tau)|^2
 +|\dot{\varphi}(\tau)|^2\right] + \dot{\eta}^2 + {\cal{O}}(g) 
 -\varepsilon\; ,
\label{presubrok} \\
\end{eqnarray} 
up to corrections of order $ \eta_0^2<<1$ where  $q_u$ is the maximum
spinodally unstable wavevector (for $\eta^2_0 <<1 \; q_u =1$.

As for the unbroken symmetry case, we can use the asymptotic value
of the effective mass squared and the conservation of the energy to
compute the first two moments of $ M^2_q(\infty) $ and to establish
similar sum rules.  

The vanishing of $ {\cal M}^2(\infty) $ implies that $  g
\Sigma(\infty) = 1 +  {\cal O}(\eta_0^2)$ and hence
\begin{equation}
\int_0^1 q^2 \, dq \; M^2_q(\infty) + {\cal O}(g, \eta_0^2)= 1\; .
\end{equation}
where we used eqs.(\ref{sigmafin}) and (\ref{aprofiR}).

Furthermore, equating the expression for the energy at $ \tau =
\infty $ [eq.(\ref{enebrok})] with its initial value
\begin{equation}
\varepsilon = \frac14+  {\cal O}(\eta_0^2)
\end{equation}
yields the second sum rule:
\begin{equation}
\int_0^1 q^4 \, dq \;
M^2_q(\infty) = \frac14 + {\cal O}(g, \eta_0^2)\; . 
\end{equation}
 
Using these sum rules allows to compute the pressure
in the $ \tau \to \infty $ limit from eq.(\ref{presubrok})
\begin{equation}
p(\infty) = \frac1{12}+  {\cal O}(g, \eta_0^2)\; .
\end{equation}
leading to the equation of state for radiation
\begin{equation}
p(\infty) = \frac13 \; \varepsilon + {\cal O}(g, \eta_0^2)
\end{equation}
as is expected since the bulk of the produced particles are massless. 

\section{Conclusions and Further Questions:}

In this article we have studied both numerically and analytically the
asymptotic non-equilibrium dynamics of relaxation in a scalar field
theory in the collisionless regime. We have focused on the relaxation 
of initial states of very large energy density that require a
non-perturbative treatment in a controlled manner, maintaining
renormalizability, energy conservation and all of the relevant
conservation laws. Detailed numerical analysis revealed the presence
of new dynamical and non-perturbative time scales and non-linear resonant
bands  that result in the power law growth of quantum fluctuations and
a power law of relaxation for the expectation value of the scalar field.
A separation between the time scales for weak coupling allowed us to  
implement a dynamical renormalization group resummation of secular terms 
via the method of multi-time scales and confirmed and completed the numerical
results. 

This {\bf dynamical} renormalization group resummation of
secular  terms leads to power law relaxation with {\bf anomalous} dynamical
exponents which are {\bf non-universal}  depending non-perturbatively on the
coupling constant.

In the unbroken symmetry phase the expectation value vanishes
asymptotically, transferring  all of the initial energy into production
of massive particles, as a result of the non-linear resonances and
despite the presence of perturbative thresholds to particle production. 
The asymptotic distribution is non-thermal
non-perturbative in the band where parametric resonance takes place at
early times. It can be interpreted as a
`semiclassical condensate' in the unbroken phase, a result of the
relaxation of the initial energy density.  
The equation of state of the produced particles interpolates between dust
and radiation domination as a function of the initial amplitude
of the expectation value of the scalar field.

In the broken phase the numerical evolution revealed a hierarchy of
time scales, and the  relaxation  of the order parameter is with
anomalous power laws. Again an implementation of the dynamical
renormalization group revealed the presence of non-linear resonances
that result in particle production after the spinodal time. 
Non-universal  power law relaxation appears with exponents  depending
non-perturbatively on the coupling constant. 
The coupling constant dependence is similar for the unbroken and broken
symmetry cases. The
asymptotic distribution of particles is localized at low momenta, is
non-perturbative and non-thermal, but the equation of state is that of
radiation. We also found the onset of a novel form  Bose condensation
in the collisionless regime and without thermalization but with
extremely long time scales.  

Although this body of results provides for a deeper understanding of
the non-equilibrium dynamics in the collisionless regime, and thus we
believe it  represents a quantum field theory example of 
{\em non-linear dynamics} there are very many  unanswered questions that
deserve further study. i) the fact that the asymptotic distributions
are of `soft' momenta and semiclassical in the sense that the
amplitude of the particle distribution inside the bands is $\propto
1/\lambda$, and that there is a separation of time scales for weak
coupling should allow a `coarse grained' description in terms of 
quasi-particle distributions \`a la Boltzmann. ii) The next order in
$1/N$ must be pursued to incorporate consistently collisional processes,
probably i) must be understood before this to separate the microscopic
from the long time scales. Collisions will compete with the collisionless
processes and affect  the onset of Bose condensation found in
the collisionless regime. Here there is the possibility that whereas thermal
equilibrium is established on collisional time scales, chemical equilibrium may
be established on much longer time scales resulting in a non-vanishing
chemical potential.  

Only a deeper understanding of i) and ii) will
lead to a complete understanding of the different regimes: collisionless,
kinetic and hydrodynamic that could account for thermalization,
chemical equilibration  and all of the important relaxational
phenomena.    
 
\section{Acknowledgements:} 
The authors thank F. Cooper, I. Egusquiza, H. J. Giacomini, E. Mottola
and  L. Yaffe, for stimulating  discussions. D. B. thanks the 
N.S.F for partial support through  grant 
awards: PHY-9605186  and LPTHE for warm hospitality.  R. H., 
is supported by DOE grant DE-FG02-91-ER40682. We thank NATO for
partial support.

\figure{}

\hbox{\epsfxsize 14cm\epsffile{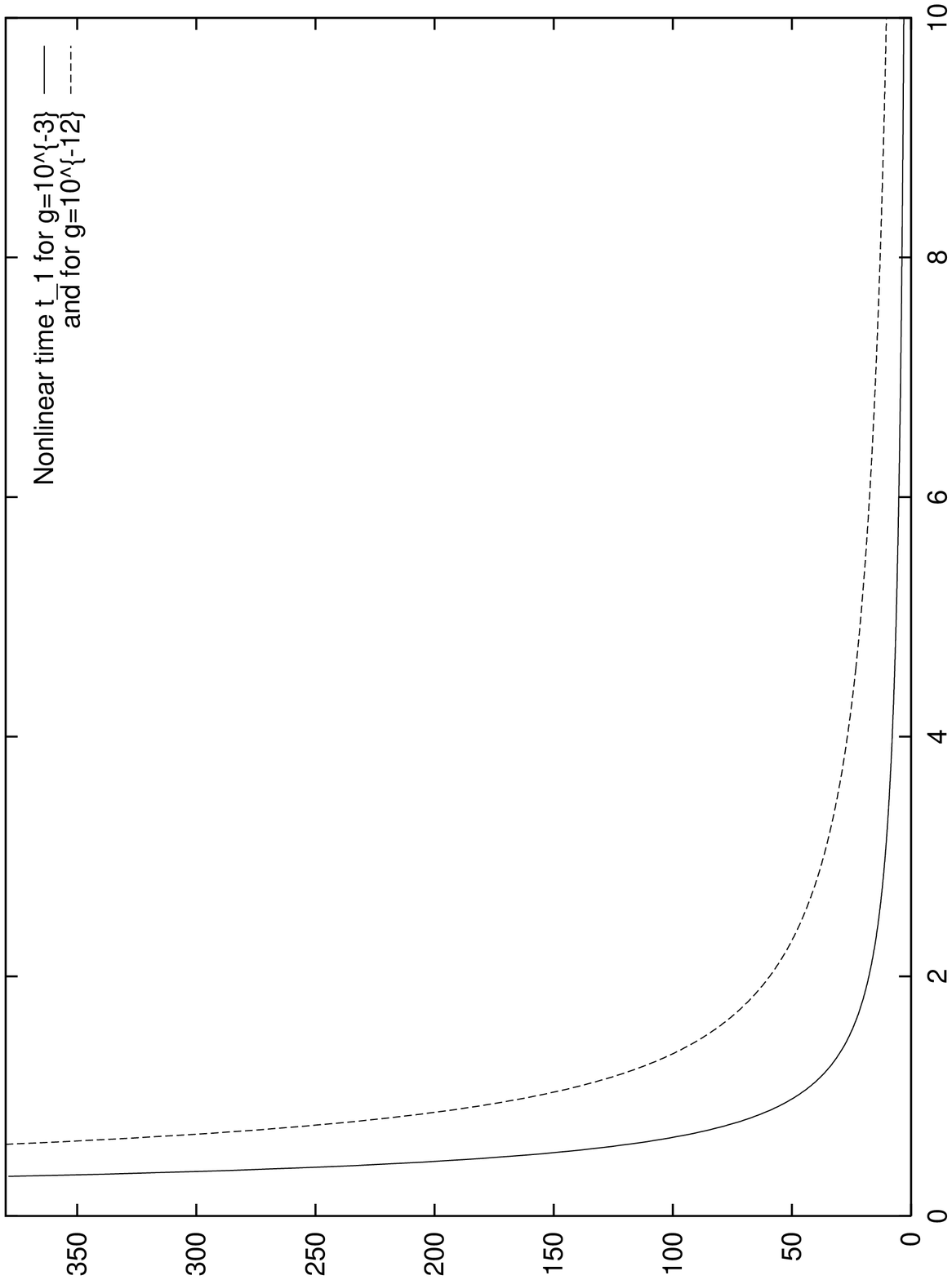}}

\figure{ {\bf Figure 1:} 

The nonlinear time for the unbroken
symmetry case as a function of $ \eta_0 $ according to
eq.(\ref{maxtime}) for $ g=10^{-3} $ and $ g=10^{-12} $. 
\label{fig1}}

\clearpage

\hbox{\epsfxsize 14cm\epsffile{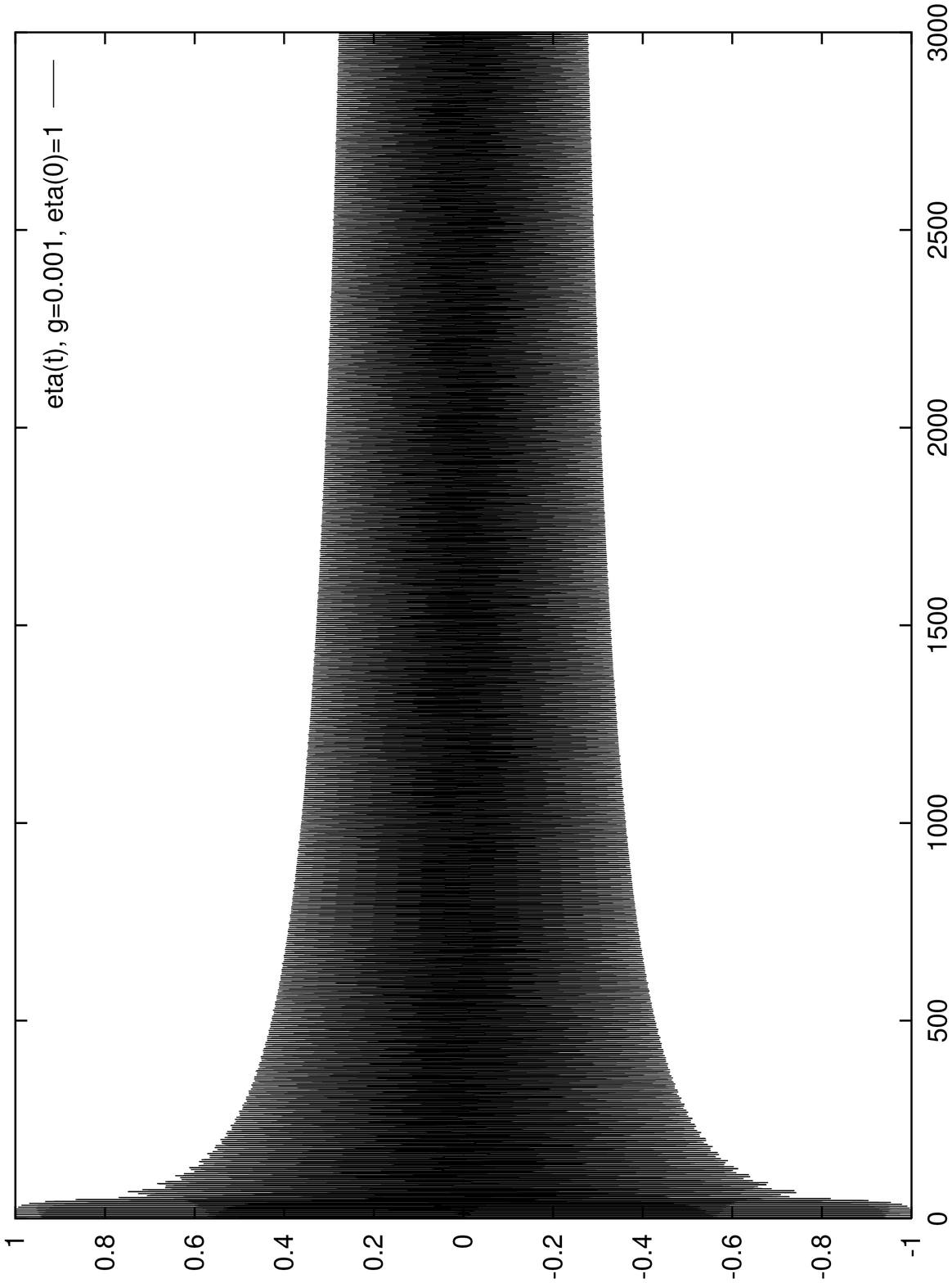}}

\figure{ {\bf Figure 2:} The zero mode $\eta(\tau)$ vs. $\tau$ for the unbroken
symmetry case with 
$ \eta_0=1 $, $ g=10^{-3} $. \label{fig2}}

\clearpage

\hbox{\epsfxsize 14cm\epsffile{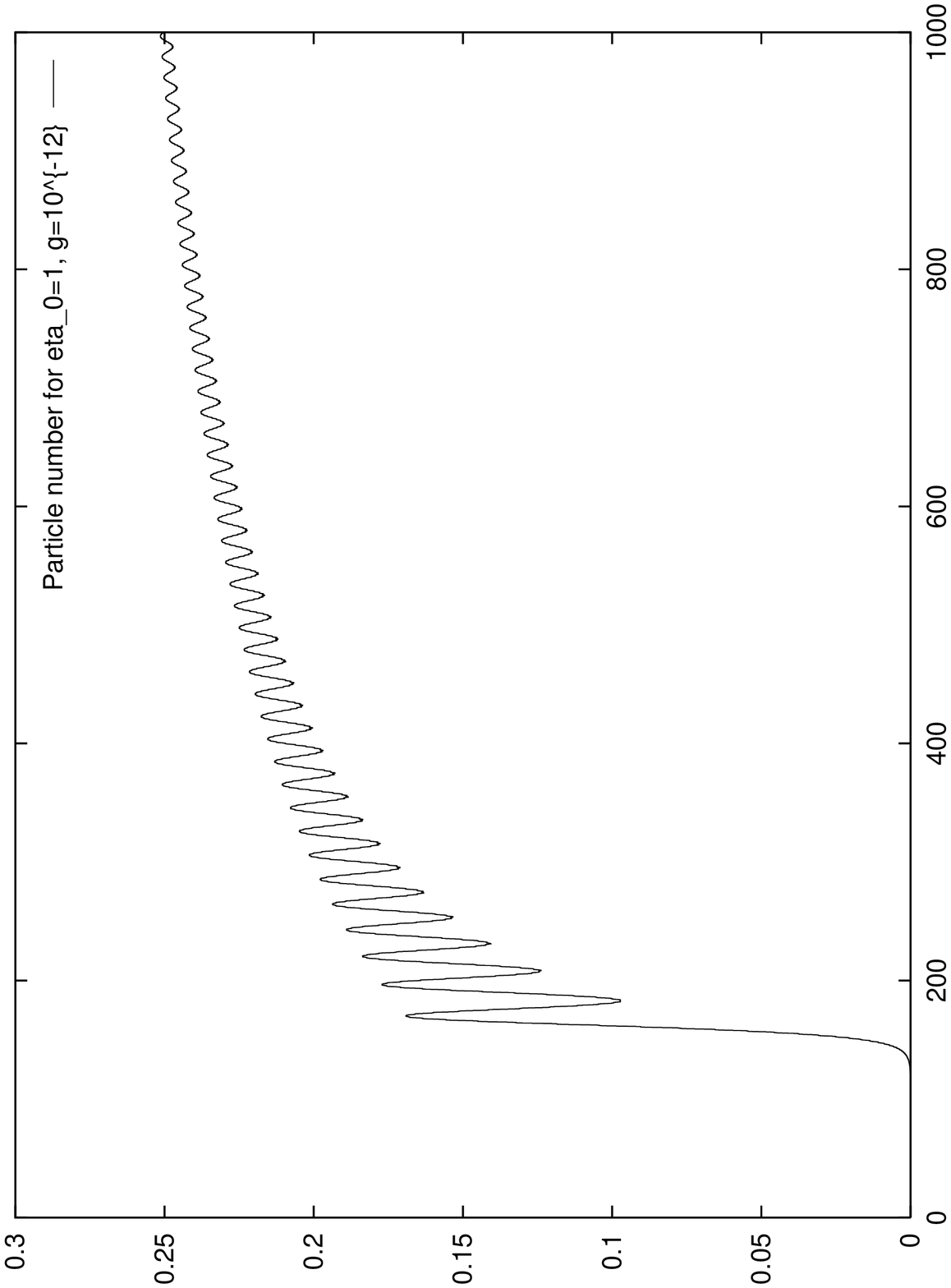}}

\figure{{\bf Figure 3:} 
The total number of produced particles as a function of time for 
$\eta_0=1$, $g=10^{-12}$. After the exponential increase around $
\tau=\tau_1 = 163.7.. $, ${\cal N}^{ad}(\tau) $ keeps growing.
For times $ \tau > 200 $, eq.(\ref{ocnum}) gives a very good
approximation to the 
numerical results (after averaging over oscillations). 
\label{fig3}}

\clearpage

\hbox{\epsfxsize 14cm\epsffile{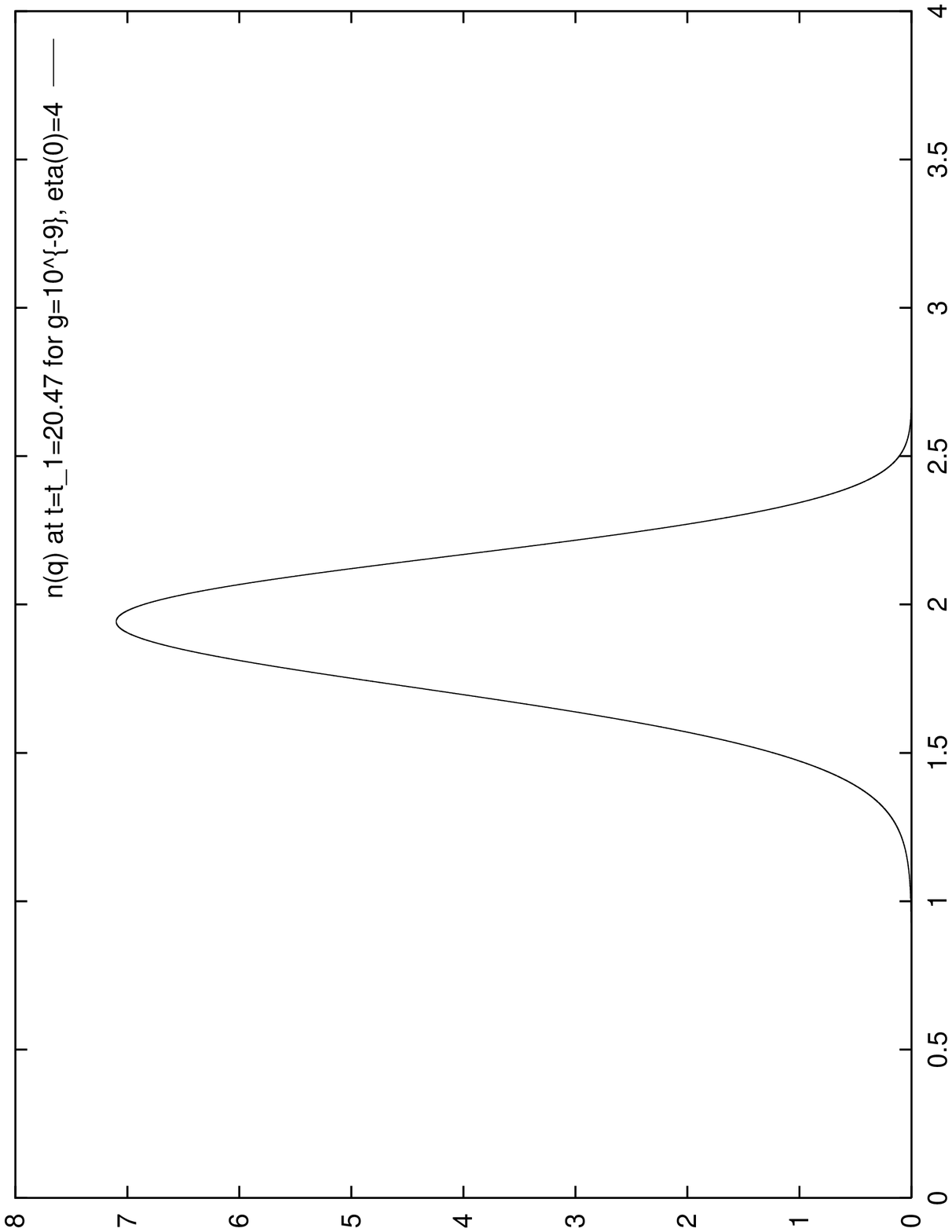}}

\figure{{\bf Figure 4:} 
Momentum distribution of the produced particles 
$ n(q) \equiv q^2 \, N^{ad}_q(\tau) $ at the nonlinear
time $ \tau = \tau_1 $ for $ \eta_0=4 $, $ g=10^{-9} $.
\label{fig4}}

\clearpage

\hbox{\epsfxsize 14cm\epsffile{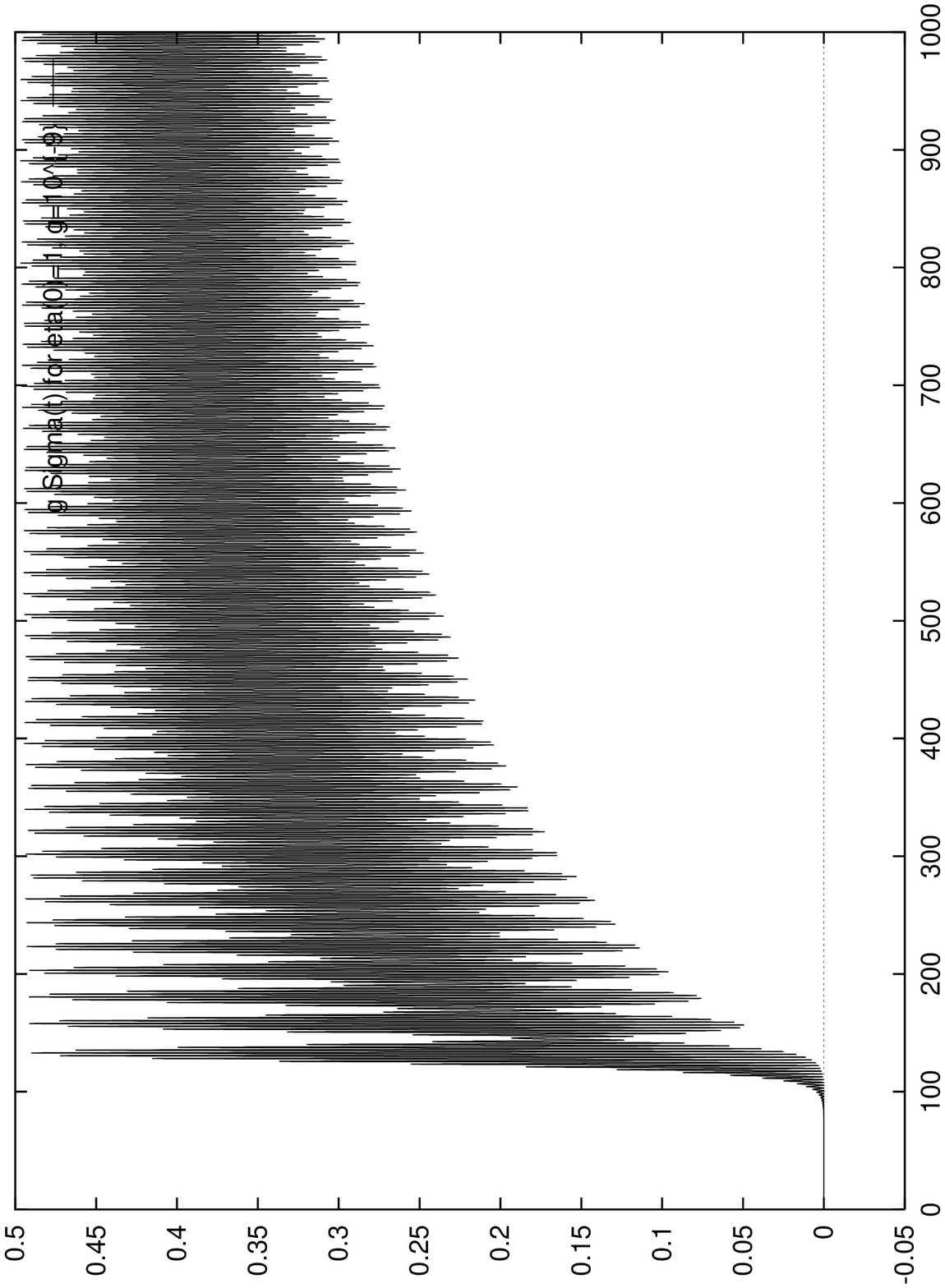}}

\figure{{\bf Figure 5:} 
The quantum fluctuations $ g \Sigma(\tau) $ as a function of time for 
$ \eta_0=1 $, $ g=10^{-9} $.
\label{fig5}}

\clearpage

\hbox{\epsfxsize 14cm\epsffile{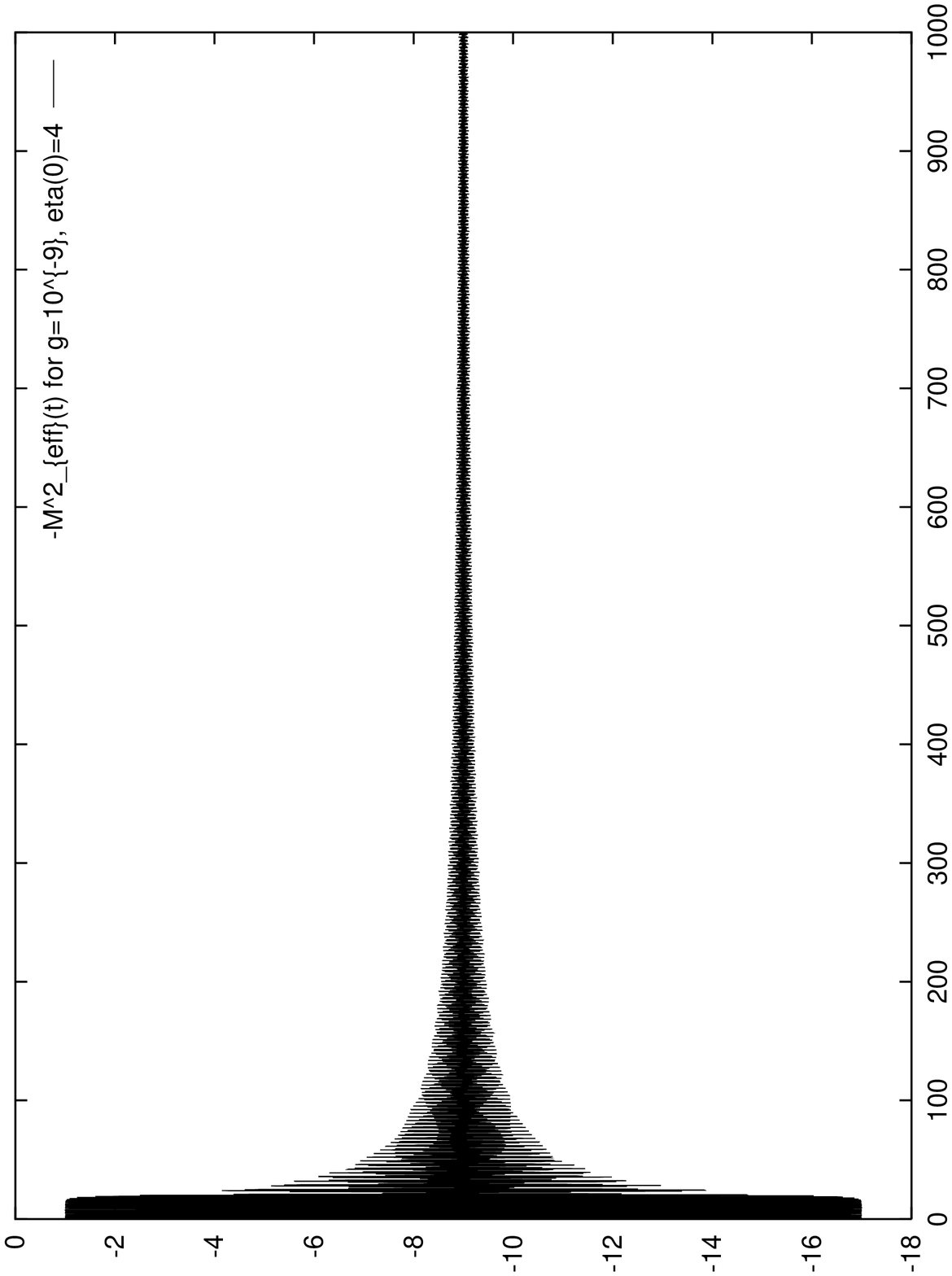}}

\figure{{\bf Figure 6a:} 
The effective mass squared as a function of time for 
$ \eta_0=4 $, $ g=10^{-9} $. Notice the asymptotic value $ \approx 1 +
\eta_0^2/2 = 9 $.
\label{fig6a}}

\clearpage

\hbox{\epsfxsize 14cm\epsffile{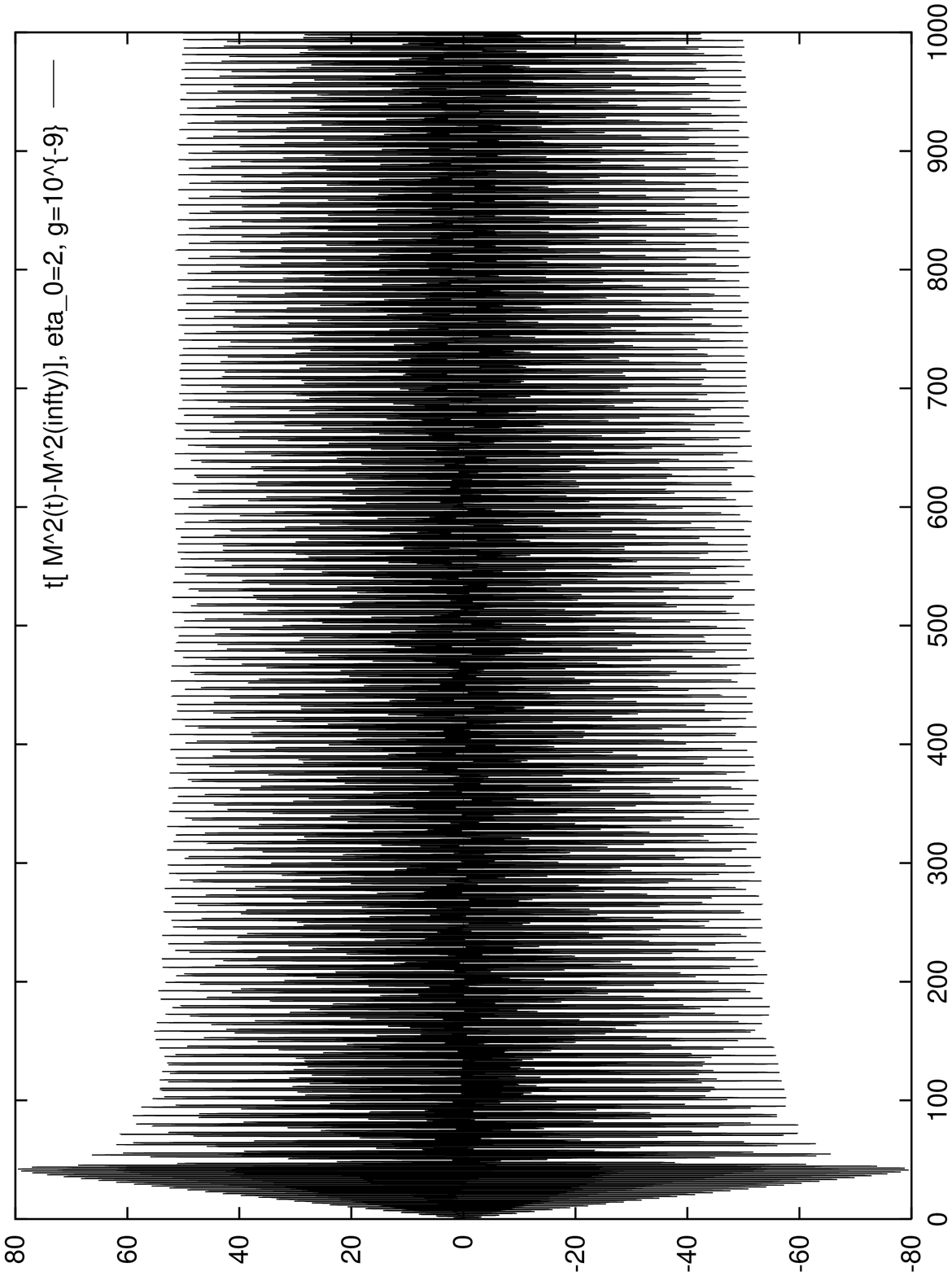}}

\figure{{\bf Figure 6b:} 
The effective mass squared minus its value at $ \tau = \infty $ times $ \tau $
as a function of time for $ \eta_0=2 $, $ g=10^{-9} $.
This function oscillates in time with constant amplitude $ K_1 + K_2 $
and frequencies $ 2 \, {\cal M}_{\infty} = 2 \sqrt{1 + \eta_0^2/2} $ and $
2 \,  {\cal M}_0 = 2 \sqrt{1 + \eta_0^2}$ [see eq.(\ref{funp1})].  
\label{fig6b}}

\clearpage

\hbox{\epsfxsize 14cm\epsffile{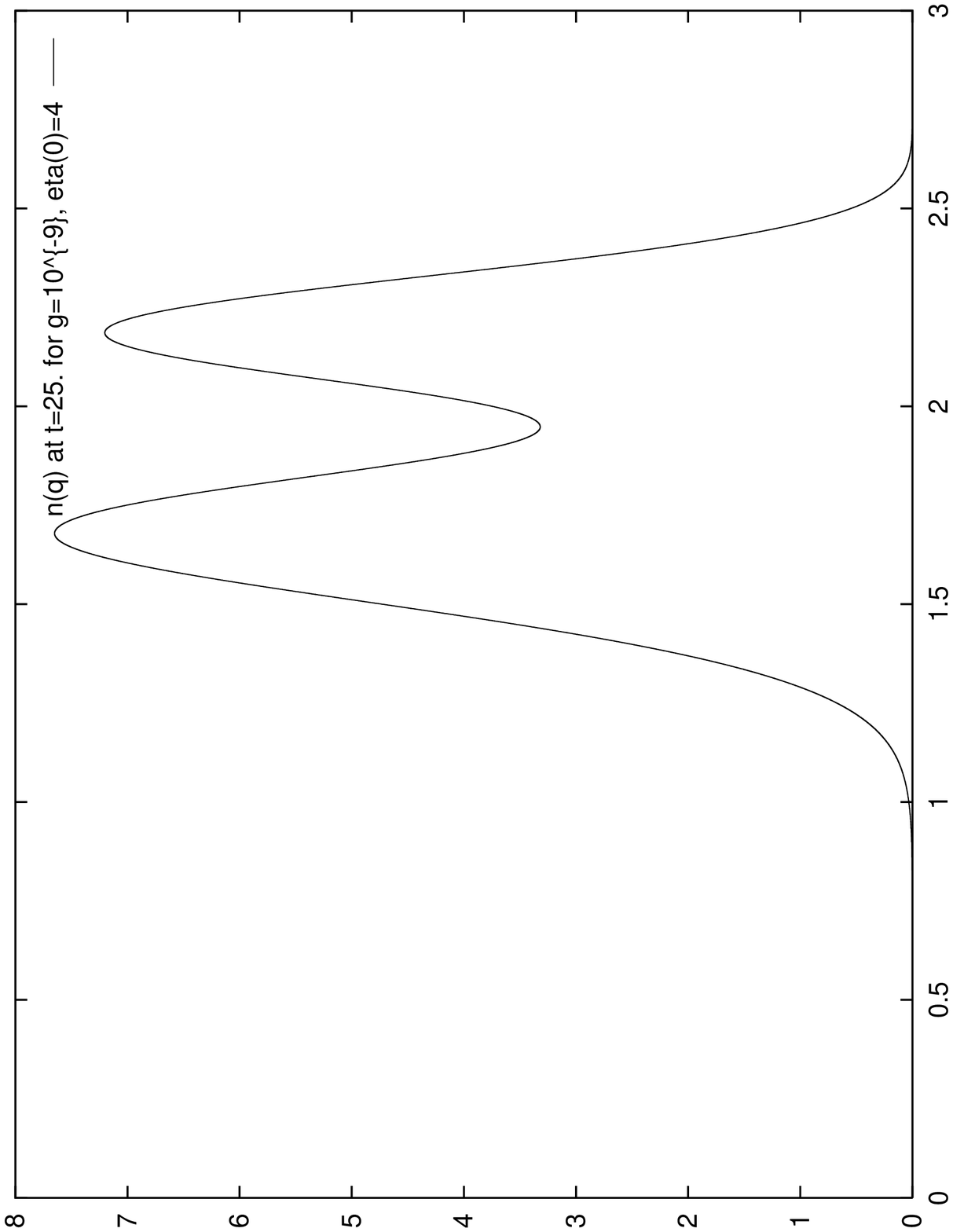}}

\figure{{\bf Figure 7a:} 
Momentum distribution of the produced particles at $ \tau = 25. $ for
$ \eta_0=4 $, $ g=10^{-9} $. Notice that the single peak present for
times $ \tau < \tau_1 $ splits into two due to the nonlinear resonances.
Many more peaks appear for subsequent times as shown in fig. 7b.
\label{fig7a}}

\hbox{\epsfxsize 14cm\epsffile{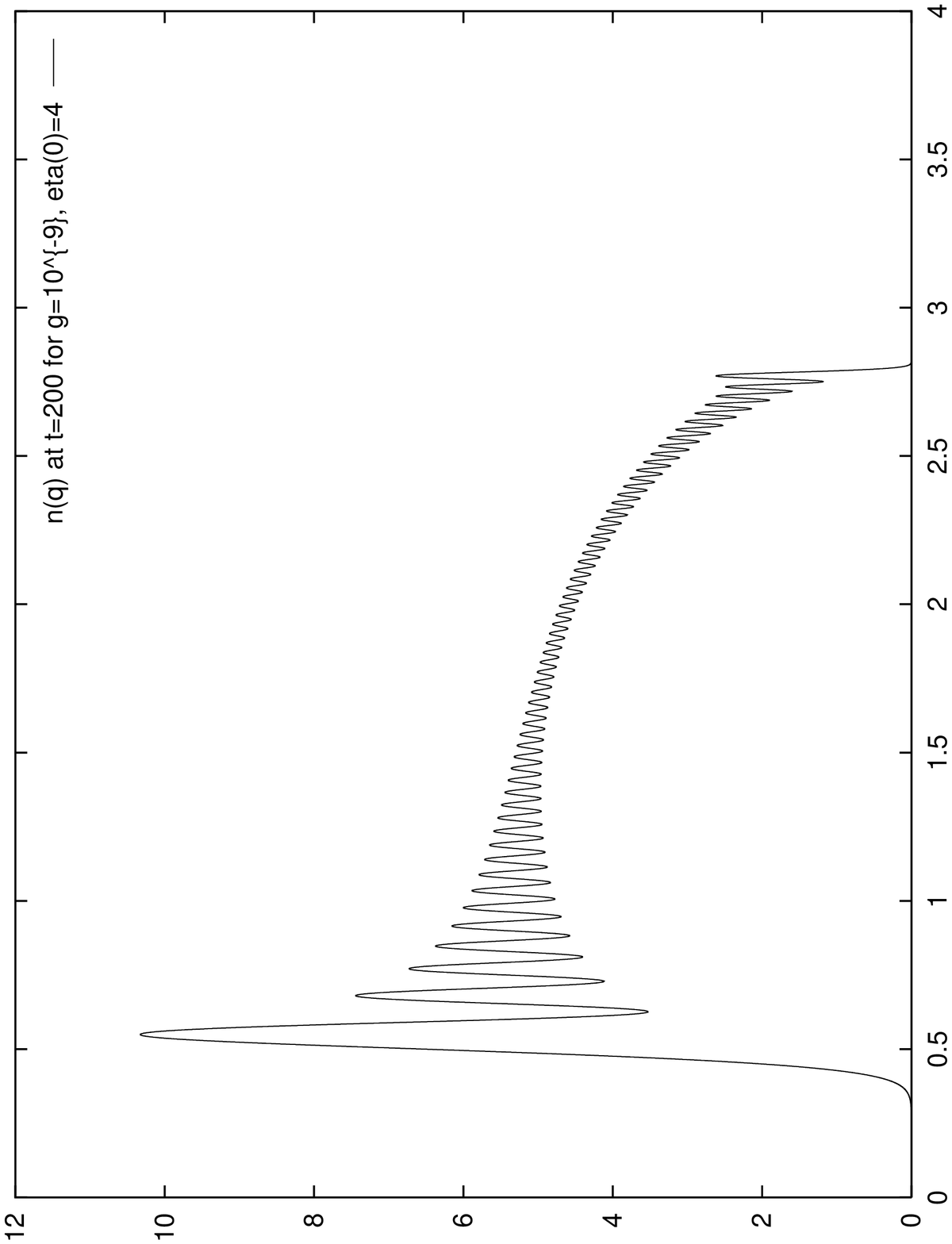}}

\figure{{\bf Figure 7b:} 
Momentum distribution of the produced particles at $ \tau = 200 $ for
$ \eta_0=4 $, $ g=10^{-9} $. Notice the main peak at $ q = 0.549 $
associated with the main non-linear resonance ($q=0$)
and the secondary peak
at $ q = 2.77 $ associated to the non-linear resonance at $
q=\eta_0/\sqrt2$. The position of both peaks are correctly  estimated
by eqs.(\ref{estpico}) and (\ref{q1tau}), respectively.
\label{fig7b}}

\clearpage

\hbox{\epsfxsize 14cm\epsffile{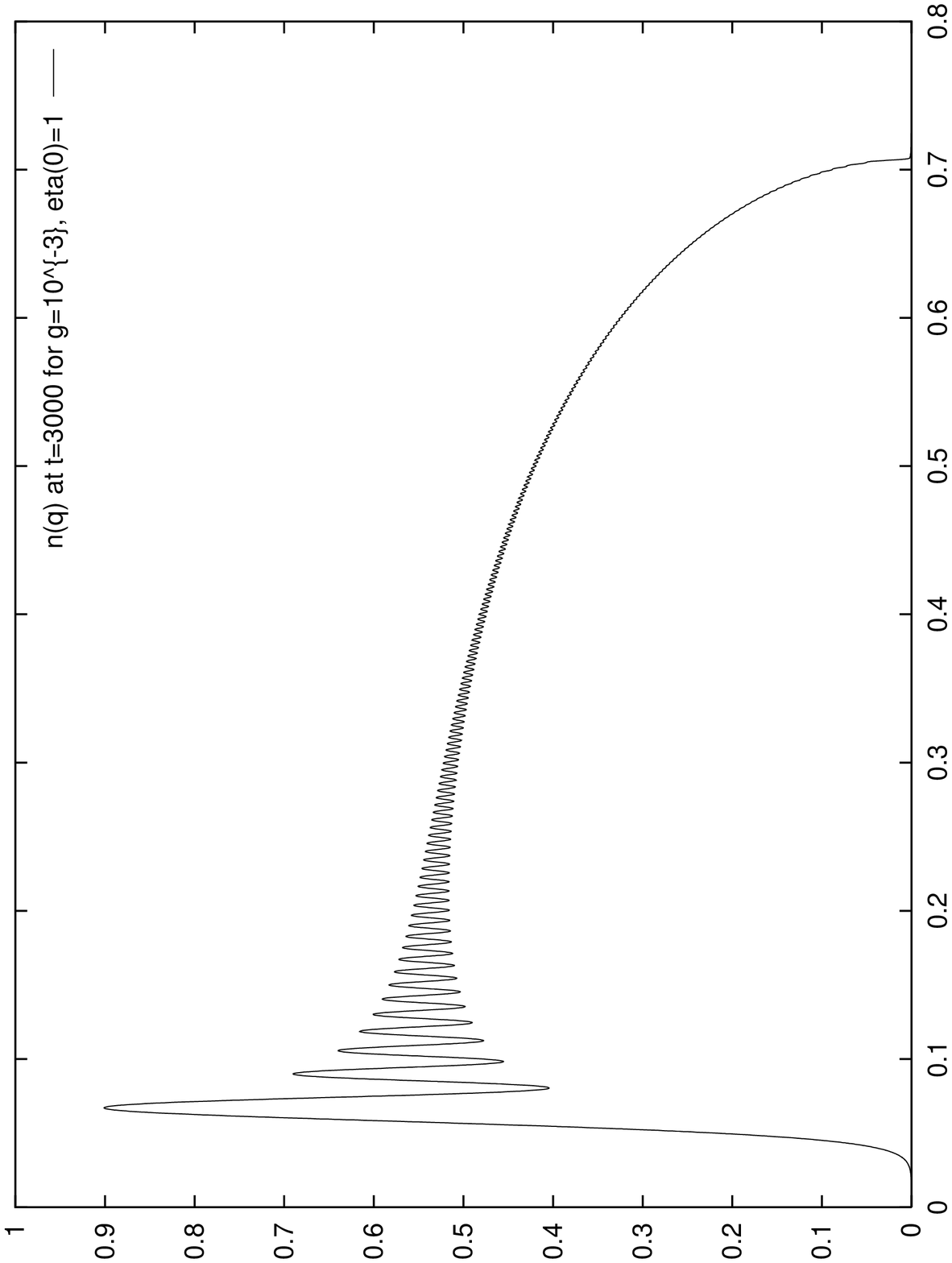}}

\figure{{\bf Figure 7c:} 
Momentum distribution of the produced particles at $ \tau = 3000 $ for
$ \eta_0=1 $, $ g=10^{-3} $. Notice the main peak at $ q=0.067 $, in
good agreement with the estimate given by eq.(\ref{estpico}).
\label{fig7c}}

\clearpage

\hbox{\epsfxsize 14cm\epsffile{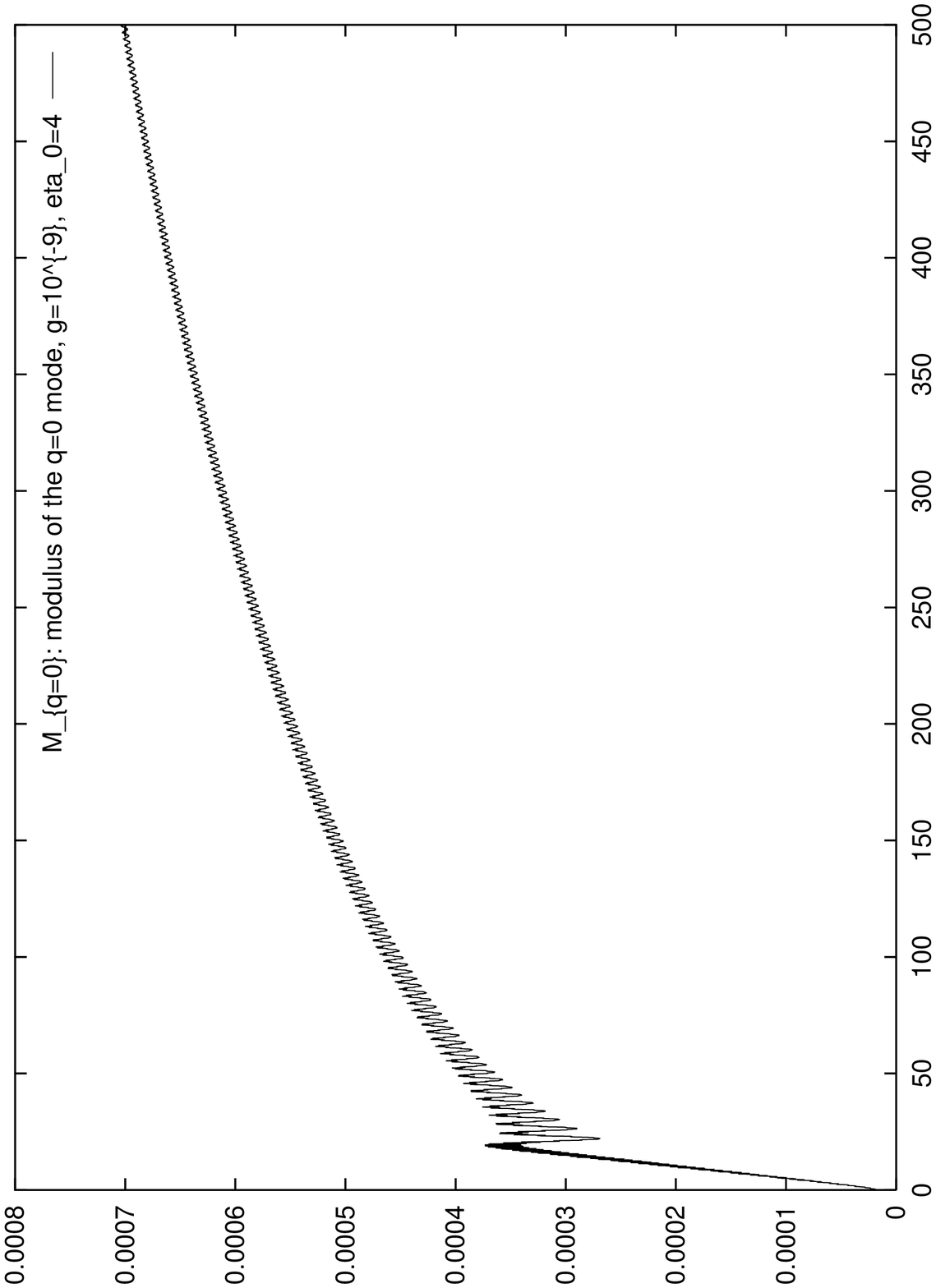}}

\figure{{\bf Figure 8a:} 

The amplitude  $ M_q(\tau) $ of the mode function $ \varphi_{q=0}(\tau) $ as a
function of time for $ \eta_0=4 $, $ g=10^{-9} $. It grows as a power 
according to eq.(\ref{modoC}).
\label{fig8a}}

\clearpage

\hbox{\epsfxsize 14cm\epsffile{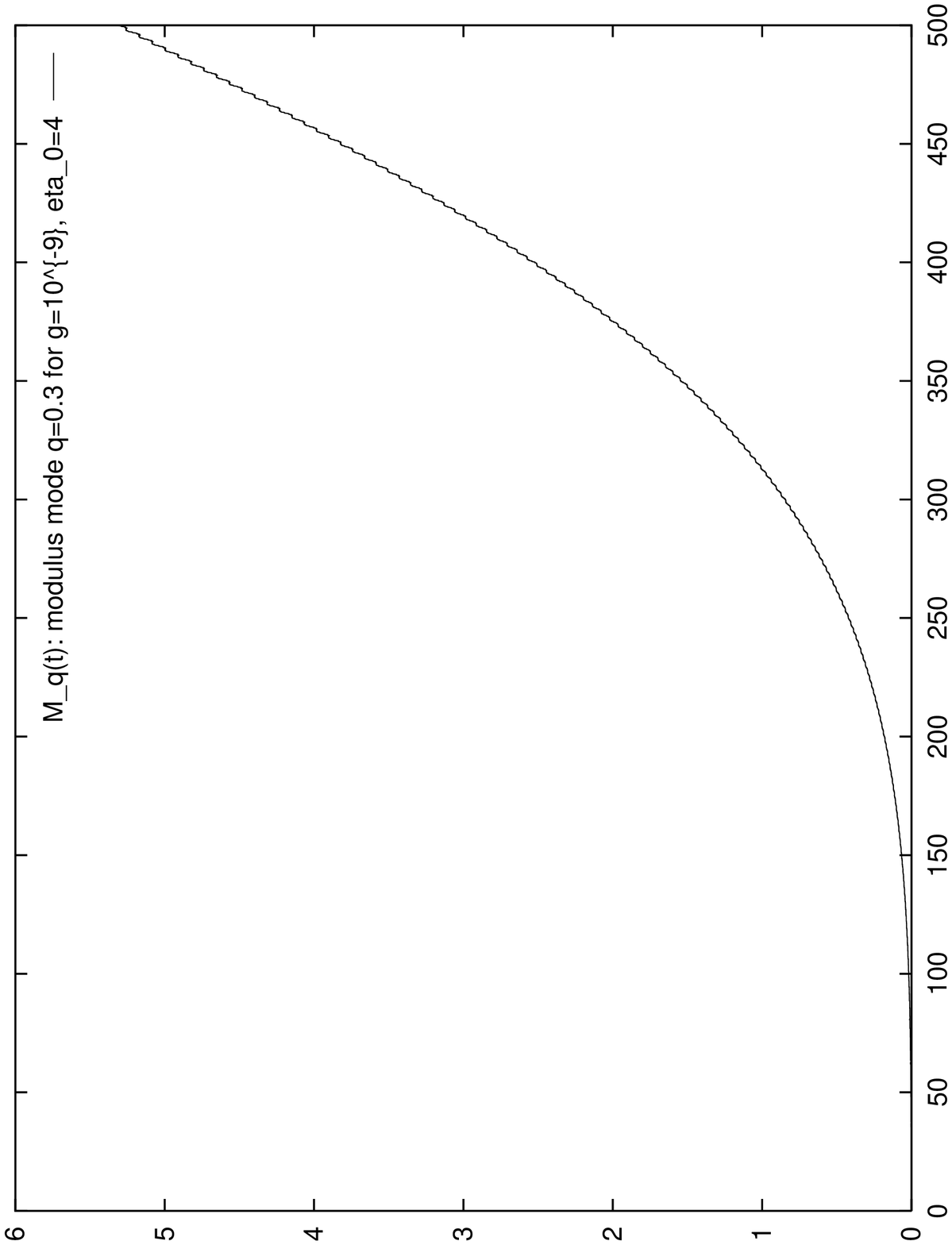}}

\figure{{\bf Figure 8b:} 

The amplitude $ M_q(\tau) $ of the mode function $ \varphi_{q=0.3}(\tau) $ as a
function of time for $ \eta_0=4 $, $ g=10^{-9} $. This amplitude grows 
faster for $ \tau > \tau_1 $
than the $ q=0 $ amplitude but much slower than the exponential growth
in the forbidden band $ 0 < q < \eta_0/\sqrt2 $
for $ \tau < \tau_1 $.
\label{fig8b}}

\clearpage

\hbox{\epsfxsize 14cm\epsffile{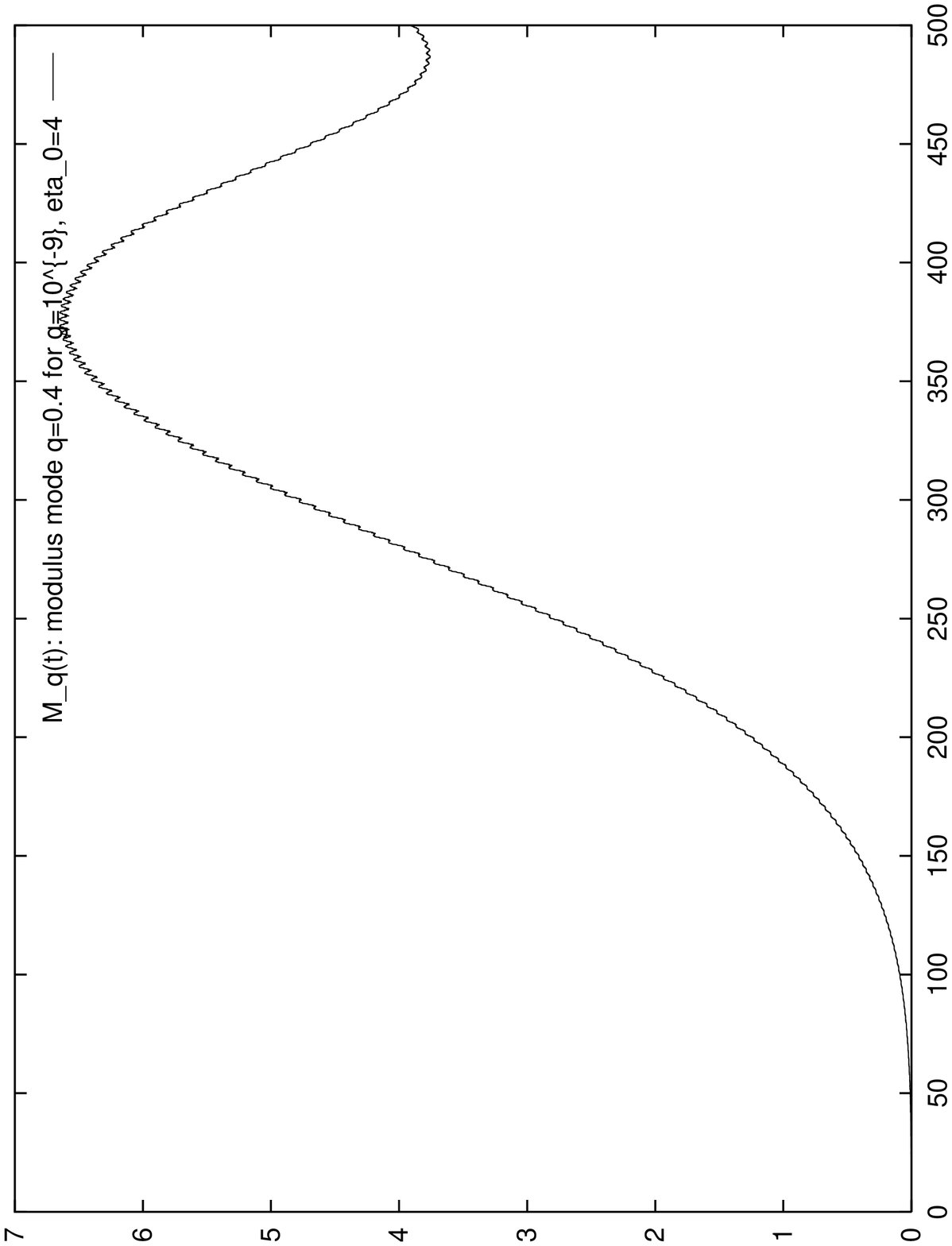}}

\figure{{\bf Figure 8c:} 

The amplitude  $ M_q(\tau) $ of the mode function $
\varphi_{q=0.4}(\tau) $ as a function of time for 
$ \eta_0=4 $, $ g=10^{-9} $. This mode grows till $ \tau \sim K_1/q^2
\sim 350. $. At such time this mode crosses out of the nonlinear
resonance band.
\label{fig8c}}

\clearpage

\hbox{\epsfxsize 14cm\epsffile{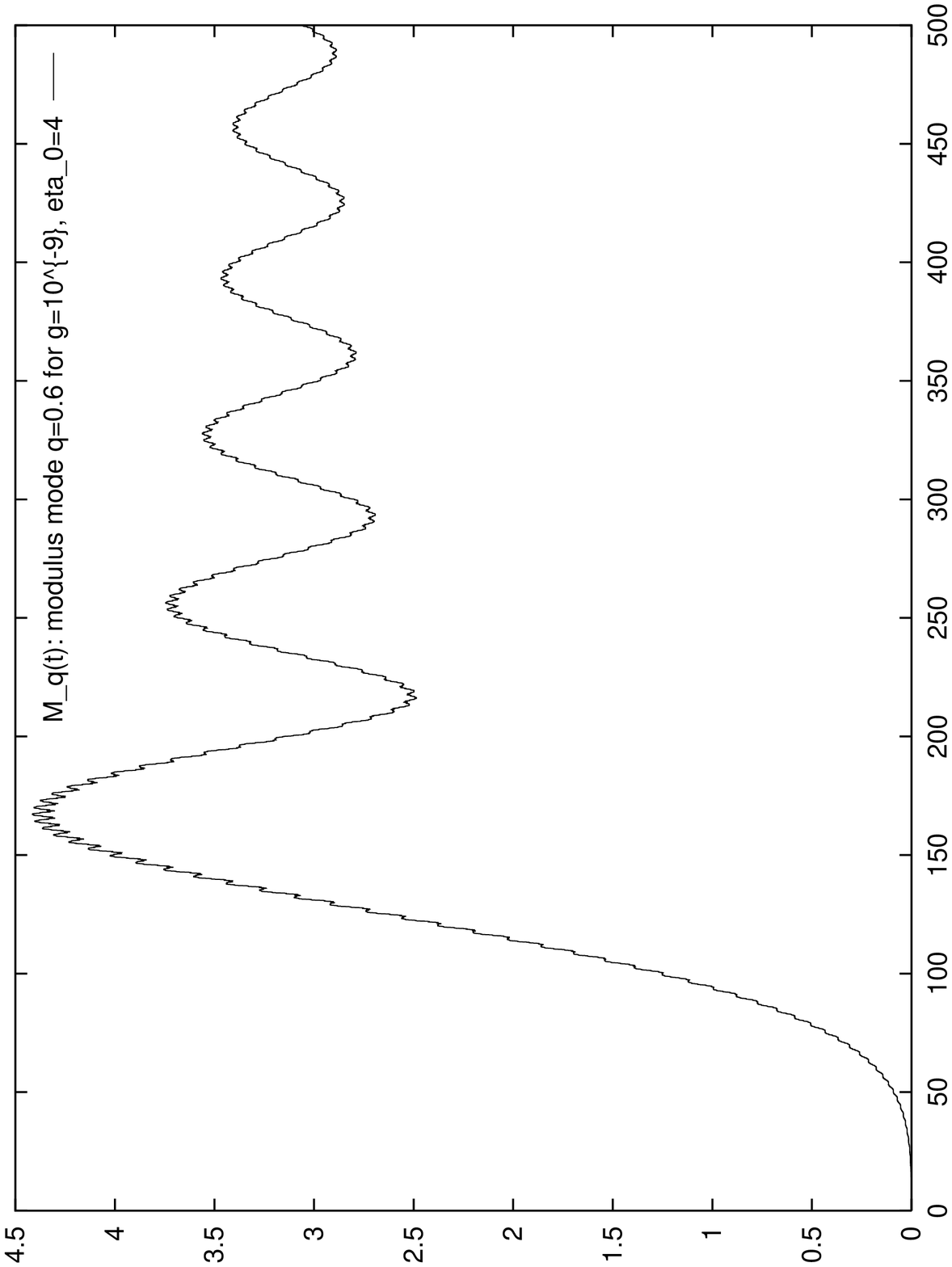}}

\figure{{\bf Figure 8d:} 

The amplitude  $ M_q(\tau) $ of the mode function $
\varphi_{q=0.6}(\tau) $ as a function of time for $ \eta_0=4 $, $
g=10^{-9} $. This mode grows till $ \tau \sim K_1/q^2
\sim 150. $. At such time the mode crosses out of the nonlinear
resonance band.
\label{fig8d}}

\clearpage

\hbox{\epsfxsize 14cm\epsffile{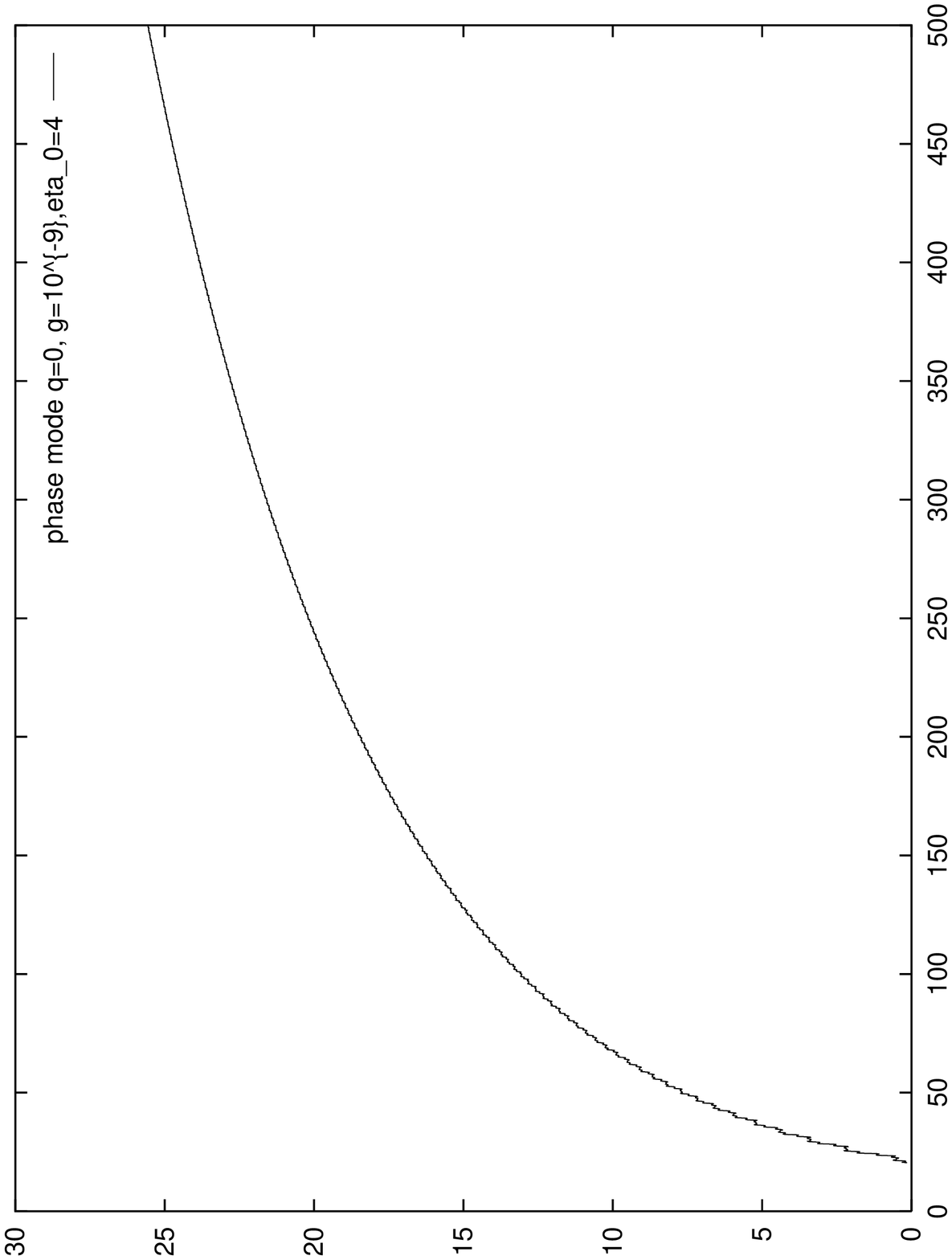}}

\figure{{\bf Figure 9a:} 

The phase $ \phi_q(\tau) $ of the mode function $ \varphi_{q=0}(\tau) $ as a
function of time for $ \eta_0=4 $, $ g=10^{-9} $. This function
follows eq.(\ref{modoC}) with a very good approximation.
\label{fig9a}}

\clearpage

\hbox{\epsfxsize 14cm\epsffile{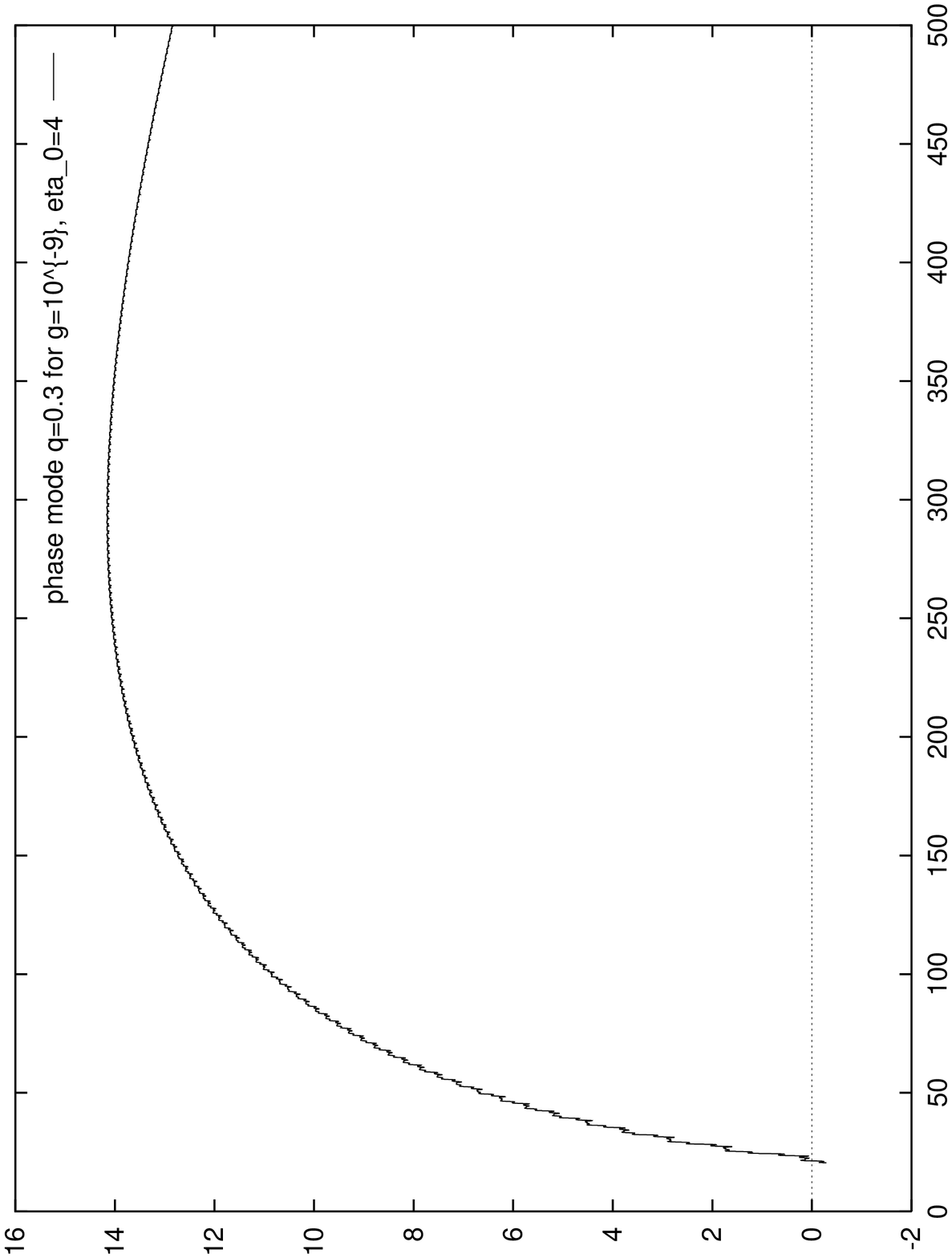}}

\figure{{\bf Figure 9b:} 

The phase $ \phi_q(\tau) $ of the mode function $ \varphi_{q=0.3}(\tau) $ as a
function of time for $ \eta_0=4 $, $ g=10^{-9} $.
\label{fig9b}}

\clearpage

\hbox{\epsfxsize 14cm\epsffile{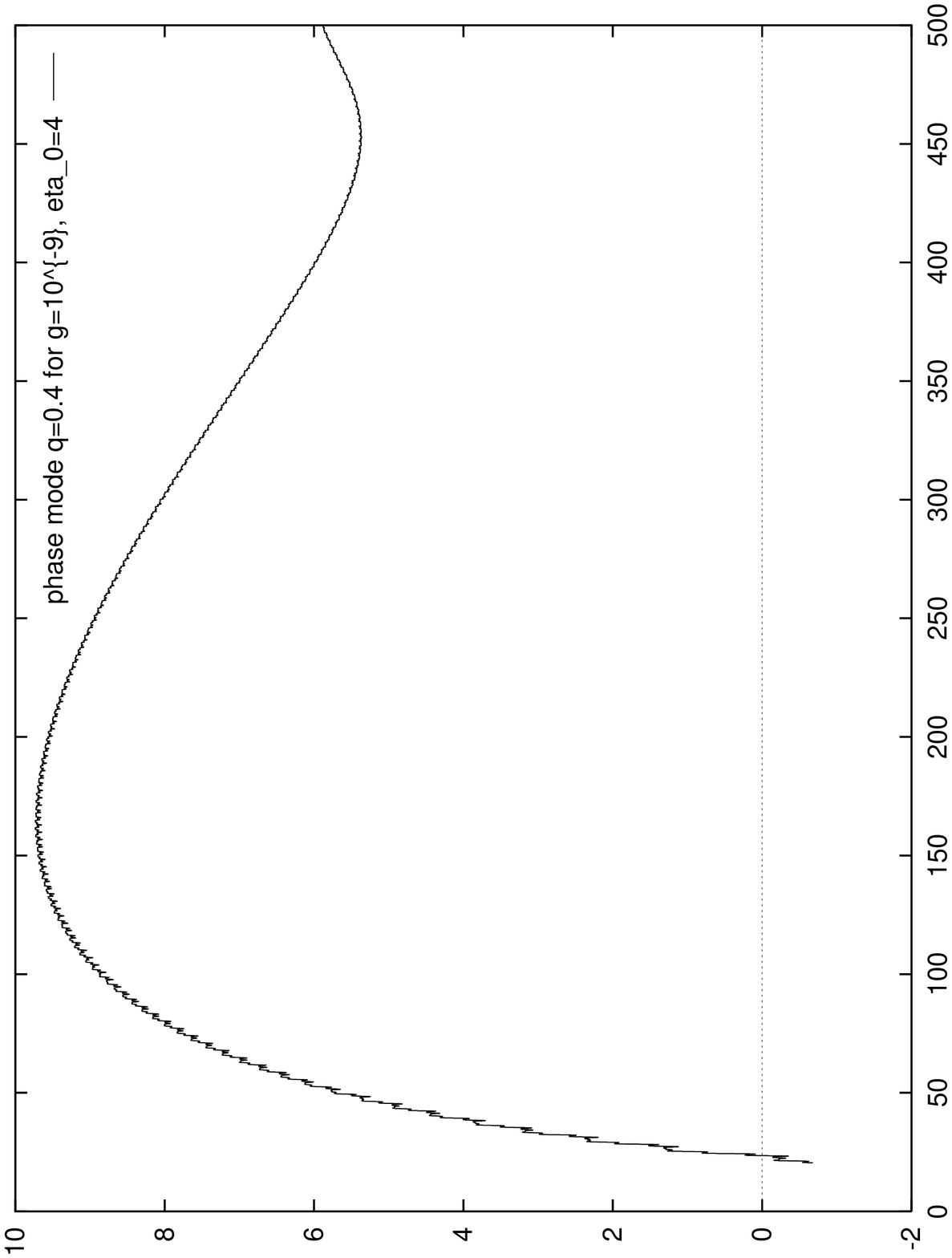}}

\figure{{\bf Figure 9c:} 

The phase $ \phi_q(\tau) $ of the mode function $ \varphi_{q=0.4}(\tau) $ as a
function of time for $ \eta_0=4 $, $ g=10^{-9} $.
\label{fig9c}}

\clearpage

\hbox{\epsfxsize 14cm\epsffile{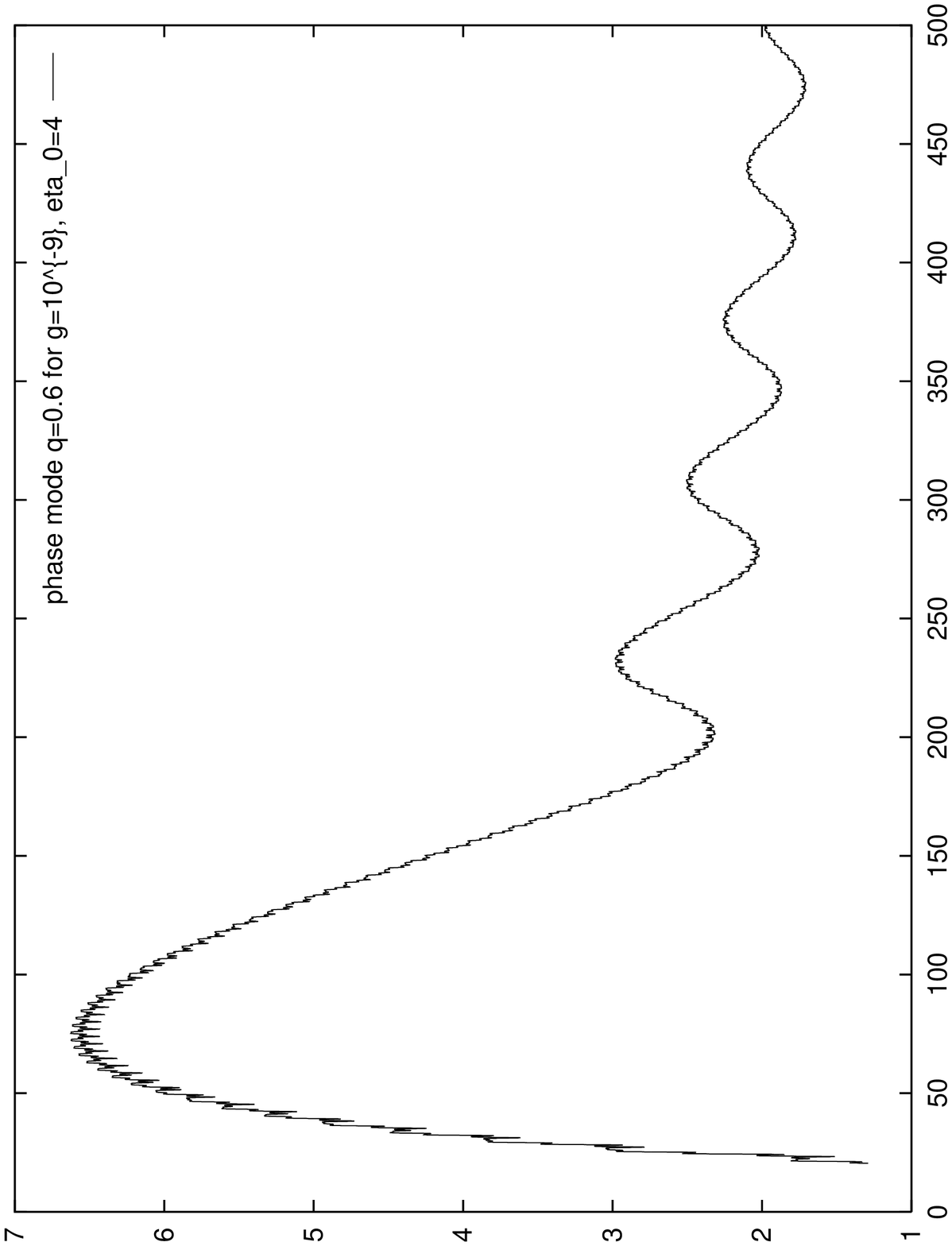}}

\figure{{\bf Figure 9d:} 

The phase $ \phi_q(\tau) $ of the mode function $ \varphi_{q=0.6}(\tau) $ as a
function of time for $ \eta_0=4 $, $ g=10^{-9} $. This phase becomes
an oscillatory function as the same time as the modulus $ M_q(\tau) $
[see fig. 8d] stops to grow.
\label{fig9d}}

\clearpage

\hbox{\epsfxsize 14cm\epsffile{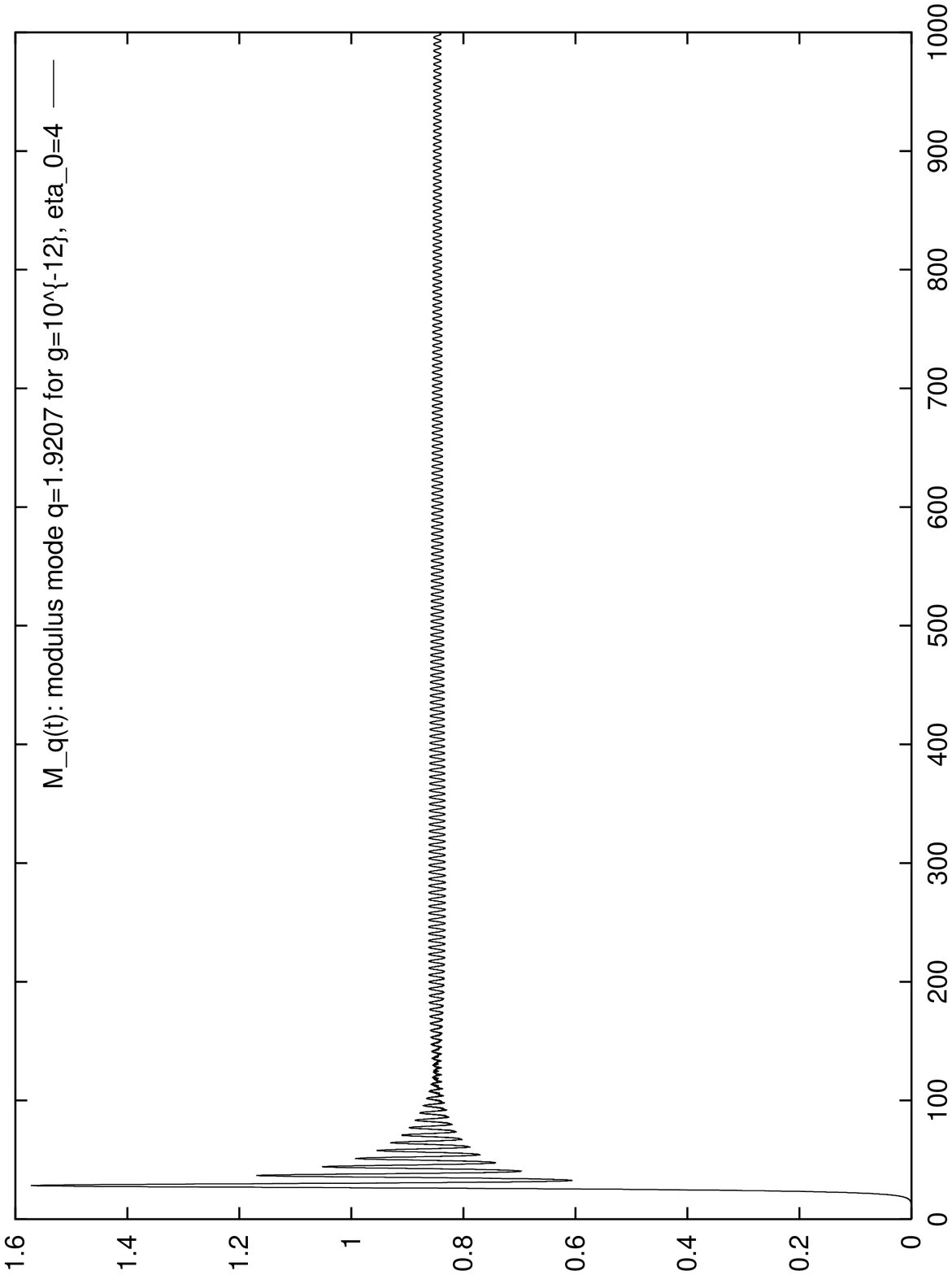}}

\figure{{\bf Figure 10a:} 

The modulus $ M_q(\tau) $ of the mode function $
\varphi_{q=1.9207}(\tau) $ as a 
function of time for $ \eta_0=4 $, $ g=10^{-12} $.
For times later than $ \tau_1 = 26.85\ldots $, 
this mode oscillates with stationary
amplitude. It lies outside both nonlinear resonance bands.
\label{fig10a}}

\clearpage

\hbox{\epsfxsize 14cm\epsffile{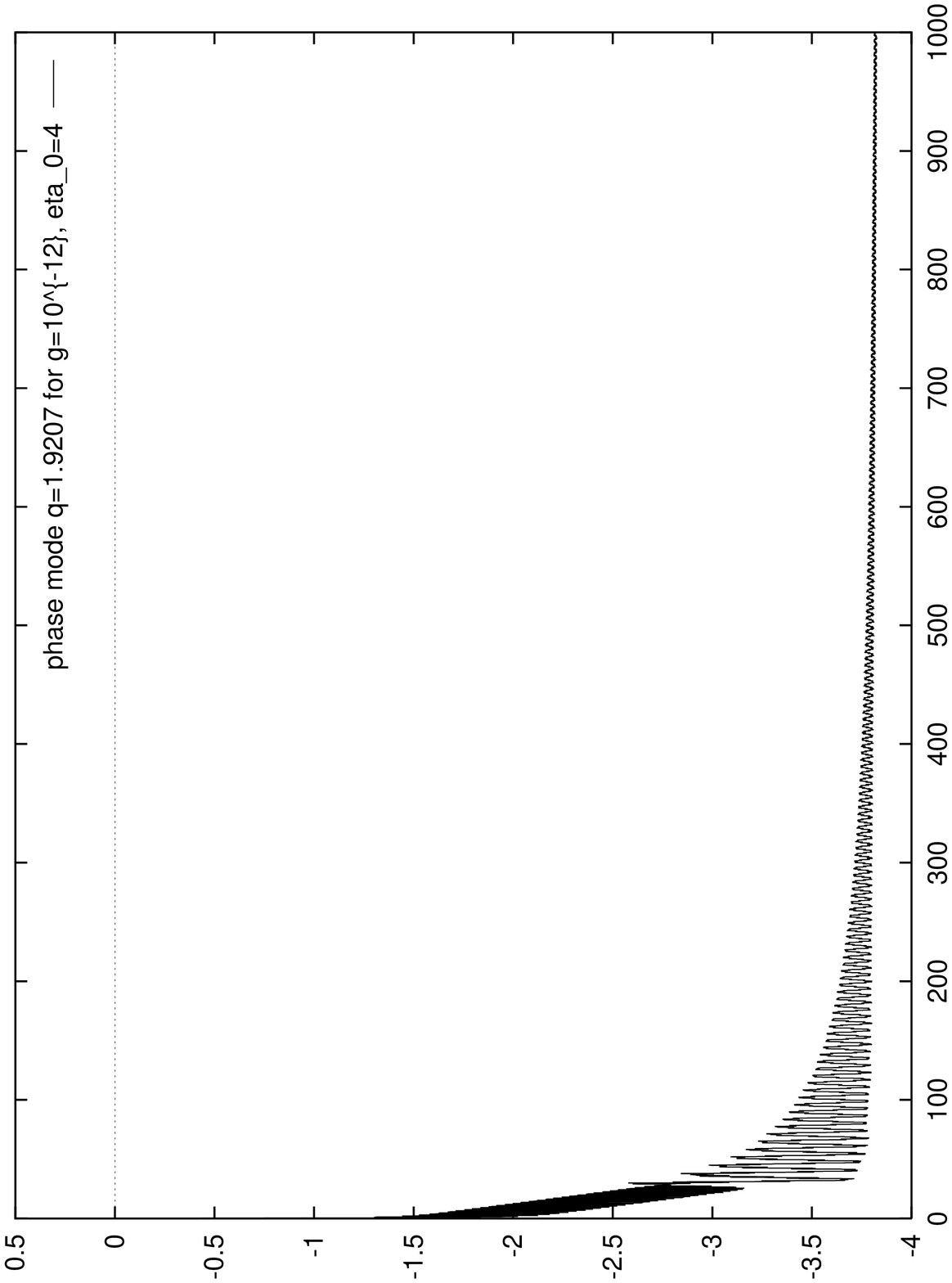}}

\figure{{\bf Figure 10b:} 

The phase $ \phi_q(\tau) $ of the mode function $
\varphi_{q=1.9207}(\tau) $ as a 
function of time for $ \eta_0=4 $, $ g=10^{-12} $.
\label{fig10b}}

\clearpage

\hbox{\epsfxsize 14cm\epsffile{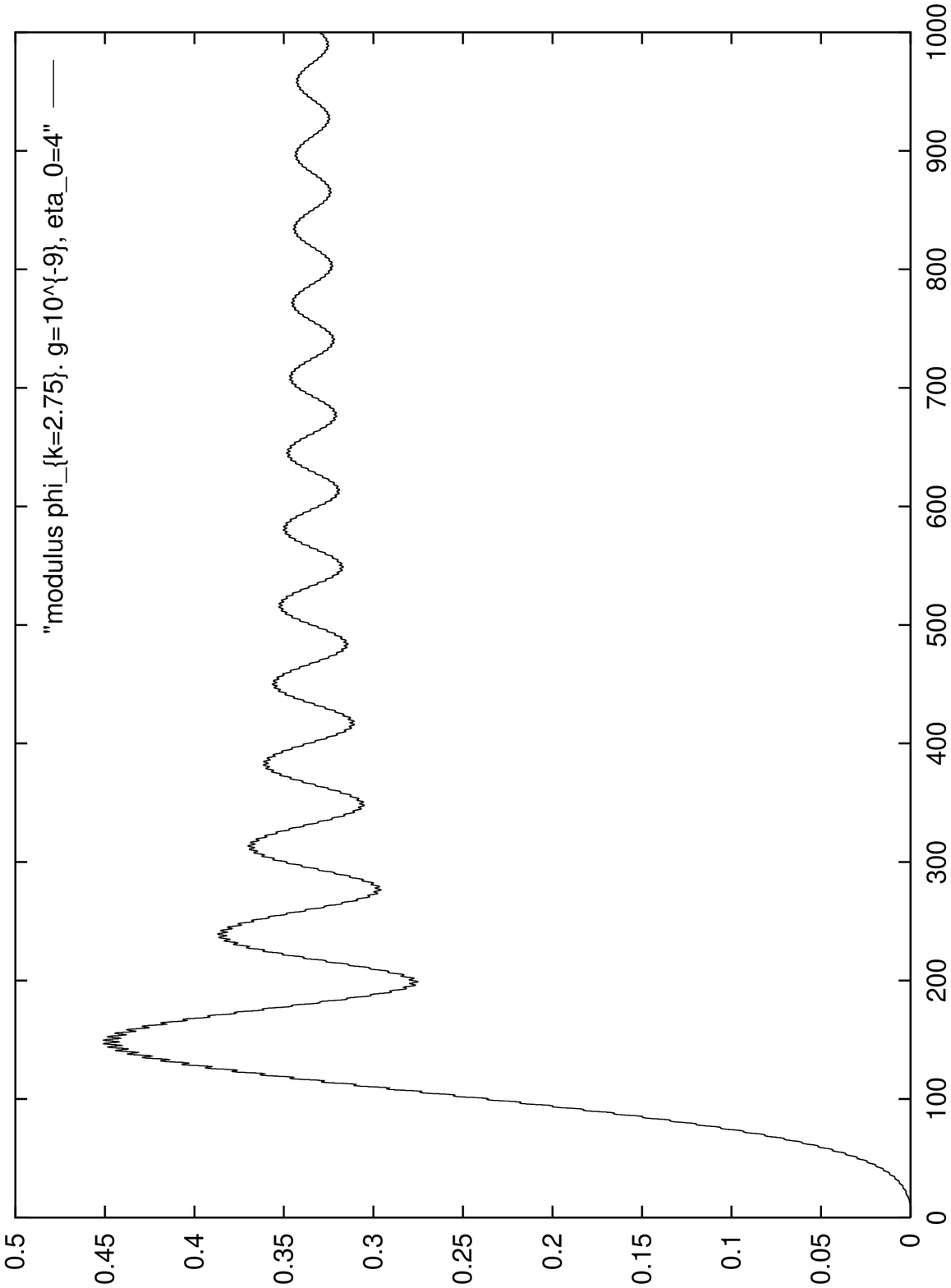}}

\figure{{\bf Figure 11a:} 

The modulus $ M_q(\tau) $ of the mode function $
\varphi_{q=2.75}(\tau) $ as a 
function of time for $ \eta_0=4 $, $ g=10^{-9} $.
This function grows till $ q = 2.75 $ gets out of the second nonlinear
resonance band. The estimate (\ref{q1tau}) ($\tau\sim 140. $) is in very good
agreement with the numerical results plotted here. 
\label{fig11a}}

\clearpage

\hbox{\epsfxsize 14cm\epsffile{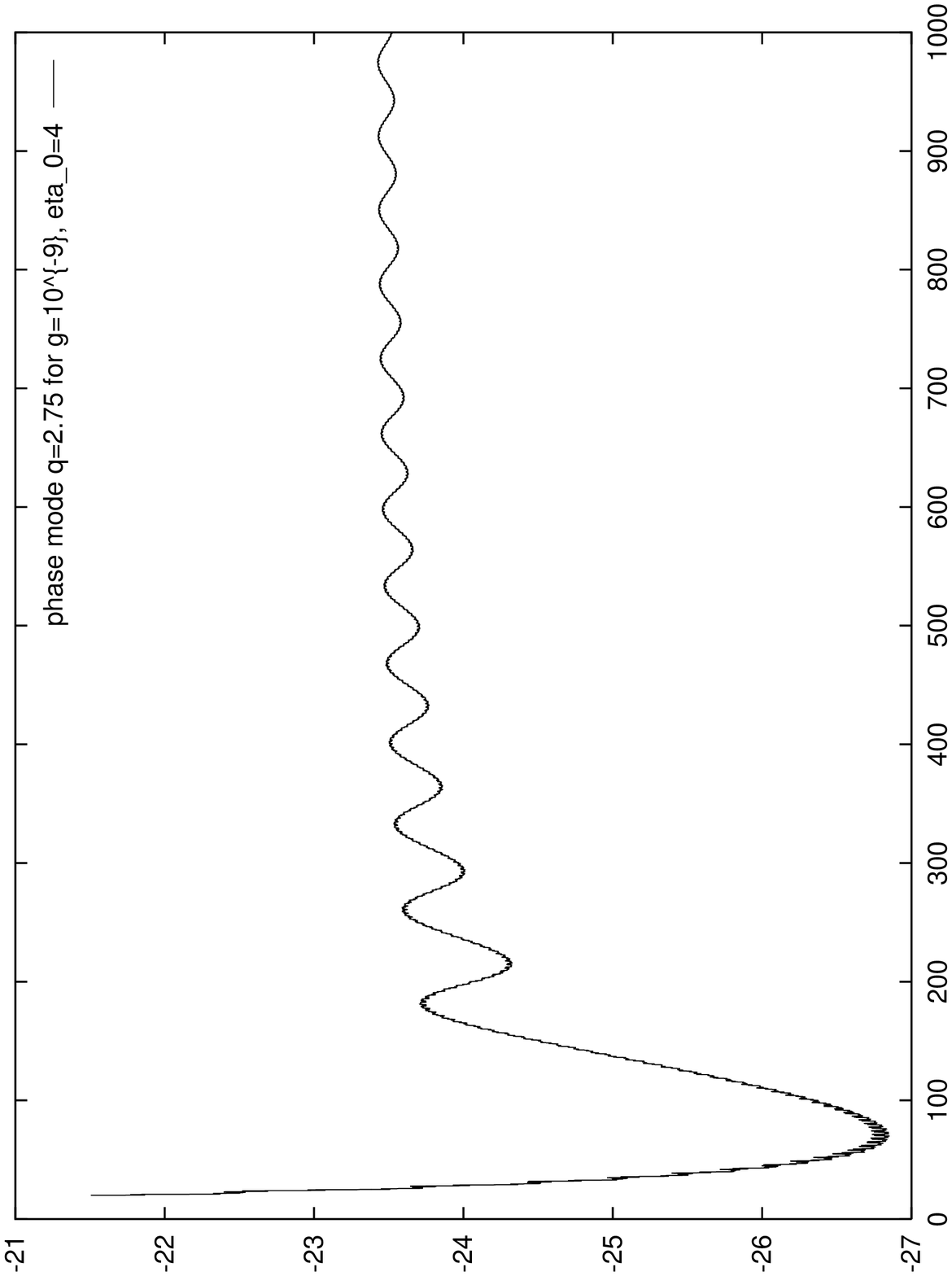}}

\figure{{\bf Figure 11b:} 

The phase $ \phi_q(\tau) $ of the mode function $
\varphi_{q=2.75}(\tau) $ as a 
function of time for $ \eta_0=4 $, $ g=10^{-9} $.
This phase changes its behaviour (around $\tau \sim 140. $ ) when 
the modulus $ M_{q=2.75}(\tau) $ ceases to grow (see fig. 11a).

\label{fig11b}}

\clearpage

\hbox{\epsfxsize 14cm\epsffile{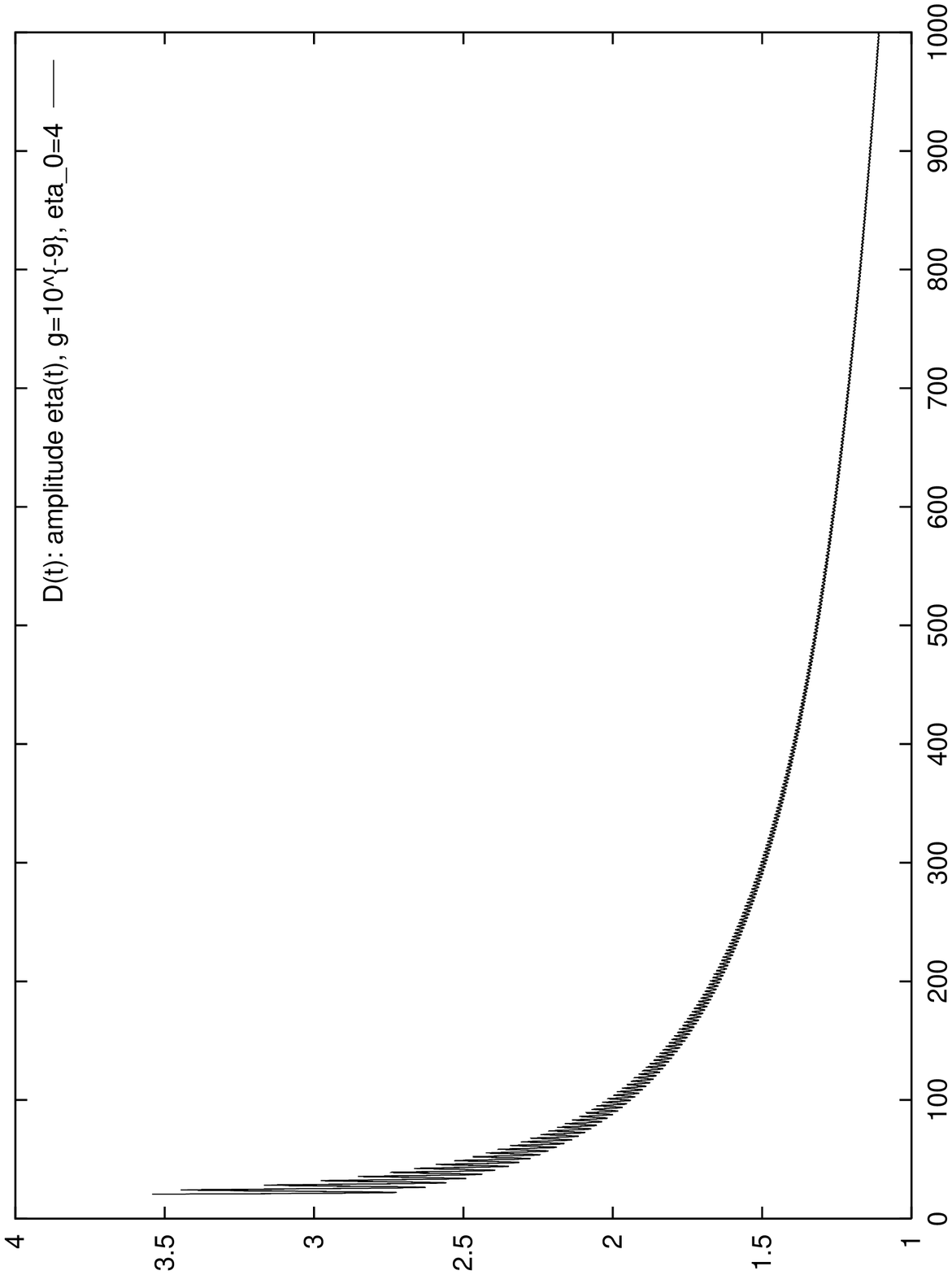}}

\figure{{\bf Figure 12a:} 

The amplitude $ D(\tau) $ of the zero mode  $ \eta(\tau) $ as a
function of time for $ \eta_0=4 $, $ g=10^{-9} $. This function
exhibits a power like decrease according to eq.(\ref{dtau}).
\label{fig12a}}

\clearpage

\hbox{\epsfxsize 14cm\epsffile{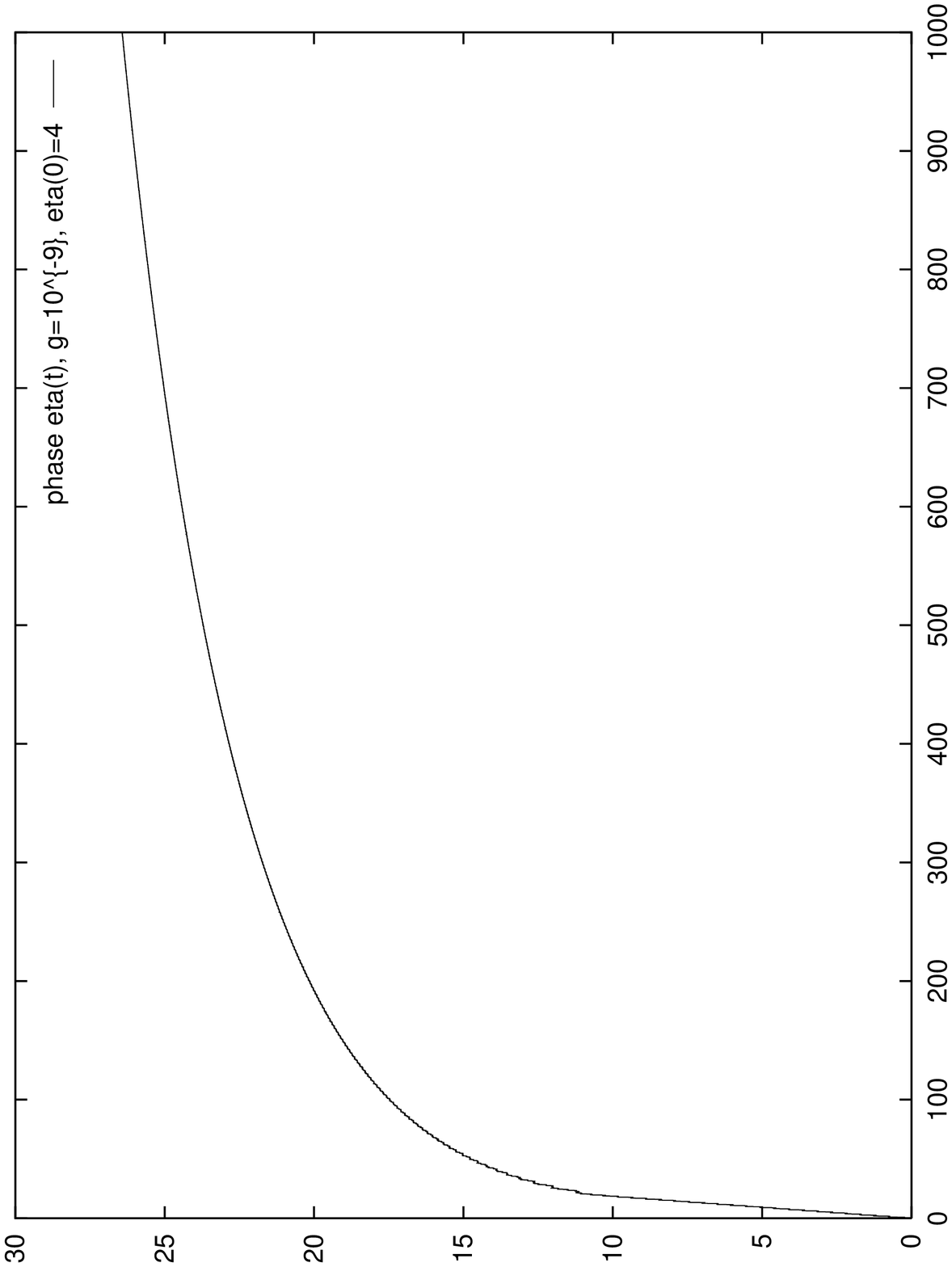}}

\figure{{\bf Figure 12b:} 

The phase $ \phi(\tau) $ of the zero mode  $ \eta(\tau) $ as a
function of time for $ \eta_0=4 $, $ g=10^{-9} $. This function
exhibits a logarithmic behavior according to  eq.(\ref{fitau}).
\label{fig12b}}

\clearpage

\hbox{\epsfxsize 14cm\epsffile{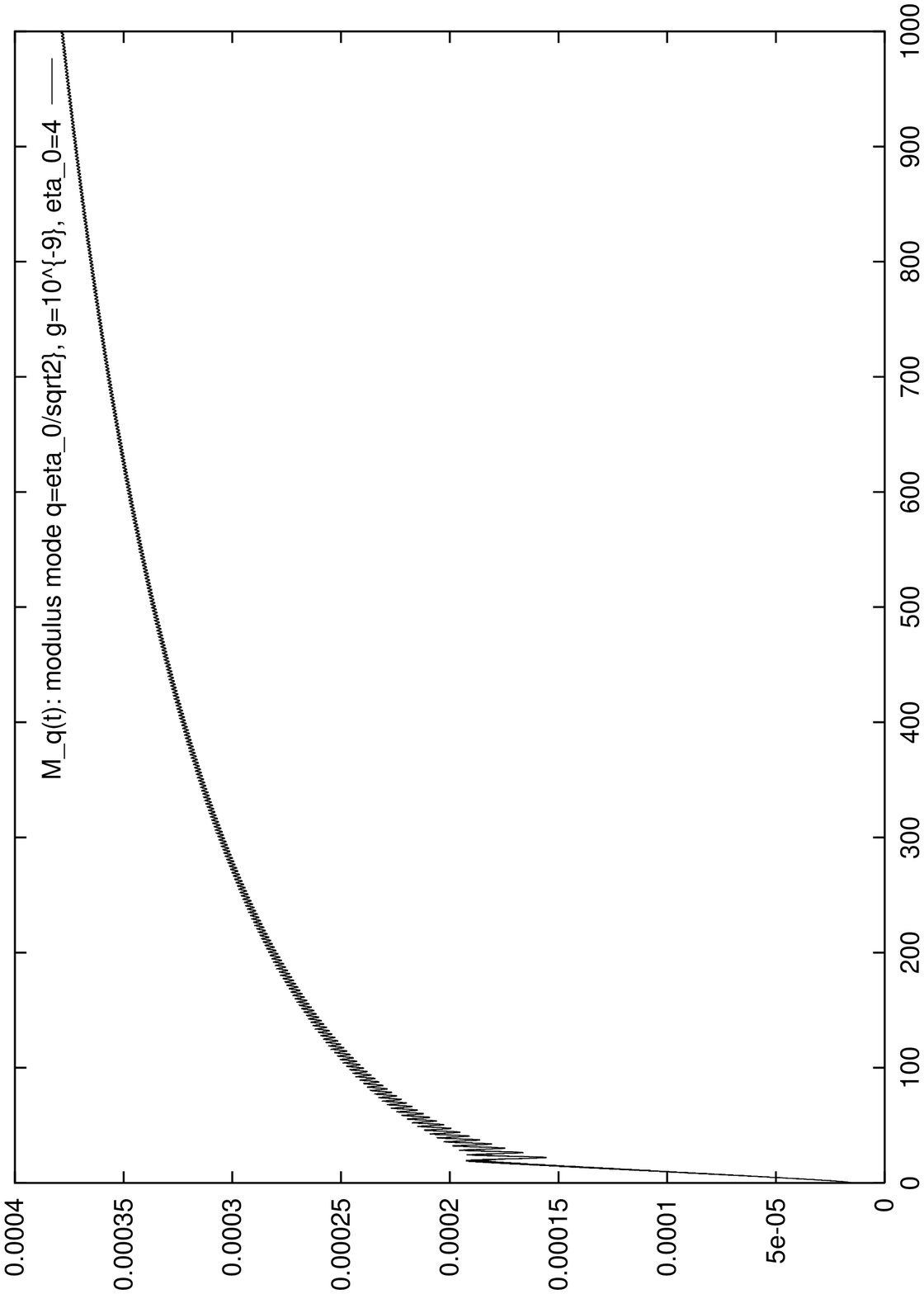}}

\figure{{\bf Figure 13a:} 

The amplitude  $ M_q(\tau) $ of the mode function $
\varphi_{q=\eta_0/\sqrt2}(\tau) $ as a function of time for 
$ \eta_0=4 $, $ g=10^{-9} $. [This corresponds to the upper border of
the forbidden band]. This  function
exhibits a power like increase according to eq.(\ref{moder2}).

\label{fig13a}}

\clearpage

\hbox{\epsfxsize 14cm\epsffile{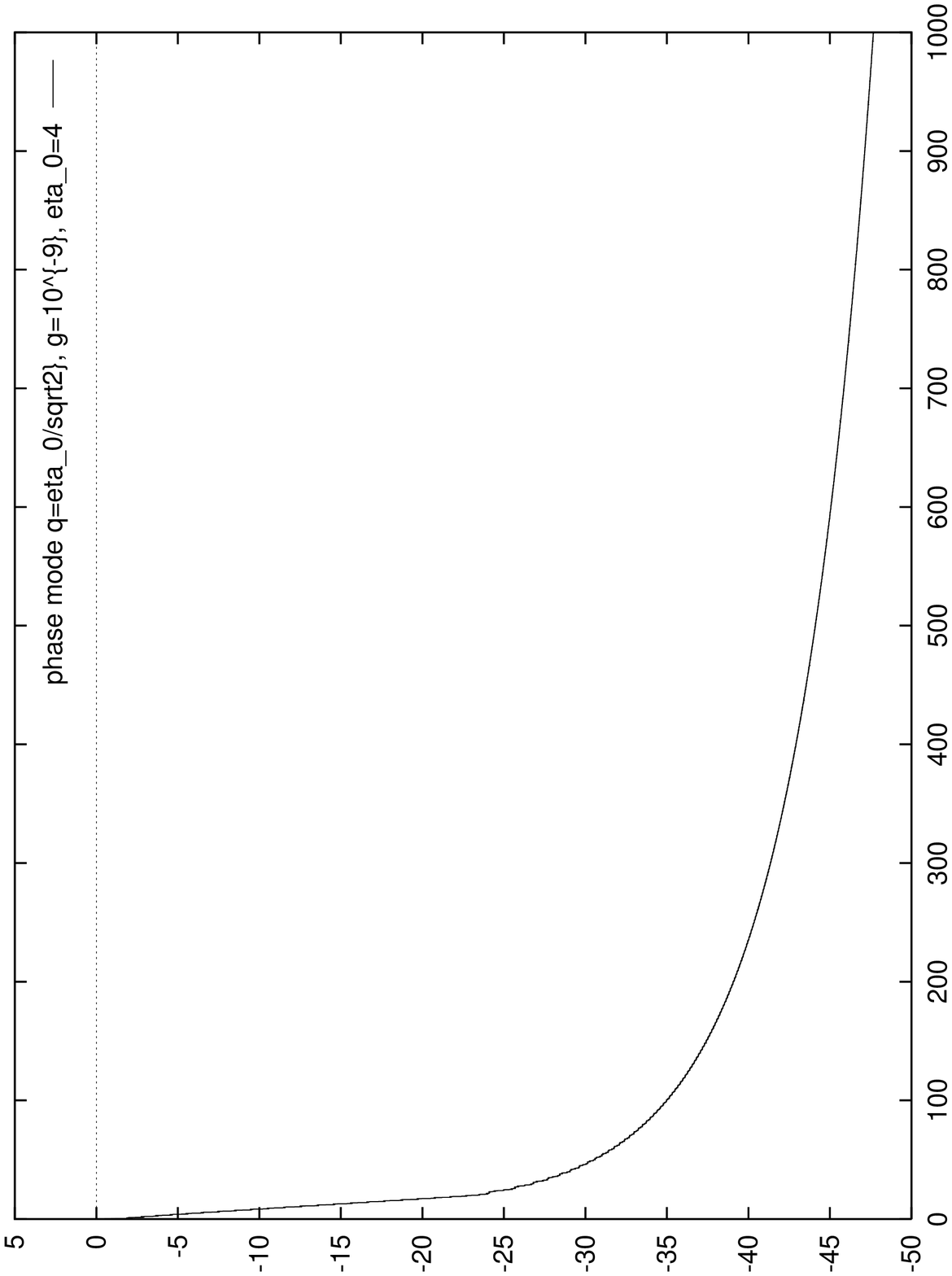}}

\figure{{\bf Figure 13b:} 

The phase $ \phi_q(\tau) $ of the mode function $
\varphi_{q=\eta_0/\sqrt2}(\tau) $ as a function of time for 
$ \eta_0=4 $, $ g=10^{-9} $. [This corresponds to the upper border of
the forbidden band]. 
This phase
exhibits a logarithmic behavior according to  eq.(\ref{moder2}).
\label{fig13b}}

\clearpage

\hbox{\epsfxsize 14cm\epsffile{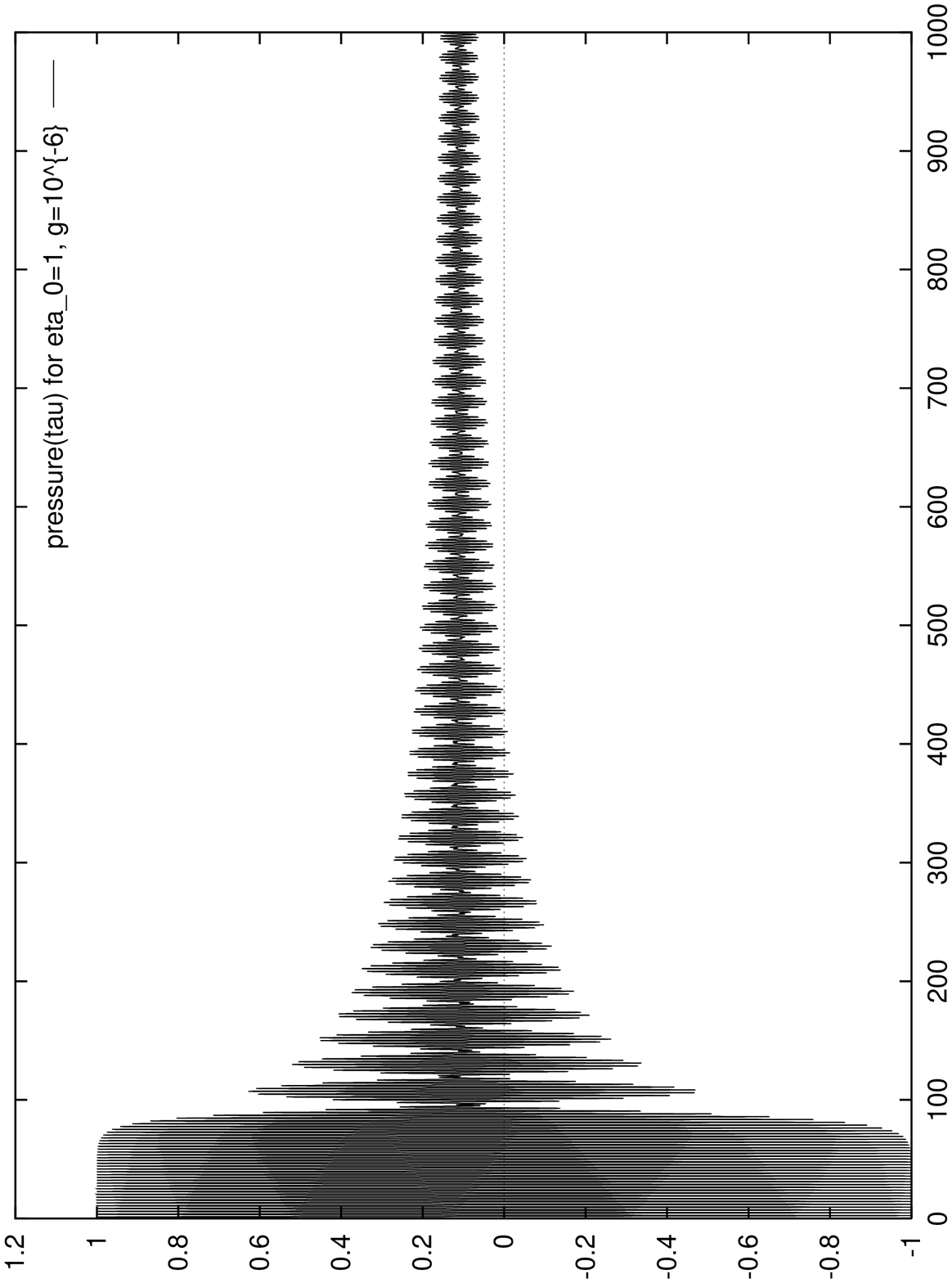}}

\figure{{\bf Figure 14a:} 

The pressure divided by the initial energy as a function of time for 
$ \eta_0=1 $, $ g=10^{-6} $. Notice the asymptotic value $
{{p(\infty)}\over {\varepsilon}} \approx { 1
\over {3 \;\left( 1 + \frac2{\eta_0^2}\right)}} = 1/9
$. \label{fig14a}}

\clearpage

\hbox{\epsfxsize 14cm\epsffile{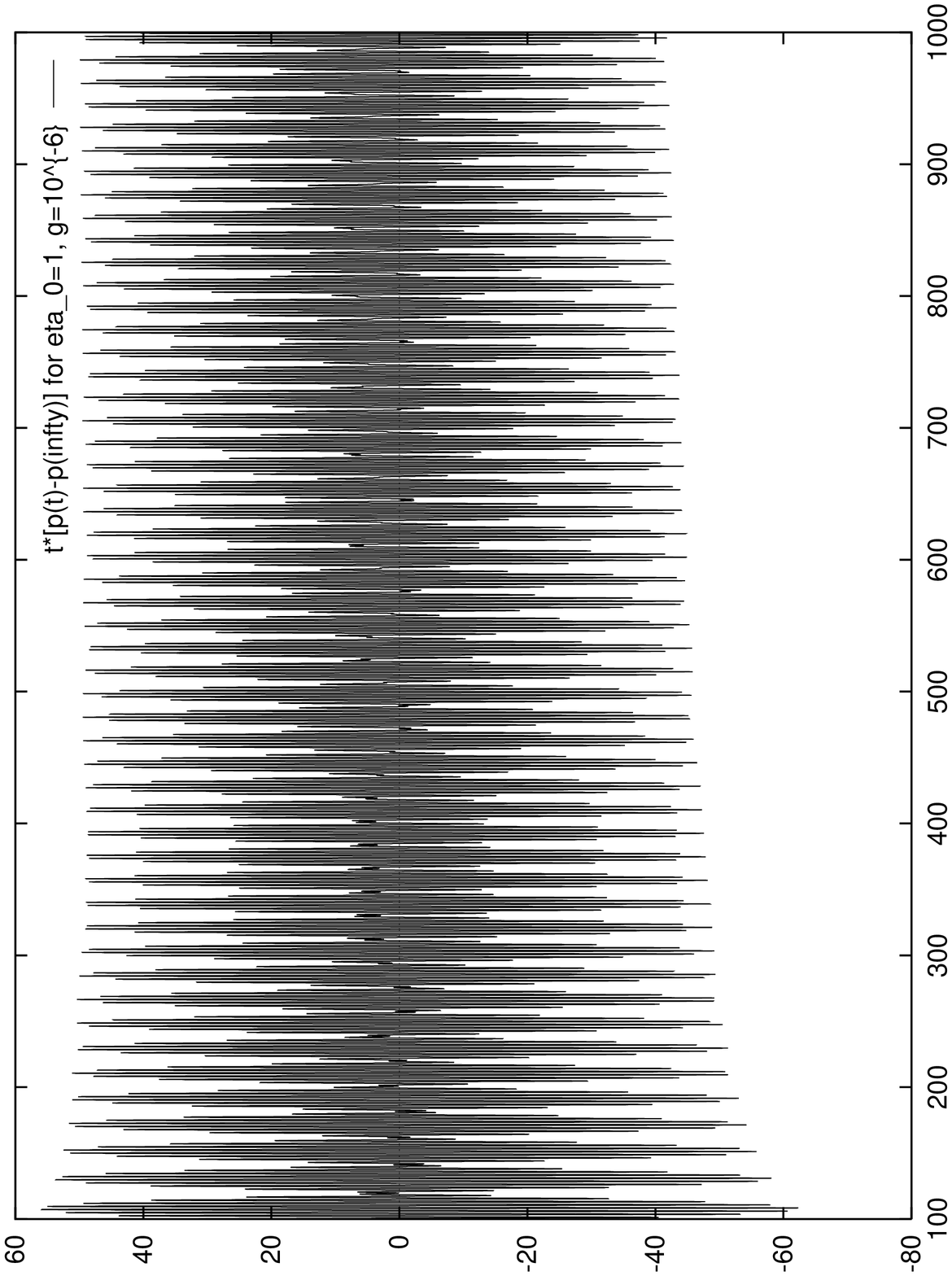}}

\figure{{\bf Figure 14b:} 

The pressure minus its value at $ \tau = \infty $ times $ \tau $
as a function of time for $ \eta_0=1 $, $ g=10^{-6} $. 
This function oscillates in time with constant amplitude and frequencies 
$ 2 \, {\cal M}_{\infty} = 2 \sqrt{1 + \eta_0^2/2} $ and $ 2 \,  {\cal
M}_0 = 2 \sqrt{1 + \eta_0^2}$  
\label{fig14b}}

\clearpage 

\hbox{\epsfxsize 14cm\epsffile{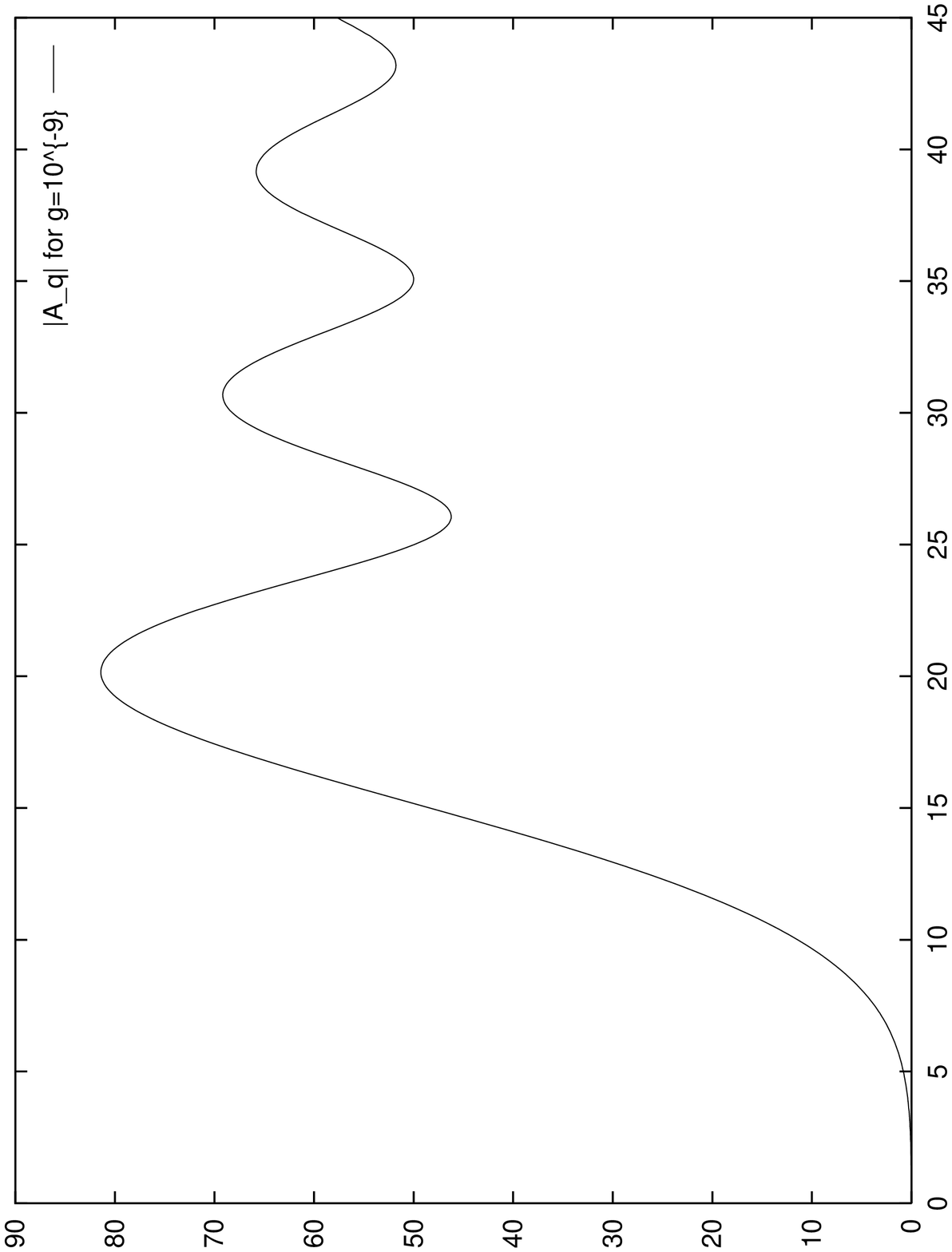}}            

\figure{{\bf Figure 15a:}

Modulus of the function $ {\cal A}_q(T_1) $ obtained through multi-time scale
analysis as a function of $ q^2 \tau /  {\cal M}_{\infty} $ . We give
in eq.(\ref{whit}) its expression in terms of Whittaker  functions.
Notice the resemblance with the full numerical solutions displayed
in figs. 8c and 8d.

\label{fig15a}}

\clearpage

\hbox{\epsfxsize 14cm\epsffile{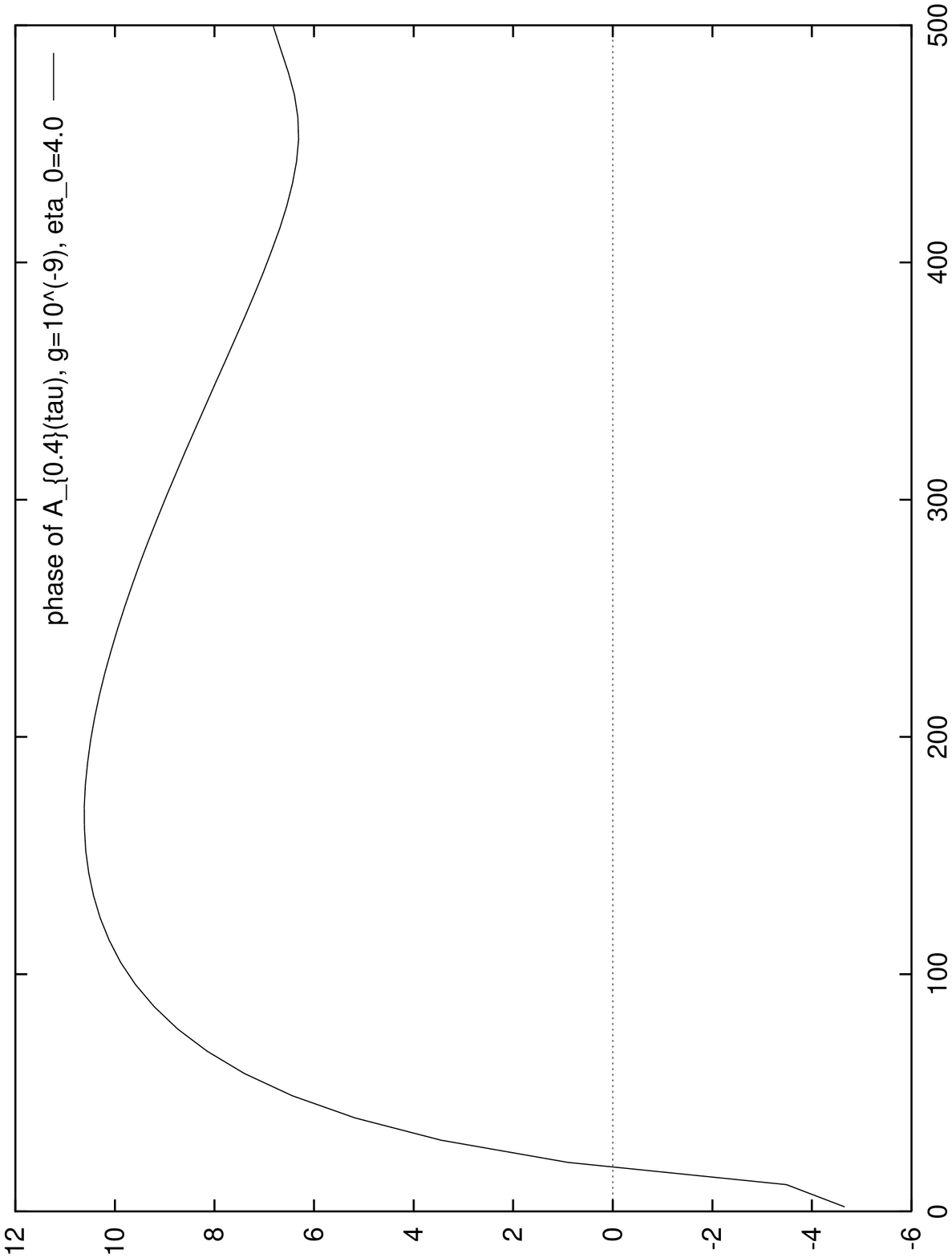}} 

\figure{{\bf Figure 15b:}

Phase of the function $ {\cal A}_q(T_1) $ obtained through multi-time scale
analysis as a function of $ q^2 \tau /  {\cal M}_{\infty} $ . We give
in eq.(\ref{whit}) its expression in terms of Whittaker  functions.
Notice the resemblance with the full numerical solutions displayed
in figs. 9c and 9d.

\label{fig15b}}

\clearpage

\hbox{\epsfxsize 14cm\epsffile{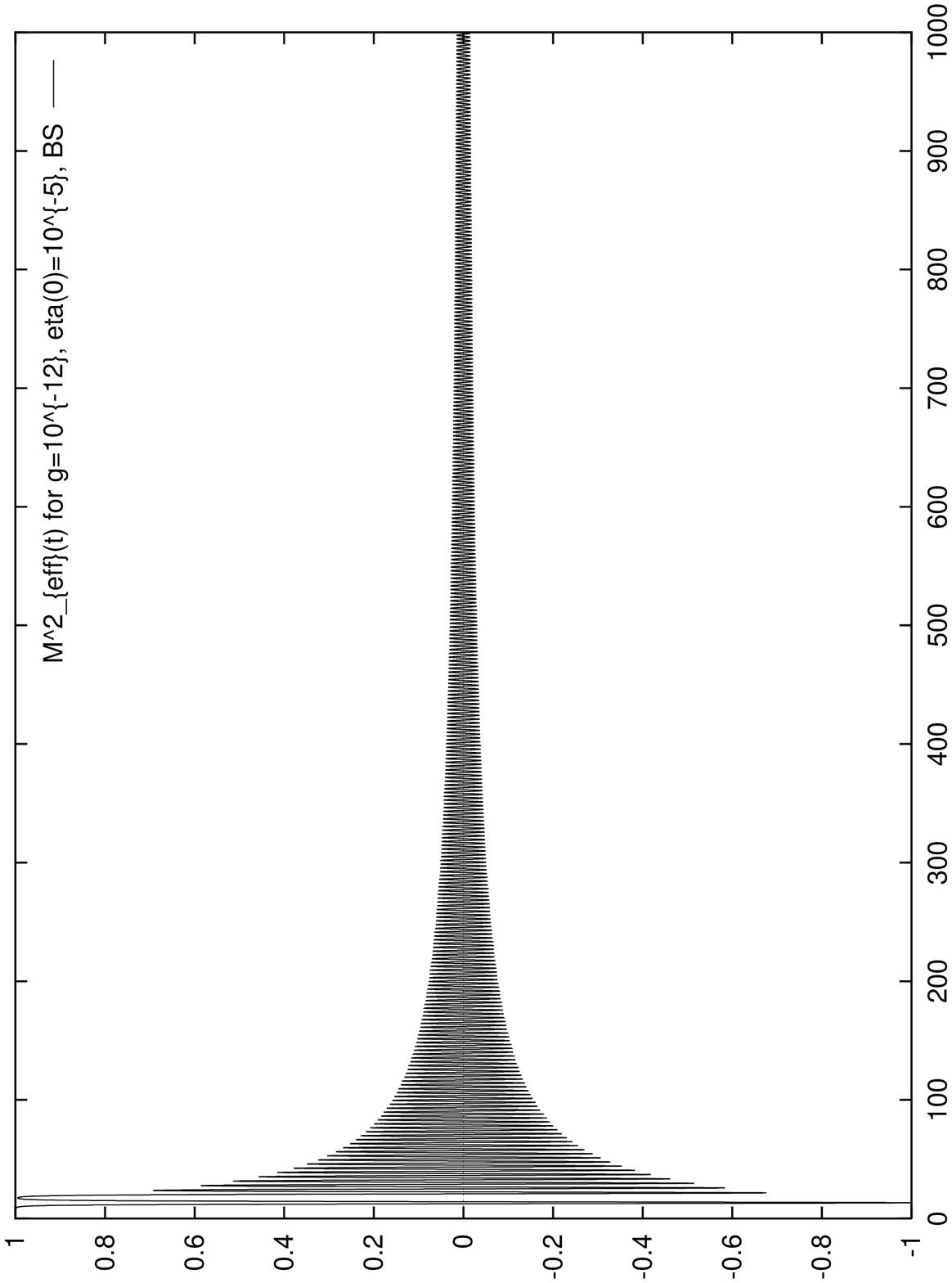}}

\figure{{\bf Figure 16:}

The effective mass squared as a function of time for broken symmetry
$ \eta_0=10^{-5} $, $ g=10^{-12} $.
This function oscillates in time as decribed by  eq.(\ref{masinR}).

\label{fig16}}

\clearpage

\hbox{\epsfxsize 14cm\epsffile{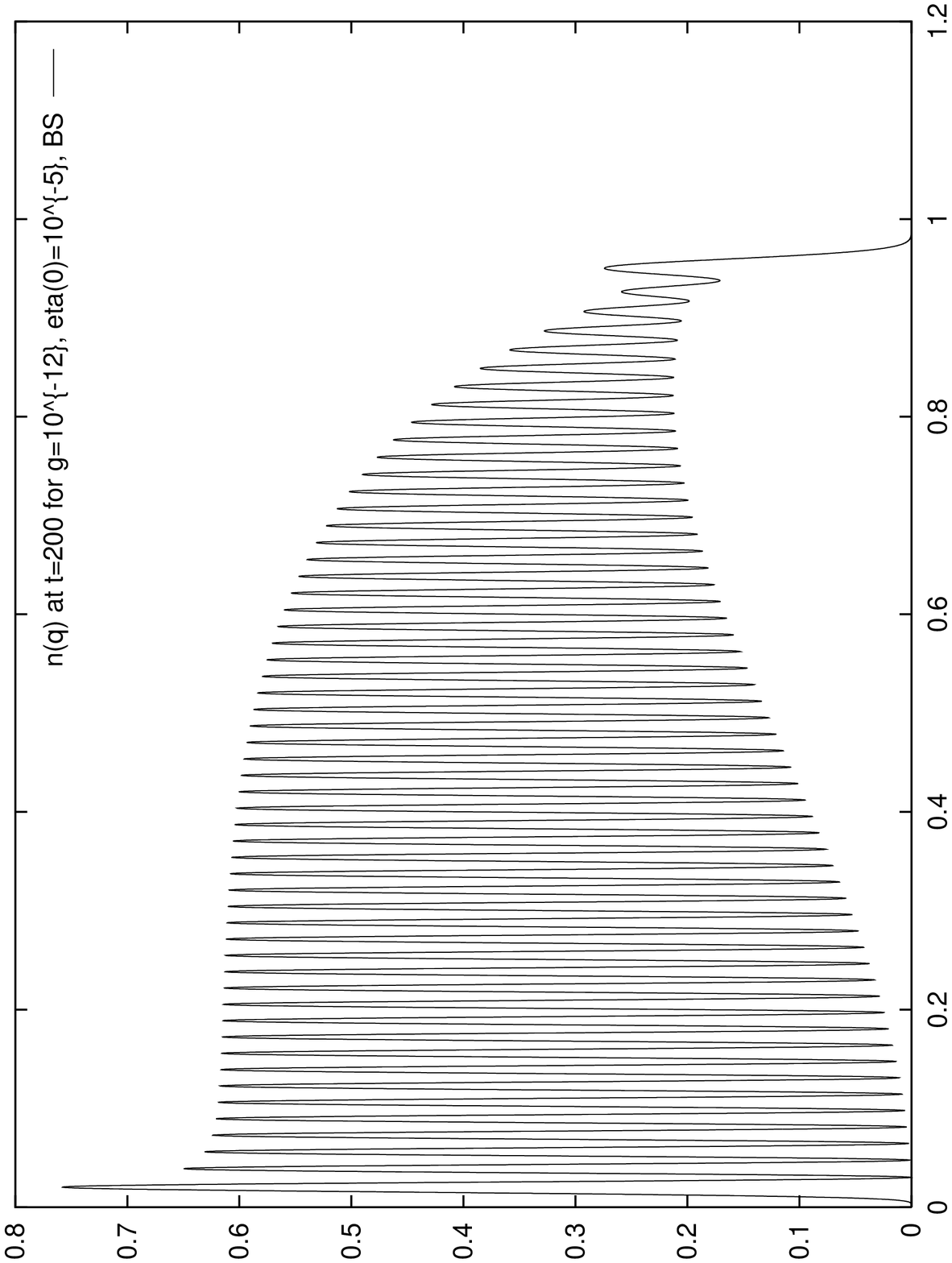}}

\figure{{\bf Figure 17:}

Momentum distribution of the produced particles at $ \tau = 200 $ for
broken symmetry$ \eta_0=10^{-5} $, $ g=10^{-12} $.
Notice the main peak at $ q = 0.020 $ and the secondary peak
at $ q = 0.946 $ associated to the non-linear resonance at $
q=1$. The position of both peaks are correctly  estimated
by eqs.(\ref{estpicoR}) and (\ref{estpico2}), respectively.
\label{fig17}}

\clearpage









\hbox{\epsfxsize 14cm\epsffile{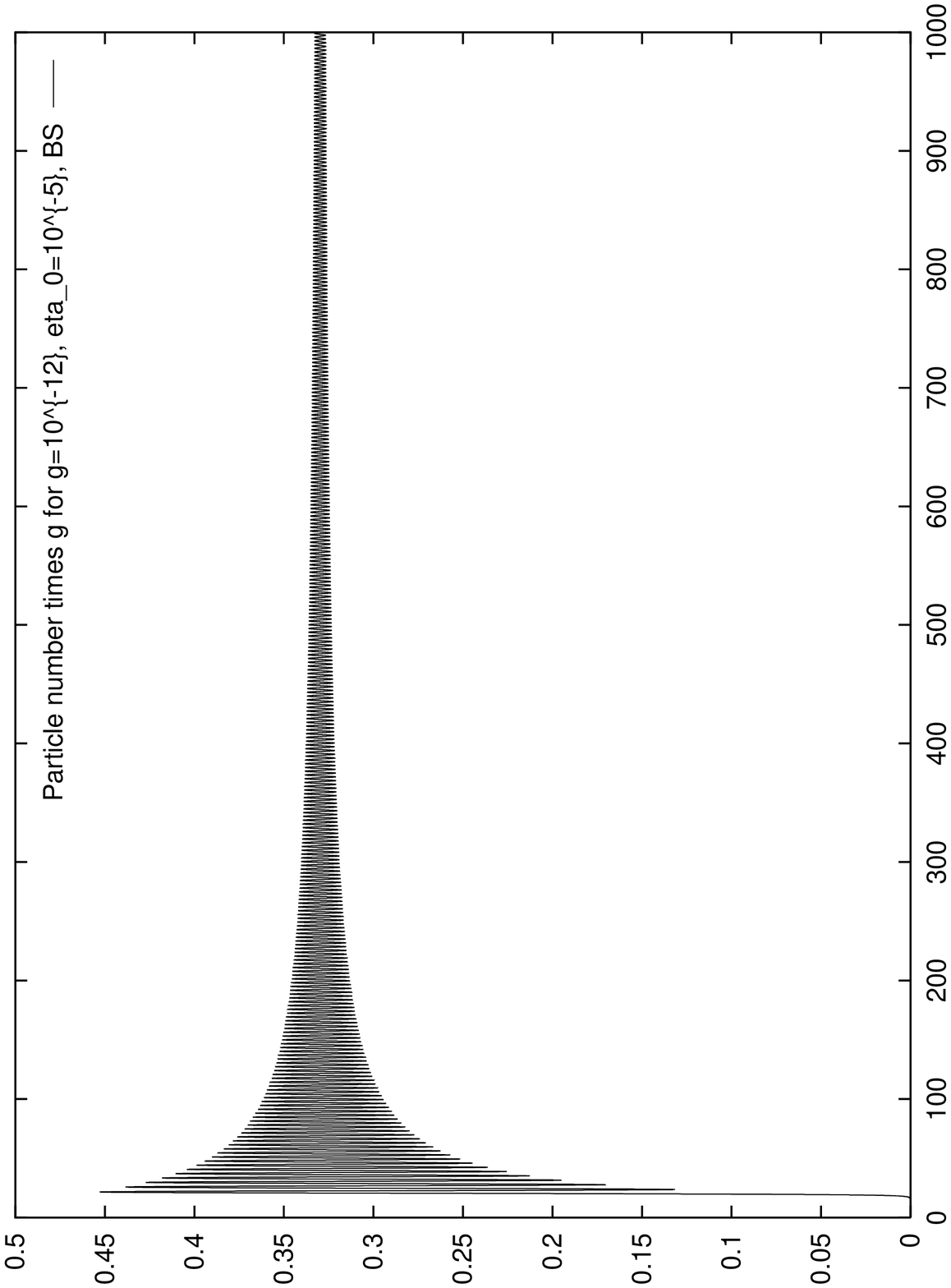}}

\figure{{\bf Figure 18:} 

The total occupation number times $g$ per  volume $ m^{-3} $
as a function of time for
broken symmetry,  $ \eta_0=10^{-5} $, $ g=10^{-12} $. Its limiting
value is $ g {\cal N}(\infty) \approx 0.33 $.
\label{fig18}}

\clearpage

\hbox{\epsfxsize 14cm\epsffile{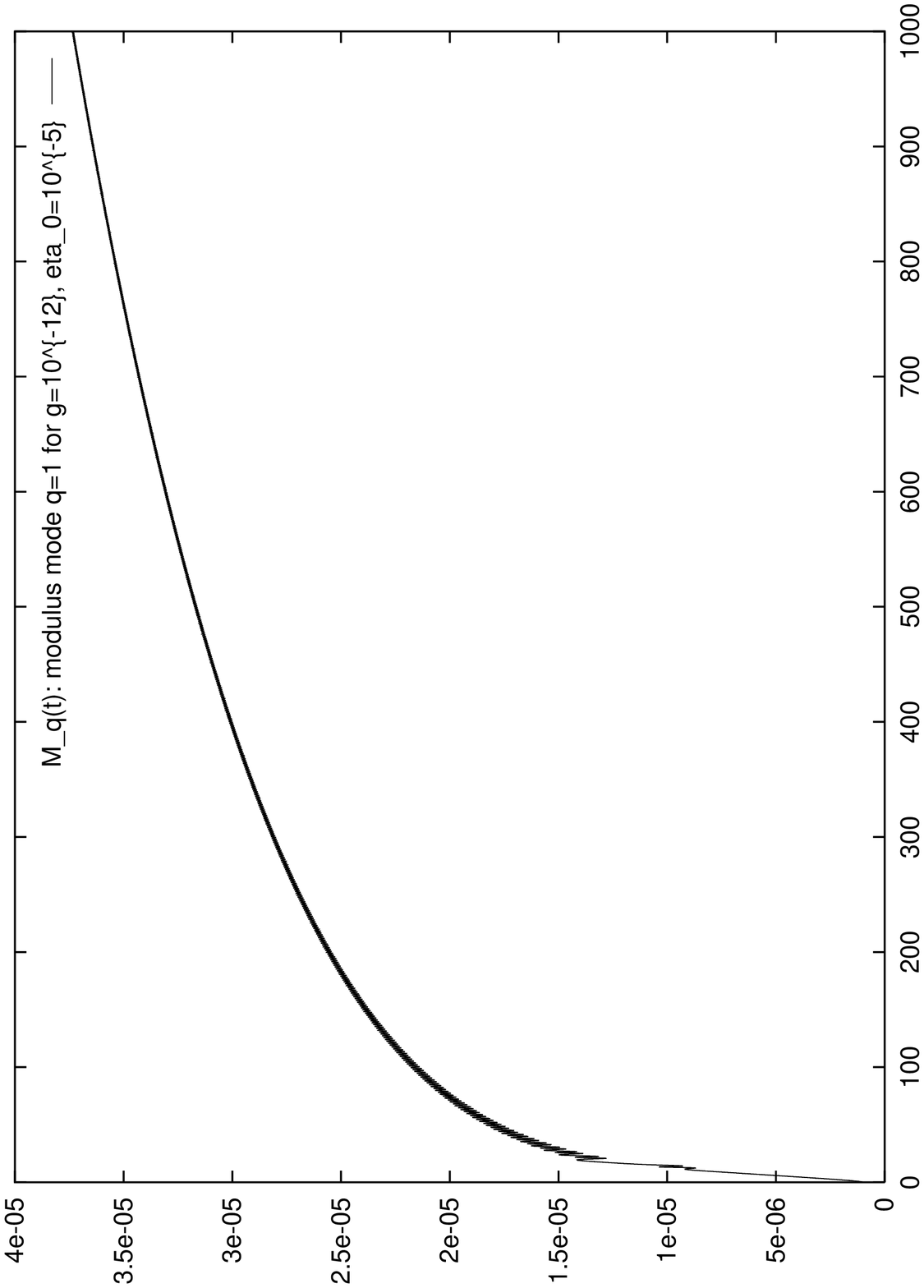}}

\figure{{\bf Figure 19a:} 

The amplitude  $ M_q(\tau) $ of the mode function $ \varphi_{q=1}(\tau) $ as a
function of time for broken symmetry, $ \eta_0=10^{-5} $, $ g=10^{-12} $.
It grows as a power according to eq.(\ref{modoq1}).
\label{fig19a}}

\clearpage

\hbox{\epsfxsize 14cm\epsffile{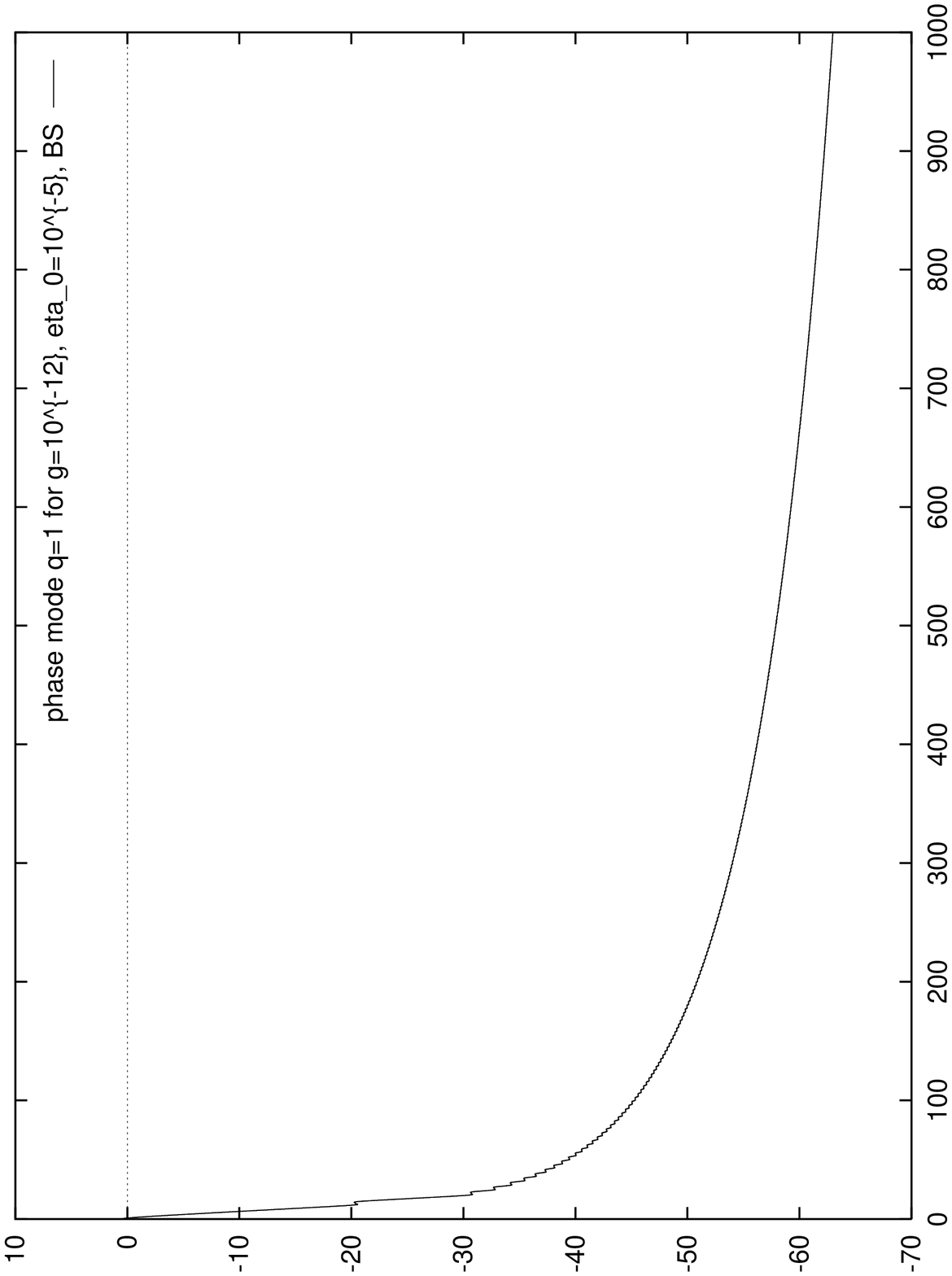}}

\figure{{\bf Figure 19b:} 

The phase $ \phi_q(\tau) $ of the mode function $ \varphi_{q=1}(\tau) $ as a
function of time for broken symmetry, $ \eta_0=10^{-5} $, $ g=10^{-12} $.
This function follows eq.(\ref{modoq1}) with a very good approximation.
\label{fig19b}}

\clearpage

\hbox{\epsfxsize 14cm\epsffile{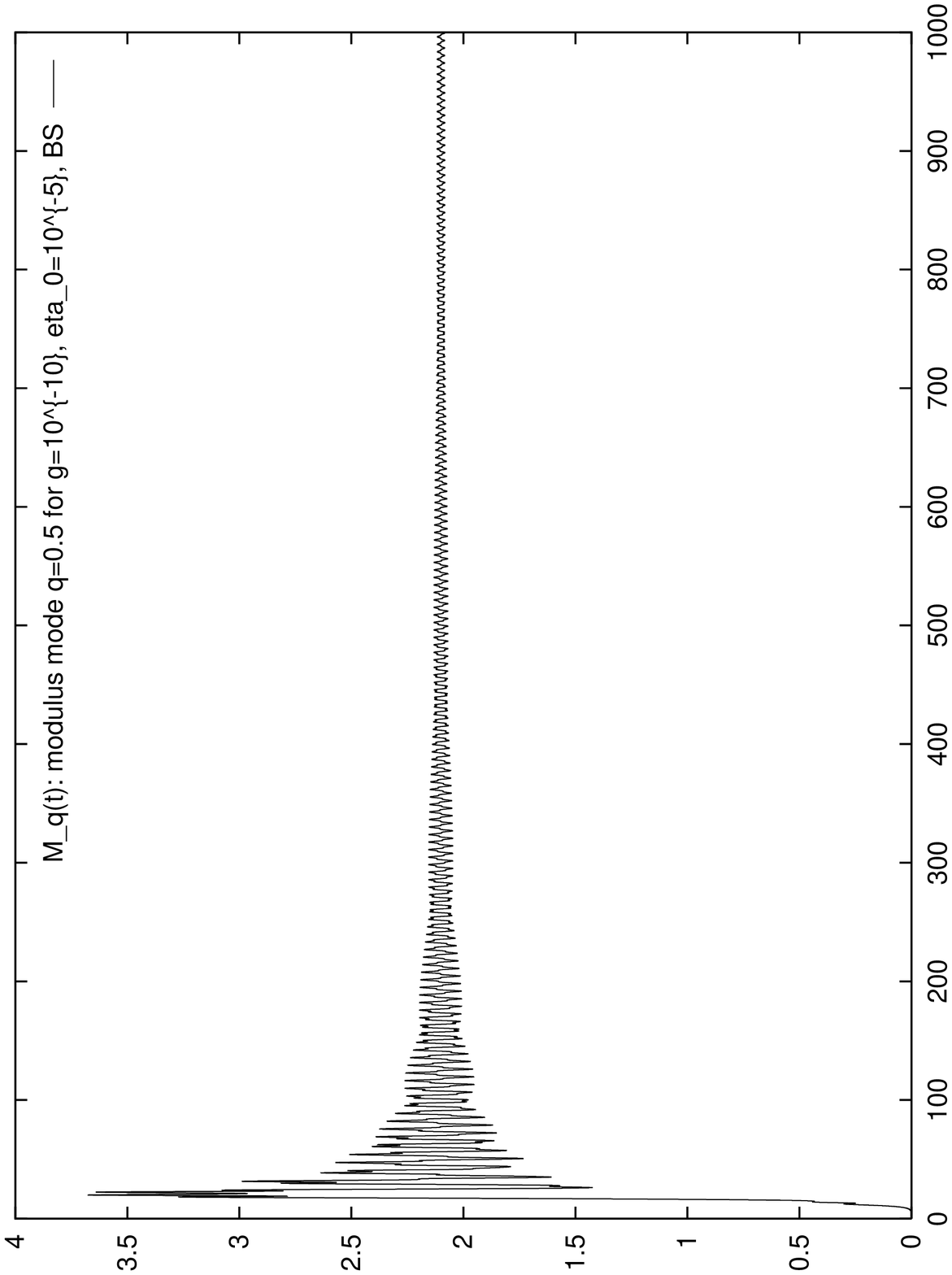}}

\figure{{\bf Figure 20a:}

The modulus $ M_q(\tau) $ of the mode function $
\varphi_{q=0.5}(\tau) $ as a 
function of time for  $ g=10^{-10} ,\; \eta_0=10^{-5} $, broken symmetry.
For times later than $ \tau_1 = 14.14\ldots $, 
this mode oscillates with stationary
amplitude. It lies outside the nonlinear resonance.
\label{fig20a}}

\clearpage

\hbox{\epsfxsize 14cm\epsffile{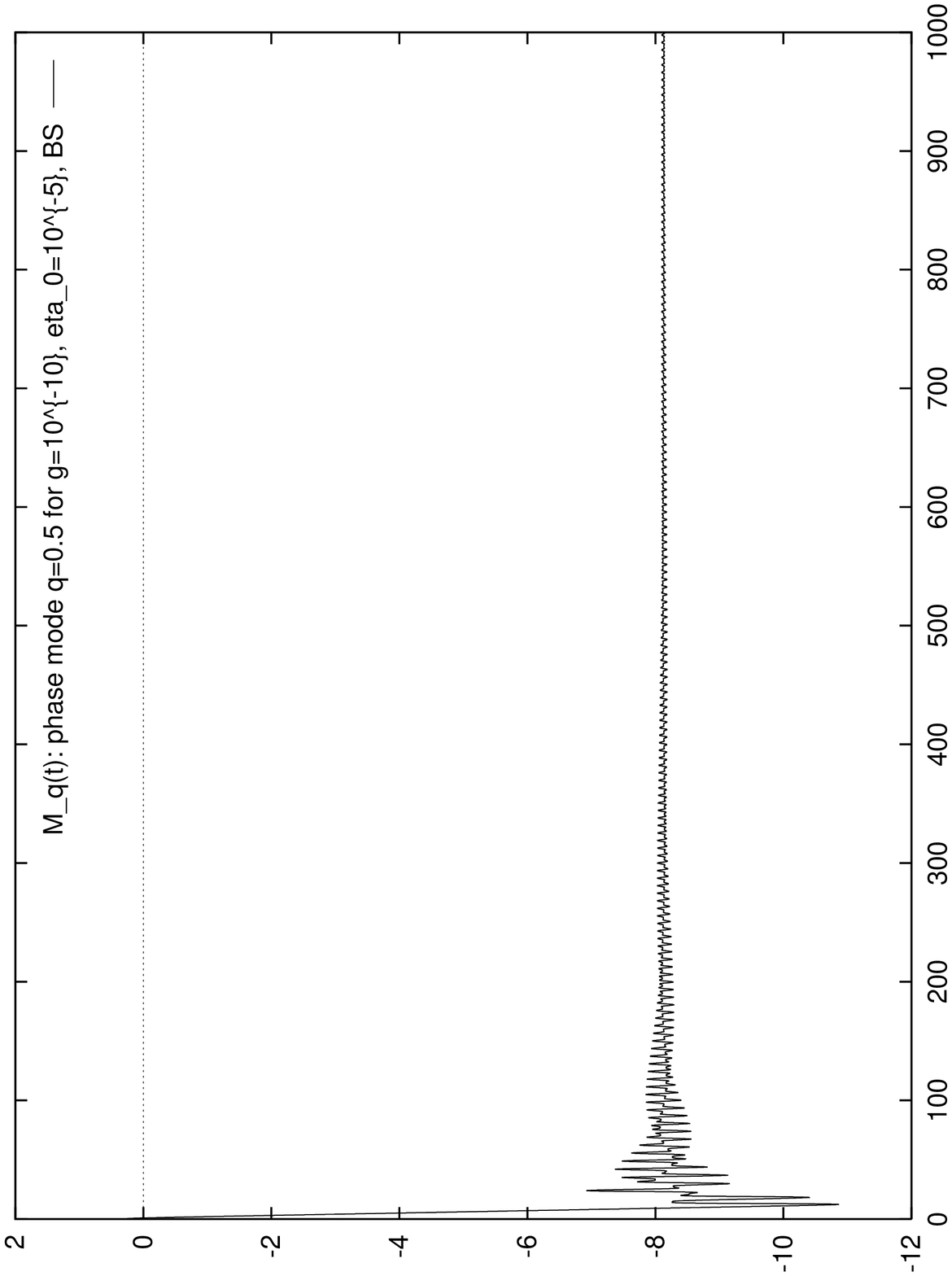}}

\figure{{\bf Figure 20b:}

The phase $ \phi_q(\tau) $ of the mode function $
\varphi_{q=0.5}(\tau) $ as a function of time
for  $ g=10^{-10} ,\; \eta_0=10^{-5} $, broken symmetry.

\end{document}

\end{document}